\documentclass[10pt,final,journal,twocolumn,singlespaced]{IEEEtran}
\usepackage{algorithm}
\usepackage{algpseudocode}
\usepackage{amsmath}
\usepackage{amssymb}
\usepackage{bm}
\usepackage{booktabs}
\usepackage{footmisc}
\usepackage{color}
\usepackage{xcolor}
\usepackage{cite}
\usepackage{url}
\usepackage{verbatim}
\usepackage{epsfig}
\usepackage{epstopdf}
\usepackage{flafter}
\usepackage{float}
\usepackage{ifthen}
\usepackage{multirow}
\usepackage{enumitem}
\usepackage{makecell} 
\usepackage{pifont}
\usepackage{subfigure}
\usepackage{times}
\usepackage{threeparttable}
\usepackage{url}  
\usepackage{subfigure,graphicx}
\usepackage{float}
\usepackage{color,multirow}
\usepackage{makecell}
\usepackage{cite} 
\usepackage{url}  
\usepackage{ifthen}  
\usepackage[T1]{fontenc}
\usepackage{graphicx}
\usepackage{ulem}
\usepackage[colorlinks,linkcolor=red,anchorcolor=gray,citecolor=green,urlcolor=blue]{hyperref}
\usepackage{graphics}
\usepackage{multirow}
\usepackage{makecell}
\usepackage{booktabs}
\usepackage{bigstrut}
\usepackage{url}
\usepackage{multirow} 
\usepackage{tabularx} 
\usepackage{booktabs} 
\usepackage{amsmath}  
\usepackage{tabu}
\usepackage{verbatim}
\usepackage{xcolor}
\definecolor{orange}{RGB}{255,107,0}

\usepackage{threeparttable}
\usepackage{footmisc} 
\usepackage{bigstrut}

\renewcommand{\algorithmicensure}{\textbf{Output:}} 

\begin{document}
	\title{Hyperspectral Denoising Using Unsupervised Disentangled Spatio-Spectral Deep Priors
		\thanks{\IEEEauthorrefmark{1}Corresponding authors: Xi-Le Zhao and Xiao Fu.}
		\thanks{This research of Xi-Le Zhao is supported by NSFC (No. 61876203, 61772003), the Key Project of Applied Basic Research in Sichuan Province (No. 2020YJ0216), the Applied Basic Research Project of Sichuan Province (No. 2021YJ0107) and National Key Research and Development Program of China (No. 2020YFA0714001). The work of X. Fu is supported by NSF ECCS-2024058 and NSF ECCS-1808159.}
		\thanks{Y.-C. Miao, X.-L. Zhao, and J.-L. Wang are with the Research Center for Image and Vision Computing, School of Mathematical Sciences, University of Electronic Science and Technology of China, Chengdu 611731, P.R.China (e-mails: szmyc1@163.com; xlzhao122003@163.com; wangjianli\_123@163.com).}
		\thanks{X. Fu is with the School of Electrical Engineering and Computer Science,
			Oregon State University (OSU), Corvallis, OR 97331, United States (e-mail: xiao.fu@oregonstate.edu).}
		\thanks{Y.-B. Zheng is with the School of Mathematical Sciences, University of Electronic Science and Technology of China, Chengdu 611731, China, and also with the Tensor Learning Team, RIKEN Center for Advanced Intelligence Project, Tokyo 103-0027, Japan (e-mail: zhengyubang@163.com).}
	}
	
	\author{Yu-Chun Miao,
		Xi-Le Zhao\IEEEauthorrefmark{1},
		Xiao Fu\IEEEauthorrefmark{1},
		Jian-Li Wang,
		and Yu-Bang Zheng}
	
	\maketitle 
	
	\begin{abstract}
		Image denoising is often empowered by accurate prior information. 
		In recent years, data-driven neural network priors have shown promising performance for RGB natural image denoising. Compared to classic handcrafted priors (e.g., sparsity and total variation), the ``deep priors'' are learned using a large number of training samples---which can accurately model the complex image generating process. However, data-driven priors are hard to acquire for hyperspectral images (HSIs) due to the lack of training data. A remedy is to use the so-called {\it unsupervised deep image prior} (DIP). Under the unsupervised DIP framework, it is hypothesized and empirically demonstrated that {\it proper} neural network {\it structures} are reasonable priors of certain types of images, and the network weights can be learned without training data. 
		Nonetheless, the most effective unsupervised DIP structures were proposed for natural images instead of HSIs. The performance of unsupervised DIP-based HSI denoising is limited by a couple of serious challenges, namely, network structure design and network complexity. This work puts forth an unsupervised DIP framework that is based on the classic spatio-spectral decomposition of HSIs. Utilizing the so-called {\it linear mixture model} of HSIs, two types of unsupervised DIPs, i.e., U-Net-like network and fully-connected networks, are employed to model the abundance maps and endmembers contained in the HSIs, respectively. This way, empirically validated unsupervised DIP structures for natural images can be easily incorporated for HSI denoising. 
		Besides, the decomposition also substantially reduces network complexity. 
		An efficient alternating optimization algorithm is proposed to handle the formulated denoising problem.
		Simulated and real data experiments are employed to showcase the effectiveness of the proposed approach.
		
	\end{abstract}
	
	\begin{figure*}[t!] 
		\centering
		\includegraphics[scale=0.47]{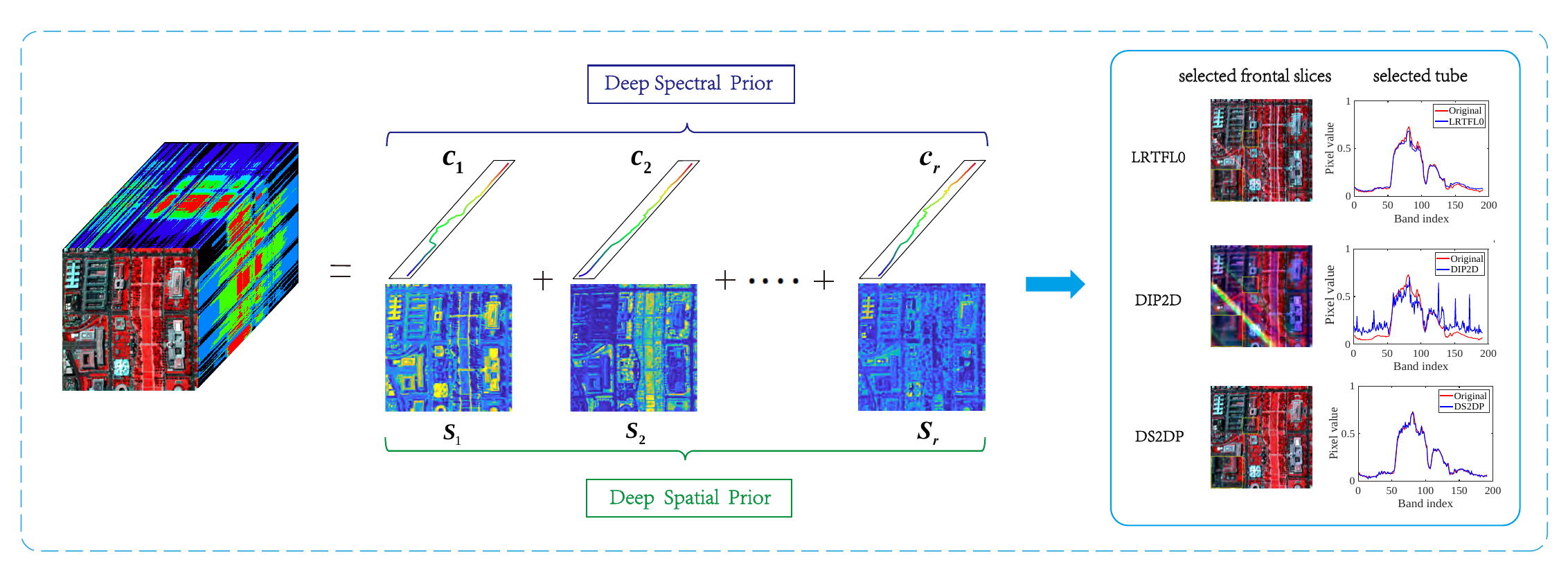}
		\caption{The LMM for HSI and the proposed  unsupervised disentangled spatio-spectral deep priors (\texttt{DS2DP}). \label{fig:lmm}}
	\end{figure*}
	
	\section{Introduction}\label{sec:Int}
	\IEEEPARstart{H}{yperspectral} images (HSIs) contain rich spectral and spatial information of areas/objects of interest. HSIs have been widely used across many disciplines, e.g., biology, ecology, geoscience, and food/medicine science \cite{Bioucas2012HUOverview}. However, the acquired HSIs are often corrupted by various types of noise. Heavy noise may affect the performance of downstream analytical tasks (e.g., hyperspectral pixel classification). In the past two decades, a plethora of HSI denoising techniques were proposed to address this challenge; see \cite{MF,TF,BTD,LBSR_HR}.
	
	At a high level, the idea of many HSI denoising methods is to fit the acquired image using an estimated image with prior information-induced priors. The rationale is that noise does not obey the HSI priors, and thus such a fitting process can effectively extract the ``clean'' HSI from the noisy version. Under this principle, early HSI denoising methods used spatial priors such as sparsity \cite{BM3D,S1,S2} and total variation (TV) \cite{chen1}. Methods that exploit spectral priors were also proposed; see \cite{PCA,MSBCRF,BM4D}. A number of denoising methods incorporated with {\it implicit} priors such as low matrix/tensor rank that is a result of multi-dimensional correlations; some examples can be found in \cite{MF,TF,BTD,Meng2018,chang1,chen2,chen3,zhang1,zhuang1}.

	More recently, data-driven priors have drawn much attention in the vision and imaging communities \cite{Deep}. In a nutshell, deep neural networks are used to learn a generative model of images from a large number of training samples. 
	Deep generative models have been successful in computer vision, see, e.g., \cite{VAE,AE,GAN}. In particular, these models are able to map low-dimensional random vectors to visually authentic images---which means that they capture the essence of the image generating process. Hence, the learned generative network is naturally a good prior of clean images. 
	This idea has also been used in HSI denoising; see, e.g., \cite{yuan1,Dong,yuan2,yuan3,chang2}.
	
	Although the methods mentioned above have attained satisfactory results for HSI denoising, these models' expressive ability is limited by the training data's adversity and quantity. 
	That is, there is a lack of training data for HSIs \cite{8693549}. This is because HSIs are, in general, much more costly to acquire relative to natural RGB images. 
	In addition, different hyperspectral sensors often admit largely diverse specifications (e.g., the frequency band used, the spectral resolution, and the spatial resolution)---data acquired from one sensor may not be useful for training deep priors for images from other sensors.
	
	Recently, Ulyanov {\it et al.} proposed an unsupervised image restoration framework, namely, {\it deep image prior} (DIP) \cite{DIP}. 
	DIP directly learns a generator network from a {\it single} noisy image---instead of learning the generator from a large number of training samples. 
	The work in \cite{DIP} showed that proper deep neural network architectures, without training on any samples, can already ``encode'' much critical information in the natural image generating process. 
	This discovery has helped design {\it unsupervised} DIPs for tasks such as image denoising, inpainting, and super-resolution.
	This work has thus attracted much attention. Since the DIP approach does not use any training data, it is particularly suitable for data-starved applications like hyperspectral imaging. Indeed, Sidorov {\it et al.} \cite{HSIDIP} extended the DIP idea to HSI denoising and observed positive results.

	Nonetheless, capitalizing on the power of DIP for HSI denoising still faces a series of challenges. Unlike RGB images that only have three spectral channels, HSIs are often measured over hundreds of spectral channels.
	Therefore, directly using the DIP method that is originally proposed for RGB images to handle HSIs may not be as promising.
	First, it is unclear if the network structures used in \cite{DIP} are still effective for HSIs.
	Second, due to the large size of HSIs, the scalability challenge is much more severe compared to the natural image cases. Indeed, as one will see in Sec.~\ref{5}, the two neural network structures used in \cite{HSIDIP} for modeling the generator of a standard HSI induce 2.150 and 2.342 million parameters, respectively---which makes the learning process challenging. 
	Third, due to the special data acquisition process of HSIs, outlying pixels and structured noise (other than Gaussian noise) often arise. The DIP denoising loss function used in \cite{DIP, HSIDIP} did not take these aspects into consideration.

	\smallskip  
	
	\noindent  
	{\bf Contributions.}  In this work, our interest lies in an unsupervised DIP-based denoising framework tailored for HSIs. Our detailed contributions are summarized as follows:

	\noindent
	$\bullet$ {\bf Disentangled Spatio-Spectral Deep Prior for HSI.} We propose an unsupervised DIP structure that is inspired by the well-established {\it linear mixture model} (LMM) for HSIs \cite{Ma2014unmixing}; see Fig.~\ref{fig:lmm}. The LMM views every hyperspectral pixel as a linear combination of spectral signatures of a number of materials ({\it endmembers}). The linear combination coefficients of different endmembers across the image give rise to the {\it abundance maps} of the endmembers \cite{Fu2016Unmixing}. Using LMM, the spatial and spectral information embedded in the HSI can be ``disentangled''. This way, the spectral and spatial priors can be designed and modeled {\it individually}. As a result, the modeling and computational complexities can be substantially reduced---which often leads to improved accuracy. By our design, empirically validated unsupervised DIP structures for natural images can be much more easily capitalized for HSI denoising. 
	
	\noindent
	$\bullet$ {\bf Structured Noise-robust Optimization.} We propose a training loss that models the structured noise (e.g., stripe-shaped or deadlines) as sparse outliers. We use an alternating optimization process to handle the formulated structured-noise robust deep prior-based denoising method, and admits simple lightweight updates.

	\noindent
	$\bullet$ {\bf Extensive Experiments.} We test the proposed approach on a large variety of simulated and real datasets. The experiments support our design---we observe substantially improved denoising performance relative to classic methods and more recent neural prior-based methods over all the datasets under test. In particular, due to our disentangled network design, the proposed method outperforms the existing unsupervised DIP-based HSI denoising methods in \cite{HSIDIP} in terms of both accuracy and memory/computational efficiency.

	\bigskip
	
	\noindent
	{\bf Notation.} A scalar, a vector, a matrix, and a tensor are denoted as $x$, $\bm{x}$, $\bm{X}$, and $\underline{\bm{X}}$, respectively. $[\bm{x}]_i$, $[\bm{X}]_{i,j}$, and $[\underline{\bm{X}}]_{i,j,k}$ denote the $i$-th, $(i,j)$-th, and $(i,j,k)$-th element of $\bm{x}\in \mathbb{R}^{I}$, $\bm{X}\in \mathbb{R}^{I\times J}$, and $\underline{\bm{X}}\in \mathbb{R}^{I\times J\times K}$, respectively. The Frobenius norms of $\bm{X}$ and $\underline{\bm{X}}$ are denoted as $\|\bm{X}\|_{F}=\sqrt{\sum_{i,j}[\bm{X}]_{i,j}^{2}}$ and $\|\underline{\bm{X}}\|_{F}=\sqrt{\sum_{i,j,k}[\underline{\bm{X}}]_{i,j,k}^{2}}$, respectively. Given $\bm{y}\in \mathbb{R}^{N}$ and a matrix $\bm{X}\in \mathbb{R}^{I\times J}$, the outer product is defined as $\bm{X}\circ \bm{y}$. In particular, $\bm X\circ \bm y \in \mathbb{R}^{I\times J\times N}$ and $[\bm X \circ \bm y]_{i,j,n} = [\bm X]_{i,j}[\bm y]_n$. 
	The matrix unfolding operator for a tensor is defined as ${\rm mat}(\underline{\bm X})$, which denotes the mode-3 unfolding of $\underline{\bm{X}}$ (see details of the unfolding of HSI in \cite{Kanatsoulis2018HSR}). The ${\rm vec}(\bm X)$ operator represents ${\rm vec}(\bm X)=[[\bm X]_{:,1}^T,\ldots,[\bm X]_{:,J}^T]^T$.

	\section{Preliminaries}
	In this section, we briefly review pertinent background information.
	
	\subsection{HSI Denoising}
	The acquired HSIs are three-dimensional arrays (i.e., tensors \cite{sidiropoulos2017tensor}). Denote $\underline{\bm X}\in\mathbb{R}^{I\times J\times K}$ as the HSI captured by a remotely deployed hyperspectral sensor, where $I\times J$ is the number of pixels presenting in the 2D spatial domain, and $K$ is the number of spectral bands. 
	Unlike natural images that are measured with the R, G, and B channels (i.e., $K=3$), HSIs are measured over tens or hundreds of frequency bands, depending on the specifications of the employed sensors.
	
	\begin{figure*}[!htbp] 
		\centering
		\includegraphics[scale=0.6]{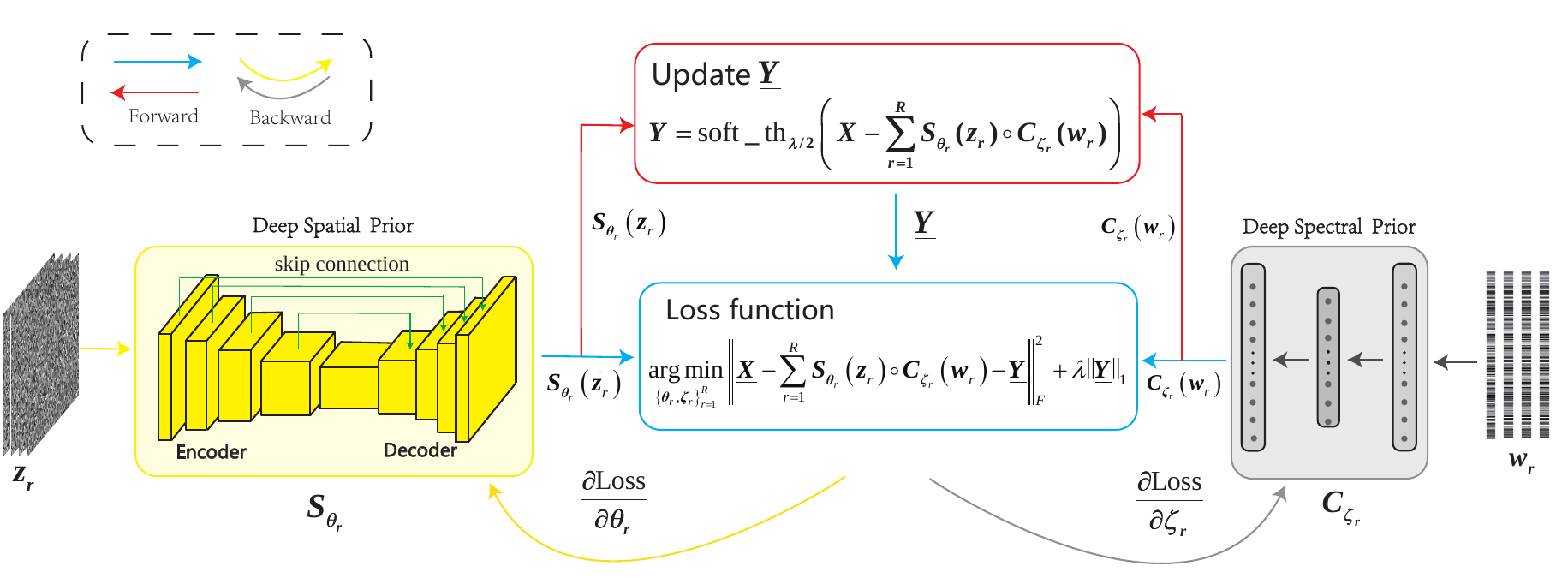}
		\caption{Illustration of the proposed \texttt{DS2DP}. The generative networks ${\cal C}_{\bm \zeta_r}$ and ${\cal S}_{\bm \theta_r}$ are applied to capture the deep spectral prior of the spectral signatures and the deep spatial prior of the abundance matrices, respectively.}	\label{detail}
	\end{figure*}

	In general, $\underline{\bm X}$ is a noise-contaminated version of the underlying ``clean'' HSI (denoted by $\underline{\bm X}_{\natural}$). There are many factors contributing to noise in the hyperspectral acquisition process, i.e., thermal electronics, dark current, and stochastic error of photon-counting. 
	If the noise is additive, we have
	\begin{equation}\label{eq:noise_model}
	\underline{\bm X} =\underline{\bm X}_\natural + \underline{\bm V},
	\end{equation}
	where $\underline{\bm V}\in\mathbb{R}^{I\times J\times K}$ denotes the noise. The objective of HSI denoising is to ``extract'' $\underline{\bm X}_\natural$ from $\underline{\bm X}$.

	\subsection{Prior-Regularization Based HSI Denoising}
	Note that even under the additive noise model in \eqref{eq:noise_model}, this problem is ill-posed---this is essentially a disaggregation problem which admits an infinite number of solutions. To overcome such ambiguity, prior information of the HSI is used to confine the solution space. A generic formulation can be summarized as follows:
	\begin{subequations}
		\begin{align}
		\widehat{\underline{\bm X}} = \arg \min_{\underline{\bm M}}&~\left\| \underline{\bm X} - \underline{\bm M} \right\|_F^2 +\lambda R\left( \underline{\bm M}\right), \label{eq:denoising_fitting}\\
		{\rm subject~to}&~\underline{\bm M} \in {\cal M}, \label{eq:denoising_constr}
		\end{align}
	\end{subequations}
	where $ \widehat{\underline{\bm X}}$ denotes the estimate for $\underline{\bm X}_\natural$ using the above estimator, $ \underline{\bm{M}} $ represents the optimization variable, ${\cal M}$ and $R(\cdot):\mathbb{R}^{I\times J\times K}\rightarrow \mathbb{R}_+$ are the constraint set and regularization function imposed according to prior knowledge about the clean image $\underline{\bm X}_\natural$, respectively, and $\lambda\geq 0$ is the regularization parameter that balances the data fidelity term (i.e., the first term in \eqref{eq:denoising_fitting}) and the regularization.
	
	\subsubsection{From Analytical Priors to Data-Driven Priors} 
	A variety of regularization/constraints have been considered in the literature. For example, in \cite{MF,TV5}, $$R(\cdot)=\|\cdot\|_{\rm TV}$$ is the TV across the two spatial dimensions, since image data exhibits certain slow changing properties over the space. In \cite{ye2014multitask,veganzones2015nonnegative}, ${\cal M}$ represents the nonnegative orthant, since HSIs are always nonnegative. In \cite{LRMR, LRTDTV,LRTF_HR,Meng2018,9372832,9314228}, low tensor and matrix rank constraints are added to $\underline{\bm M}$ through low-rank parameterization, respectively. Such prarameterization-based regularization can be written as
	\begin{align}\label{eq:udip}
	\widehat{\bm z }= \arg \min_{\bm z}&~\left\| \underline{\bm X} - {\cal G}\left( {\bm z}\right) \right\|_F^2,
	\end{align}
	where ${\cal G}:\mathbb{R}^N \rightarrow \mathbb{R}^{I\times J\times K}$ is a pre-specified parameterization function that represents the $I\times J\times K$ HSI using $N$ parameters, i.e., $\bm z$, and $ {\cal G}\left( {\bm z}\right) $ represents the estimation for the underlying clean HSI generated by ${\cal G}$  with parameters $\bm z$.
	For example, if ${\rm mat}\left(\underline{\bm X}\right)$ is believed to be a low-rank matrix, $ {\rm mat}\left({\cal G}\left( {\bm z}\right)\right)=\bm A\bm B^T$  and $\bm z=[{\rm vec}(\bm A)^T,{\rm vec}(\bm B)^T]^T$. After estimating the parameters $\bm z$, the clean image can be simply estimated via $$\widehat{\underline{\bm X}}={\cal G}(\widehat{\bm z}).$$
	Classic priors are useful but often insufficient to capture the complex nature of the underlying structure of HSIs. 
	
	A number of works used deep neural networks to parameterize the regularization---i.e., these works use a deep neural network ${\cal G}_{\bm \theta}(\cdot):\mathbb{R}^{N} \rightarrow \mathbb{R}^{I\times J\times K}$ whose network weights are collected in $\bm \theta\in\mathbb{R}^D$ to act as the regularization in \eqref{eq:denoising_fitting} \cite{yuan1,Dong,yuan2,yuan3,chang2}. 
	Instead of having an analytical expression, such regularizers are ``trained'' using a large number of training samples. As deep neural networks are universal function approximators, such learned ``deep priors'' are believed to be able to approximate complex generative processes of HSIs and thus are more effective priors for denoising. 
	\begin{align}\label{eq:supdip}
	\widehat{\bm z }= \arg \min_{\bm z}&~\left\| \underline{\bm X} - {\cal G}_{\bm \theta}\left( {\bm z}\right) \right\|_F^2,
	\end{align}

	However, unlike natural RGB images that have tens of thousands of training samples for learning ${\cal G}_{\bm \theta}$, HSI (especially remotely sensed HSI) datasets are relatively rare due to their costly acquisition process. Without a large amount of (diverse) HSIs, training such a regularizer may be out of reach.

	\subsubsection{Unsupervised Deep Image Prior} 	Very recently, Ulyanov {\it et al.} proposed the so-called DIP \cite{DIP} to circumvent the lack of training samples. The major discovery in \cite{DIP} is that a proper neural network {\it architecture} (without knowing the neural network weights $\bm \theta$) can already encode much prior information of images. As a result, tasks such as image denoising can be done by learning a neural network ${\cal G}_{\bm \theta}(\bm z)$ to fit $\underline{\bm X}$ with a {\it random} but ${\it known}$ $\bm z$. 
	
	With this idea, the denoising problem can be formulated as follows:
	\begin{align}\label{eq:udip}
	\widehat{\bm \theta }= \arg \min_{\bm \theta}&~\left\| \underline{\bm X} - {\cal G}_{\bm \theta}\left( {\bm z}\right) \right\|_F^2,
	\end{align}
	and the denoised image can be estimated via
	\begin{equation}
	\widehat{\underline{\bm X}} =  {\cal G}_{\widehat{\bm \theta}}\left( {\bm z}\right).
	\end{equation}
	The idea of DIP is quite different compared to the supervised deep prior-based approaches such as those in \cite{yuan1,Dong,yuan2,yuan3} [cf. Eq.~\eqref{eq:supdip}]. In DIP, the network weights $\bm \theta$ is learned from a single degraded image in an unsupervised manner, and $\bm z$ is given instead of learned. 
	
	At first glance, it may be surprising that an untrained neural network can be used for image denoising (and also inpainting and super-resolution as revealed in \cite{DIP}). 
	The key rationale behind this approach may be understood as follows: First, some carefully designed neural network structures (e.g., convolutional neural network with proper modifications) are able to capture much information in the generating process of some types of images of interest. 
	That is, {\it not all} neural network structures could work well for all types of images. Different structures may need to be carefully {\it handcrafted} for different types of images.
	The handcrafted neural network structure is analogous to the handpicked priors such as the $L_1$ norm, Tikhonov regularization, and TV regularization---which are also not learned from training samples.
	In the original paper \cite{DIP}, the U-Net-like "hourglass" architecture was shown to be powerful in natural RGB image restoration tasks under the DIP framework. In \cite{HSIDIP}, various network structures (namely, \texttt{DIP2D} and \texttt{DIP3D}) were experimented for HSI denoising---and the results can be quite different, as one will also see in Sec.~\ref{4}.
	Second, in image restoration tasks, the degraded (noisy) $\underline{\bm X}$ still contains much information in the underlying image. Hence, the fitting loss in \eqref{eq:udip} also ``forces'' the ${\cal G}_{\bm \theta}$ to faithfully capture the essential information in $\underline{\bm X}$. In particular, since ${\cal G}_{\bm \theta}$ has a structured underlying generative process (by construction), the learned ${\cal G}_{\bm \theta}$ is more likely to capture the ``structured signal part'' (i.e., the clean image $\underline{\bm X}_\natural$) in $\underline{\bm X}$ other than the random noise part.
	
	Since the DIP procedure does not use any training examples, it is particularly attractive to data-starved applications such as hyperspectral imaging. In addition, although it involves careful structure handcrafting, DIP still inherits many good properties of neural networks, e.g., being capable of modeling complex generative processes. Consequently, it often exhibits more appealing image restoration performance compared to classic regularizer/parameterization based methods (e.g., TV and low matrix/tensor rank); see \cite{DIP,HSIDIP}.

	\subsection{Challenges}
	The unsupervised DIP-based approaches are attractive since they are effective without using any training data. However, finding a proper network structure to serve as prior of HSIs and learning the corresponding $\bm \theta$ is by no means a trivial task. 
	A couple of notable new challenges that arise in the domain of hyperspectral imaging are as follows:
	
	\subsubsection{Challenge - 1 Integrating Unsupervised DIP and HSI}
	Since HSIs are quite different compared to natural RGB images (in terms of sensors, sensing processes, resolutions, and frequency bands used), directly using the neural network structure in \cite{DIP} in hyperspectral imaging may not be best practice. The work in \cite{HSIDIP} proposed two structures crafted for this, but it is not clear if these two structures are ``optimal'' due to the lack of extensive experiments. 
	In fact, as we will show in Sec.~\ref{4}, these two unsupervised DIP structures are sometimes not as promising as some classic models (e.g., low-rank tensor decomposition-based denoising) in terms of denoising performance. 
	Hence, the first challenge lies in if we could {\it circumvent} designing a new unsupervised DIP network architecture from scratch---which could involve much trial-and-error and time/resource consuming. In particular, can we leverage some underlying structures of the HSIs to avoid exhaustively searching through ad-hoc DIP architectures, but utilize some existing DIP network structures (e.g., those in \cite{DIP}) to effectively assist our HSI denoising task? We will answer this question.
	
	\subsubsection{Challenge - 2 Network Size}
	Another challenge that arises in unsupervised DIP-based HSI denoising is that the HSIs are large-scale images due to the large number of spectral bands contained in the pixels. Directly modeling the generative process of a large-scale 3D image (or a third-order tensor) inevitably leads to an overly sized neural network ${\cal G}_{\bm \theta}$. Although the work in \cite{HSIDIP} employed a number of tricks for network size reduction, the final constructions still yield a large number of network parameters. This leads to a computationally heavy optimization problem [cf. Eq.~\eqref{eq:udip}]. Since the problem is already nonconvex and challenging, the excessive scale of the optimization problem only makes the denoising procedure less efficient. The challenging nature of numerical optimization may also affect the denoising performance since "bad" local minima may be easier to happen.
	
	\begin{figure*}[!htb]
		\centering\scriptsize
			\renewcommand\arraystretch{1}
			\setlength{\tabcolsep}{1pt}
			\begin{tabular}{cccccccc}
				\includegraphics[width=0.5\linewidth]{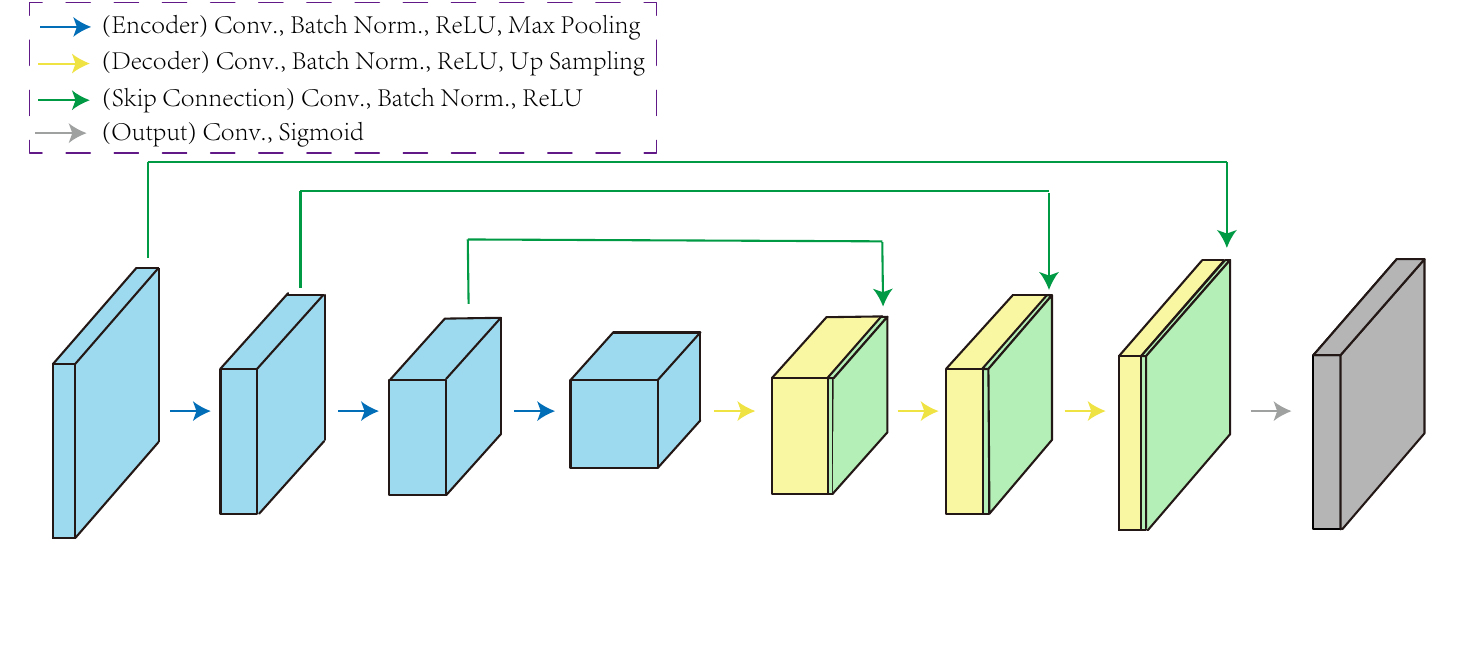}&
				\includegraphics[width=0.5\linewidth]{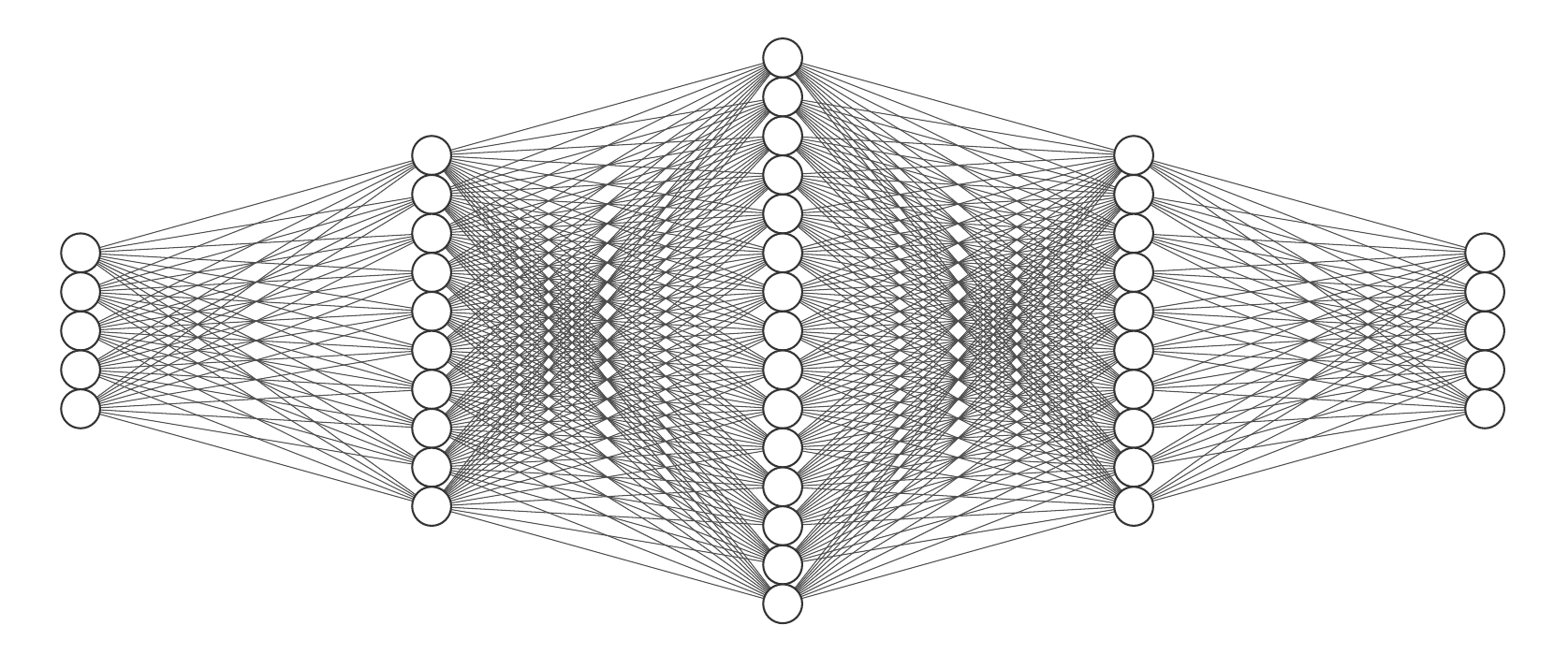}\\
				(a) & (b)\\
			\end{tabular}
			\caption{The detailed network structures. The two figures conrespond to: (a) the unsupervised DIP for abundance maps; and (b) the unsupervised DIP for endmembers}
			\label{net}
	\end{figure*}
	
	\section{Proposed Approach}
	To circumvent the challenges, we will leverage the well-established LMM of HSI to come up with our customized unsupervised DIPs in the next section. As will be seen, using the LMM to disentangle the spatial and spectral modalities of the HSIs allows us to use well-established/simple DIP structures to model each modality, which spares the agnostic pain of searching for a new DIP to model the high-dimensional hyperspectral data. The disentanglement also effectively reduces the model complexity. To this end, we briefly review the main idea of LMM.
	
	\subsection{Linear Mixture Model of HSI}
	The LMM of $\underline{\bm X}$ is as follows (when the noise is absent):
	\begin{equation}\label{eq:LMM}
	\underline{\bm X} = \sum_{r=1}^R \bm S_r\circ \bm c_r,
	\end{equation}
	where $\bm S_r\in\mathbb{R}^{I\times J}$ and $\bm c_r\in\mathbb{R}^K$ represent the $r$-th endmember's abundance map and the spectral signature, respectively, and $R$ is the number of endmembers contained in the HSI. The LMM can also be expressed as $$[\underline{\bm X}]_{i,j,k} =\sum_{r=1}^R [\bm S_r]_{i,j}[\bm c_r]_k;$$ see \cite{Bioucas2012HUOverview,Ma2014unmixing}.
	Physically, it means that every pixel is a non-negative combination of the spectral signatures of the constituting endmembers in the HSI. 
	Note that 
	\[ \bm S_r\geq\bm 0,~\bm c_r\geq \bm 0 \]
	according to their physical meanings---and thus the model in \eqref{eq:LMM} is often related to non-negative matrix factorization (NMF) \cite{Wycoff2013HSR}. An illustration of the LMM can be found in Fig.~\ref{fig:lmm}. The LMM model with a relatively small $R$ can often capture around 98\% of the energy of the HSI \cite{HySime}. Hence, it is a reliable model for HSIs. Indeed, the LMM has been utilized for a large variety of hyperspectral imaging tasks, e.g., hyperspectral unmixing \cite{Yokoya2012HSR,Aggarwal2016HU,Qian2017Unmixing,Fu2016Unmixing,Xiong2019HUBTD,Bioucas2012HUOverview}, hyperspectral super-resolution \cite{7410766}, pansharpening \cite{8075419}, compression and recovery \cite{7494936}, and denoising \cite{8077483}, just to name a few.  In this work, we propose to use the LMM to help design unsupervised DIP neural network structures and denoising algorithms.

	\subsection{LMM-Aided Unsupervised DIP for HSI}
	
	Notably, the LMM disentangles the spectral and spatial information into two sets of latent factors, i.e., $\{\bm S_r\}_{r=1}^R$ and $\{ \bm c_r \}_{r=1}^R$. Our motivations for using the LMM representation to design  unsupervised DIP for HSIs are as follows:
	\begin{itemize}[itemindent= 10 pt, leftmargin = 0 pt]
		\item First, LMM disentanglement allows using known effective DIP structures for natural images for HSI. The physical meaning of the latent factors entails the opportunity to employ known effective neural network structures of unsupervised DIP. The abundance matrix $\bm S_r$ can be understood as how the material $r$ spreads over space.
		The hypothesis is that the abundance maps exhibit similar properties to natural images that focus on capturing and conveying spatial information. Under this hypothesis, it is reasonable to use unsupervised DIP neural network structures that are known to work well for natural images to model $\bm S_r$. 
		Moreover, the $ \bm c_r $ vector can be understood as the spectral signature of the $r$-th material, which is the variation of reflectance or emittance of material over different wavelengths. It is known that fully connected neural networks (FCNs) can approximate such relatively simple 1-D continuous smooth functions well.
		
		\item Second, LMM disentanglement effectively reduces network complexity. By disentanglement and LMM, the model size of the HSI is substantially reduced. Instead of directly imposing unsupervised DIP on the whole HSI, we employ two types of unsupervised DIPs (i.e., the deep spatial and spectral priors) to model abundance maps and spectral signatures, respectively. Since the number of endmembers is often not large, the computational complexity is substantially reduced.
	\end{itemize}
	Following the above argument, we model the HSI using the following:
	\begin{equation}
	\underline{\bm X} =  \sum_{r=1}^R {\cal S}_{\bm \theta_r}(\bm z_r) \circ {\cal C}_{\bm \zeta_r}(\bm w_r ),
	\end{equation}
	where ${\cal S}_{\bm \theta_r}(\cdot):\mathbb{R}^{N_a} \rightarrow \mathbb{R}^{I\times J}$ is the unsupervised DIP neural network of the $r$-th endmember's abundance map, and $\bm \theta_r$ collects all the corresponding network weights; similarly, ${\cal C}_{\bm \zeta_r}(\cdot):\mathbb{R}^{N_s}\rightarrow \mathbb{R}^{K}$ and $\bm \zeta_r$ denote the unsupervised DIP of the $r$-th endmember and its corresponding network weights, respectively; the vectors $\bm z_r\in\mathbb{R}^{N_a}$ and $\bm w_r\in\mathbb{R}^{N_s}$ are low-dimensional random vectors that are responsible for generating the $r$-th abundance map and endmember, respectively. 
	Our detailed design for ${\cal S}_{\bm \theta_r}$ and ${\cal C}_{\bm \zeta_r}$ are as follows:
	
	\subsubsection{Unsupervised DIP for Abundance Maps} As mentioned, the abundance maps capture the spatial information of the corresponding materials. We propose to employ the U-Net-like ``hourglass'' architecture in \cite{DIP} for modeling ${\cal S}_{\bm \theta_r}$. Note that this network architecture was shown to be able to capture the spatial prior of nature images. The U-Net is an asymmetric autoencoder \cite{unet} with skip connections, whose structure is shown in Fig. \ref{net} (left).
	
	\subsubsection{Unsupervised DIP for Endmembers} The endmembers are relatively simple to model---since they can be understood as one-dimensional smooth functions. Hence, we employ FCNs as the unsupervised DIP for $ {\cal C}_{\bm \zeta_r} $. We use FCNs with three layers; also see Fig. \ref{net} (right).
	
	\bigskip
	
	Besides the above unsupervised DIP design,
	in this work, we also take into consideration of impulsive noise and grossly corrupted pixels (outliers) that often arise in HSIs. 
	Unlike natural images whose sensing environment can be well controlled, remotely sensed HSIs often suffer from heavily corrupted pixels or spectral bands due to various reasons; see \cite{LRTDTV, LRTF_HR}. If not accounted for, the HSI denoising performance could be severely hindered by such noise.
	To this end, we consider a noisy data acquisition model as follows:
	\begin{equation}
	\underline{\bm X} =\underbrace{ \sum_{r=1}^R {\cal S}_{\bm \theta_r}(\bm z_r) \circ {\cal C}_{\bm \zeta_r}(\bm w_r ) }_{\underline{\bm X}_\natural} + \underline{\bm Y} + \underline{\bm V},
	\end{equation}
	where $\underline{\bm V}$ represents ubiquitous noise, e.g., the Gaussian noise, and $\underline{\bm Y}$ denotes the impulsive noise or outliers. Accordingly, We propose the following denoising criterion:
	\begin{equation}\label{eq:proposed}{
		\mathop{\arg\min}_{\{ {\bm \theta _r},{\bm \zeta _r}\} _{r = 1}^R} ~ \left\| {\underline{\bm  X}  - \sum\limits_{r = 1}^R {{\bm S_{{\bm \theta_r}}}} ({\bm z_r}) \circ {\bm C_{{\bm \zeta _r}}}({\bm w_r}) - \underline{\bm  Y }} \right\|_F^2 + \lambda {\left\| {\underline{\bm  Y} } \right\|_1},}
	\end{equation}
	where $\lambda\geq 0$ and $\|\underline{\bm Y}\|_1 = \sum_{i=1}^I\sum_{j=1}^J\sum_{k=1}^K |[\underline{\bm Y}]_{i,j,k}|$ is used for imposing the sparsity prior on $\underline{\bm Y}$, since outliers happen sparsely. 

	\subsection{Optimization Algorithm}
	Let us denote the objective function in \eqref{eq:proposed} using the following shorthand notation:
	\begin{equation}\label{eq:shorthand}
	\mathop{\arg\min}_{\{ {\bm \theta _r},{\bm \zeta _r}\} _{r = 1}^R,\underline{\bm  Y}}~{\sf Loss}\left( \left\{\bm \theta_r,\bm \zeta_r \right\}_{r=1}^R,\underline{\bm Y} \right).
	\end{equation}
	We propose the following algorithmic structure:
	\begin{align}
	&\{\bm \theta_r^{t+1},\bm \zeta_r^{t+1}\}_{r=1}^R \nonumber\\ 
	&\quad\quad\ \leftarrow \mathop{\arg\widetilde{\min}}_{ \{\bm \theta_r,\bm \zeta_r\}_{r=1}^R }~{\sf Loss}\left(\{\bm \theta_r,\bm \zeta_r\}_{r=1}^R,\underline{\bm  Y }^t \right) \label{eq:networkupdate}\\
	&\underline{\bm  Y }^{t+1} \leftarrow \arg\min_{ \underline{\bm  Y }}~{\sf Loss}\left(\{\bm \theta_r^{t+1},\bm  \zeta_r^{t+1}\}_{r=1}^R,\underline{\bm  Y } \right), 
	\label{eq:Yupdate}
	\end{align}
	where the superscript ``$t$'' is the iteration index. In \eqref{eq:networkupdate}, we use $\widetilde{\min}$ to denote {\it inexact} minimization since exactly solving the subproblem w.r.t. the network parameters may not be possible---due to its large size and nonconvexity.
	
	\subsubsection{Solution for \eqref{eq:networkupdate}} Note that the subproblem w.r.t. $\{\bm \theta_r,\bm \zeta_r\}_{r=1}^R$ is nothing but a regression problem using neural models. Hence, any off-the-shelf neural network optimizer can be employed for updating $\{\bm \theta_r,\bm \zeta_r\}_{r=1}^R$. In this work, we use the (sub-)gradient descent\footnote{Since the ReLU activation functions used in the U-Net and the FCN are not differentiable at one point, the algorithm is subgradient based. Nonetheless, we use $\nabla$ (usually for denoting gradient) to denote subgradient for notation simplicity.} algorithm with momentum that has been proven effective in complex network learning problems \cite{Adam}:
	\begin{subequations}
		\begin{align}
		\bm \theta_r^{t+1} &\leftarrow \bm \theta_r^t - \alpha^t \nabla_{\bm \theta_r}~{\sf Loss}\left(\{\bm \theta_r,\bm \zeta_r^{t}\}_{r=1}^R,\underline{\bm Y}^t \right)\\
		\bm \zeta_r^{t+1} &\leftarrow \bm \zeta_r^t - \alpha^t \nabla_{\bm \zeta_r}~{\sf Loss}\left(\{\bm \theta_r^{t},\bm \zeta_r\}_{r=1}^R,\underline{\bm Y}^t \right),
		\end{align}
	\end{subequations}
	for all $r=1,\ldots,R$. Note that the gradient w.r.t. $\bm \theta_r$ and $\bm \zeta_r$ can be computed by the standard back-propagation algorithm \cite{BP}. Here, $\alpha^t$ is the step size (i.e., learning rate) of iteration $t$. There are multiple ways of determining $\alpha^t$. In this work, we use the step size rule advocated in the Adam algorithm \cite{Adam}. 

	\subsubsection{Solution for \eqref{eq:Yupdate} } 
	The subproblem~\eqref{eq:Yupdate} is convex---whose solution is the well-known soft-thresholding proximal operator \cite{soft}. Hence, the update of $\underline{\bm Y}$ can be expressed as
	\begin{equation}\label{eq:Y_update_prox}
	\underline{\bm  Y }^{t+1}={\rm soft\_th}_{\lambda /2}\left(\underline{\bm  X }-\sum_{r=1}^{R} \widehat{\bm S}_{r}^{t+1}  \circ \widehat{\bm c}_{r}^{t+1}\right).
	\end{equation}
	where $$ \widehat{\bm S}_{r}^{t+1} = \mathcal{S}_{\bm \theta_r^{t+1}}(\bm z_r),~ \widehat{\bm c}_{r}^{t+1} = \mathcal{C}_{\bm \zeta^{t+1}_r}(\bm w_r)$$ and 	${\rm soft\_th}_{\lambda/2}(\cdot)$ applies soft-thresholding to every entry of its input, in which the entry-wise thresholding is defined as
	\begin{equation}
	{\rm soft\_th}_{\delta}(x)=\operatorname{sgn}(x) \max (|x|-\delta, 0).
	\end{equation}
	
	\begin{algorithm}[h]
		\renewcommand\arraystretch{1.2}
		\caption{DS2DP for HSI Denoising.}
		\begin{algorithmic}[1]
			\renewcommand{\algorithmicrequire}{\textbf{Input:}} %
			\renewcommand{\algorithmicensure}{\textbf{Output:}}
			\Require the HSI $ \underline{\bm  X } \in \mathbb{R}^{I \times J \times K} $, the regularization parameter $\lambda$, and  the number of endmembers $R$.
			\State sample random $ \bm z_r $ and $ \bm w_r $ from uniform distribution;
			\For {$t=1$ to $T$}   (repeat until convergence)
			\State $ \widehat{\bm S}_{r} = \mathcal{S}_{\bm \theta_r^{t-1}}(\bm z_r)$, $ \widehat{\bm c}_{r} = \mathcal{C}_{\bm \zeta^{t-1}_r}(\bm w_r)$;
			\State update $\bm \theta_r,\bm \zeta_r$ for all $r$; using the Adam \cite{Adam};
			\State update $ \underline{\bm  Y } $ according to \eqref{eq:Yupdate};
			\EndFor
			\State
			$ \widehat{\underline{\bm  X }}=\sum_{r=1}^{R} \widehat{\bm S}_{r} \circ \widehat{\bm c}_{r}$;
			\Ensure
			the denoising HSI $\widehat{\underline{\bm  X }}$.
		\end{algorithmic}
		\label{algo:proposed}
	\end{algorithm}

	The algorithm is summarized in Algorithm~\ref{algo:proposed}, which we name as the {\it unsupervised disentangled spatio-spectral deep prior} (\texttt{DS2DP}) algorithm. The algorithm falls into the category of inexact block coordinate descent \cite{Meisam}. Under some relatively mild conditions, the algorithm produces a solution sequence that converges to a stationary point of the optimization problem in \eqref{eq:proposed}; see detailed discussions in \cite{Meisam}.
	
	\begin{table*}[!htbp]
		\scriptsize
		\setlength{\tabcolsep}{2.0pt}
		\renewcommand\arraystretch{1.8}\vspace{-0.22cm}
		\caption{Quantitative comparison of the denoising results by different methods. The \textbf{best} and \underline{second} best values are highlighted in bold and underlined, respectively.}\vspace{0.2cm}
		\begin{center}
			\begin{tabular}{cl|ccc|ccc|ccc|ccc|ccc|ccc}
				\Xhline{1.0pt}
				
				\multicolumn{2}{c}{Case}       &\multicolumn{3}{c}{Case 1}  &\multicolumn{3}{c}{Case 2}  & \multicolumn{3}{c}{Case 3}  &\multicolumn{3}{c}{Case 4}  &\multicolumn{3}{c}{Case 5}  & \multicolumn{3}{c}{Case 6} \\
				
				\Xhline{0.8pt}
				
				Dataset & Method        &PSNR    & SSIM     &SAM      &PSNR   & SSIM   &SAM     &PSNR     & SSIM    &SAM   &PSNR    & SSIM     &SAM     &PSNR     & SSIM    &SAM     &PSNR     & SSIM    &SAM\\
				
				\Xhline{0.8pt}
				\multirow{6}[1]{*}{WDC Mall} 
				&DIP2D         & 30.408 & 0.871 & 0.122  & 26.540 & 0.770 & 0.163   & 24.043 & 0.708 & 0.228  & 22.679 & 0.678 & 0.271     & 23.366 & 0.696 & 0.227   & 21.759 & 0.594 & 0.282\\
				
				&DIP3D         & *   &    *  &    * & *   &    *  &    *  & *   &    *  &    *  & * & * & *    & * & * & *  & * & * & *\\
				
				&LRMR   & 34.954 & 0.951 & 0.130  & 34.954 & 0.951 & 0.130  & 32.422 & 0.933 & 0.156 & 32.058 & 0.925 & 0.148   & 32.358 & 0.920 & 0.159   & 29.815 & 0.907 & 0.210 \\
				
				&LRTDTV         & 35.293 & 0.952 & 0.106 & 35.087 & 0.950 & 0.106  & 33.307 & 0.925 & 0.148  & 33.024 & 0.919 & 0.136   & 33.464 & 0.914 & 0.113  & 31.691 & 0.894 & 0.136\\
				
				&LRTFL0         & \underline{36.043} & \underline{0.964} & \underline{0.112} & \underline{35.796} & \underline{0.961} & \underline{0.111}  &  \underline{34.151} & \underline{0.948} & \underline{0.133}  & \underline{35.278} & \underline{0.941} & \underline{0.115}    & \underline{34.296} & \underline{0.949} & \underline{0.123}  & \underline{33.224} & \underline{0.943} & \underline{0.163}\\
				
				&DS2DP      & \textbf{36.439} & \textbf{0.965} & \textbf{0.102} & \textbf{35.926} & \textbf{0.969} & \textbf{0.093}  & \textbf{34.562} & \textbf{0.951} & \textbf{0.116}  & \textbf{35.887} & \textbf{0.954} & \textbf{0.100}    & \textbf{35.087} & \textbf{0.962} & \textbf{0.100} & \textbf{34.352} & \textbf{0.967} &\textbf{0.116}\\

				\Xhline{0.8pt}
				\multirow{6}[1]{*}{Pavia Centre} 
				&DIP2D         & 31.965 & 0.897 & 0.068  & 29.603 & 0.876  & 0.072   & 25.319 & 0.758  & 0.186  & 23.587 & 0.728  & 0.232    & 24.885 & 0.768  & 0.164   & 22.175 & 0.551  & 0.180 \\
				
				&DIP3D         & 26.969 & 0.694 & 0.075 & 26.338 & 0.691  & 0.078  &25.421 & 0.651  & 0.094  & 23.445 & 0.637  & 0.104    & 24.173 & 0.672  & 0.091  & 23.039 & 0.627  & 0.131\\
				
				&LRMR   &  33.293 & 0.926 & 0.090  & 33.293 & 0.926  & 0.090   & 30.398 & 0.816  & 0.052 & 32.398 & 0.916  & 0.142    & 31.409 & 0.901  & 0.106   & 24.667 & 0.742  & 0.724 \\
				
				&LRTDTV         & 33.511 & 0.921 & 0.095 & 33.608 & 0.923  & 0.065  & 31.465 & 0.901  & 0.104  & 33.096 & 0.903  & 0.147    & 31.415 & 0.881  & 0.104 & 31.882 & 0.894  & 0.101\\
				
				&LRTFL0      & \underline{33.833} & \underline{0.923} & \underline{0.088} &  \underline{33.310} & \underline{0.935}  & \underline{0.089}  & \underline{31.751} & \underline{0.917}  & \underline{0.096}  & \underline{32.756} & \underline{0.927}  & \underline{0.089}   & \underline{32.676} & \underline{0.928}  & \underline{0.090}  & \underline{32.003} & \underline{0.920}  & \underline{0.101} \\
				
				&DS2DP      & \textbf{35.211} & \textbf{0.947}  & \textbf{0.062} & \textbf{34.336} & \textbf{0.941} & \textbf{0.058}  & \textbf{32.545} & \textbf{0.926}  & \textbf{0.094}   & \textbf{33.682} &\textbf{0.934}  & \textbf{0.066}    & \textbf{33.836} & \textbf{0.936}  & \textbf{0.064}  & \textbf{32.523} & \textbf{0.924}  & \textbf{0.086}\\

				\Xhline{0.8pt}
				\multirow{6}[1]{*}{Pavia University} 
				&DIP2D         & 33.103 & 0.852 & 0.107  & 25.818 & 0.770 & 0.177   & 25.157 & 0.727 & 0.223  & 24.047 & 0.714 & 0.269    & 24.024 & 0.719 & 0.283   & 21.549 & 0.574 & 0.382 \\
				
				&DIP3D         & 30.070 & 0.804 & 0.111 &24.968 & 0.705 & 0.151  &  25.307 & 0.701 & 0.156  & 24.198 & 0.683 & 0.166    & 24.265 & 0.701 & 0.166  & 23.509 & 0.640 & 0.173 \\
				
				&LRMR   & 33.063 & 0.862 & 0.113  & 31.582 & 0.787 & 0.149  & 31.155 & 0.860 & 0.119 & 31.858 & 0.861 & 0.115    & 31.385 & 0.829 & 0.139   & 27.615 & 0.747 & 0.240 \\
				
				&LRTDTV         & 33.136 & 0.875 & 0.108 & 32.223 & 0.861 & 0.110  & 31.497 & 0.841 & 0.151  & 32.190 & 0.866 & 0.112    & 32.123 & 0.851 & 0.136  & 31.027 & 0.830 & 0.187\\
				
				&LRTFL0      & \underline{34.312} & \underline{0.890} & \underline{0.092} & \underline{33.724} & \underline{0.879} & \underline{0.099}  & \underline{32.972} & \underline{0.867} & \underline{0.123}  & \underline{33.642} & \underline{0.877} & \underline{0.103}    & \underline{33.146} & \underline{0.863} & \underline{0.124}  & \underline{32.735} & \underline{0.858} & \underline{0.126}\\
				
				&DS2DP     & \textbf{35.202} & \textbf{0.928} & \textbf{0.068} &  \textbf{34.600} & \textbf{0.917} & \textbf{0.073}   & \textbf{33.916} & \textbf{0.915} & \textbf{0.085}   & \textbf{34.600} & \textbf{0.917} & \textbf{0.073}    & \textbf{34.467} & \textbf{0.918} & \textbf{0.074} & \textbf{33.795} & \textbf{0.915} & \textbf{0.081}\\

				\Xhline{0.8pt}
				\multirow{6}[1]{*}{CAVE} 
				&DIP2D         & 29.643 & 0.636 & 0.339  & 23.839 & 0.589 & 0.421   & 23.204 & 0.562 & 0.449  & 21.955 & 0.526 & 0.506    & 22.416 & 0.538 & 0.484   &22.416 & 0.539 & 0.484 \\
				
				&DIP3D         & 28.960 & 0.709 & 0.332 & 23.397 & 0.571 & 0.447  & 23.377 & 0.566 & 0.449  & 22.157 & 0.534 & 0.471    & 22.435 & 0.549 & 0.460  & 21.405 & 0.509 & 0.501\\
				
				&LRMR   & 30.633 & 0.661 & 0.418  & 30.633 & 0.661 & 0.418  & 27.724 & 0.607 & 0.466 & 31.809 & 0.807 & 0.334    & 29.015 & 0.680 & 0.445   & 26.404 & 0.659 & 0.536 \\
				
				&LRTDTV        & \underline{35.529} & \underline{0.883} & \underline{0.165 }& \underline{34.769} & 0.877 & \underline{0.210}  & 32.792 & 0.843 &0.260  & \underline{34.036} & \underline{0.862} & \underline{0.232}   & 31.779 & 0.772 & 0.361   & \underline{31.063} & 0.773 & 0.430\\
				
				&LRTFL0      & 33.241      & 0.877     & 0.233 & 33.191      &\underline{0.891}     & 0.262  &  \underline{32.978}      & \underline{0.846}    & \underline{0.209}  & 33.743      & 0.852     & 0.264    & \underline{32.139}      & \underline{0.781}     & \underline{0.352}   &  30.956 & \underline{0.855} & \underline{0.301}\\
				
				&DS2DP      & \textbf{36.043} & \textbf{0.923} & \textbf{0.142} &  \textbf{35.603} & \textbf{0.907} & \textbf{0.146} & \textbf{33.892} & \textbf{0.956} & \textbf{0.165} &\textbf{35.682} & \textbf{0.914} & \textbf{0.155}    & \textbf{32.775} & \textbf{0.862} & \textbf{0.187} & \textbf{32.588} & \textbf{0.848} & \textbf{0.213}\\
				
				\Xhline{1.0pt}
			\end{tabular}
		\end{center}\vspace{-0.3cm}
		\label{tab}
	\end{table*}

	\begin{figure*}[!htb]
		\footnotesize
		\setlength{\tabcolsep}{1.2pt}
		\centering
		\begin{tabular}{c}
			\includegraphics[width=0.8\linewidth]{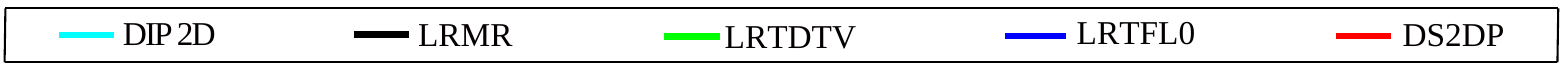}\\
		\end{tabular}
		\begin{tabular}{cccccccc}
			\includegraphics[width=0.16\linewidth]{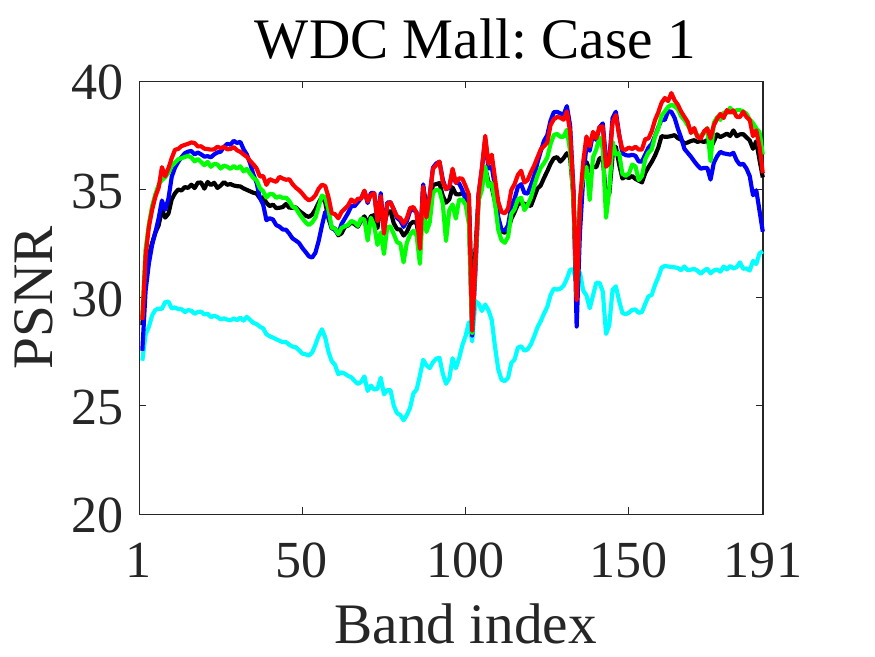}&			
			\includegraphics[width=0.16\linewidth]{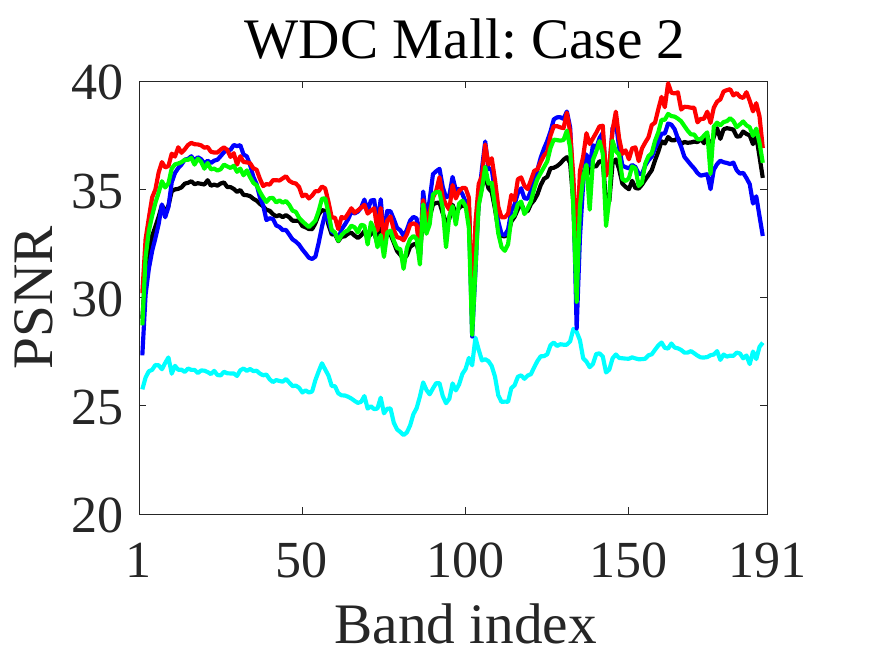}&
			\includegraphics[width=0.16\linewidth]{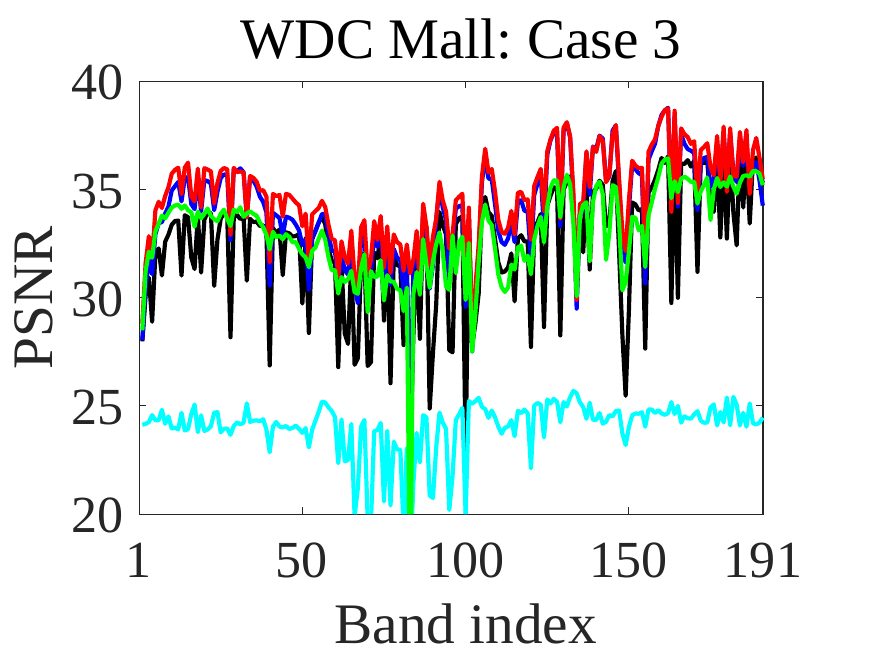}&
			\includegraphics[width=0.16\linewidth]{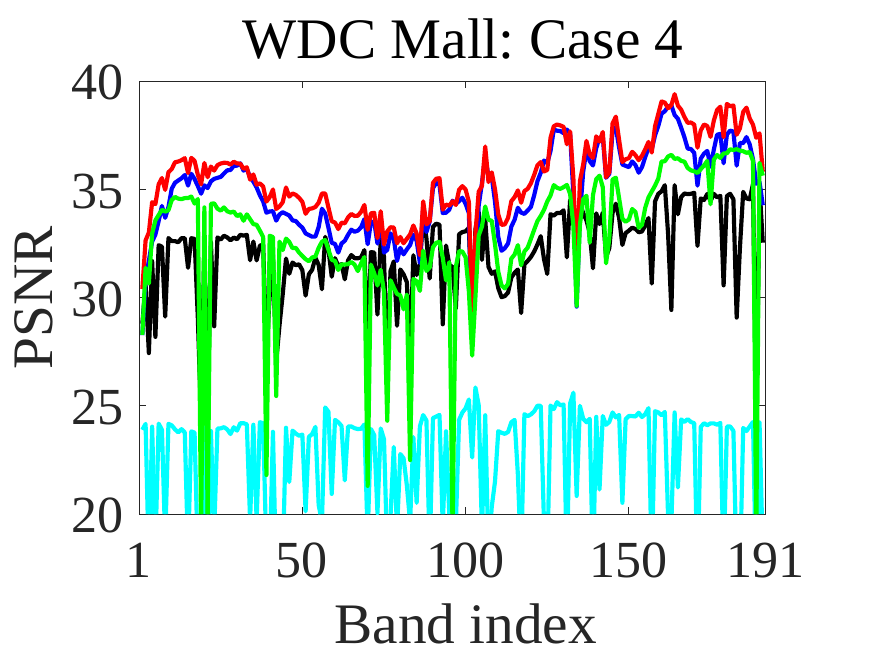}&
			\includegraphics[width=0.16\linewidth]{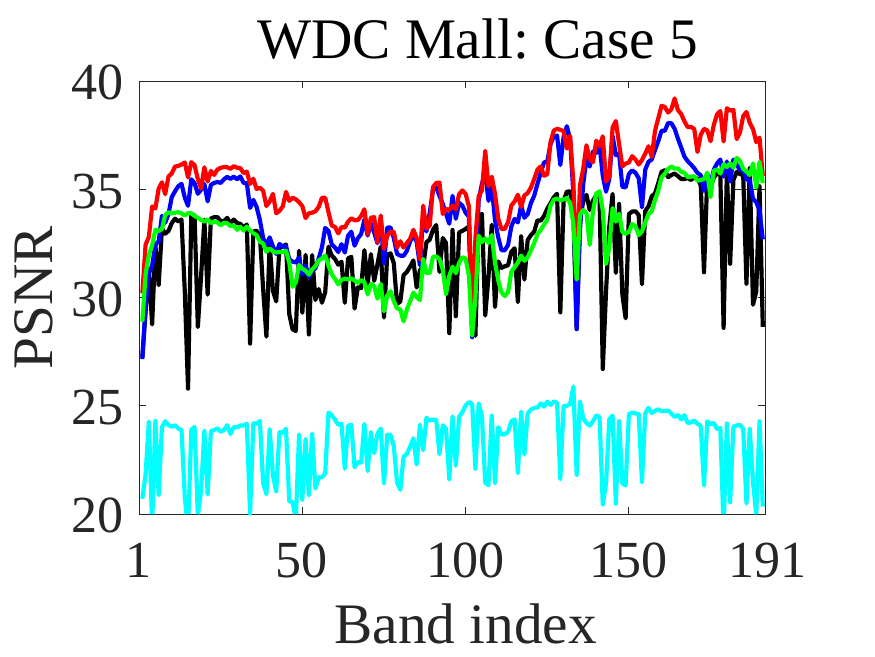}&
			\includegraphics[width=0.16\linewidth]{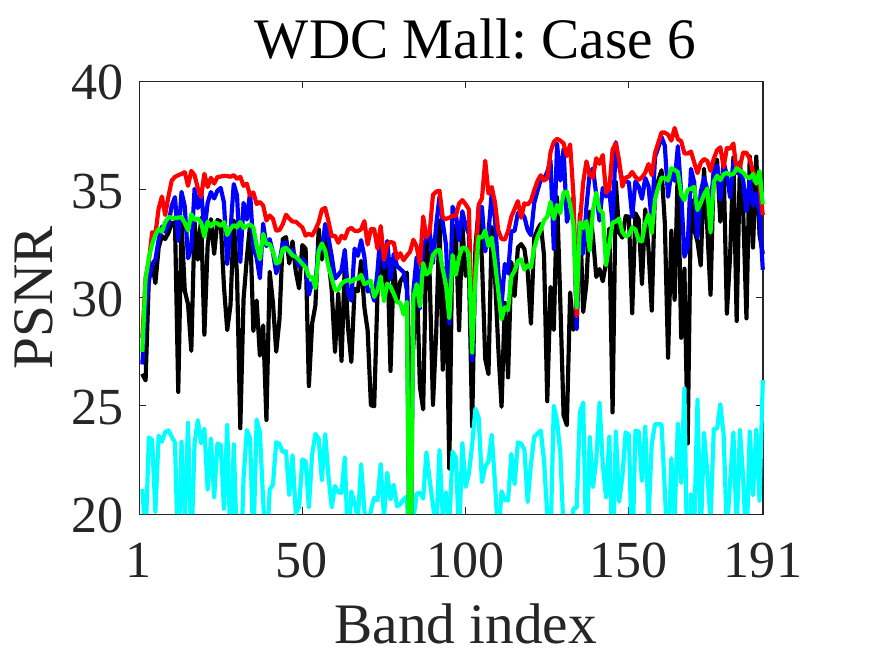}&\\
			\includegraphics[width=0.16\linewidth]{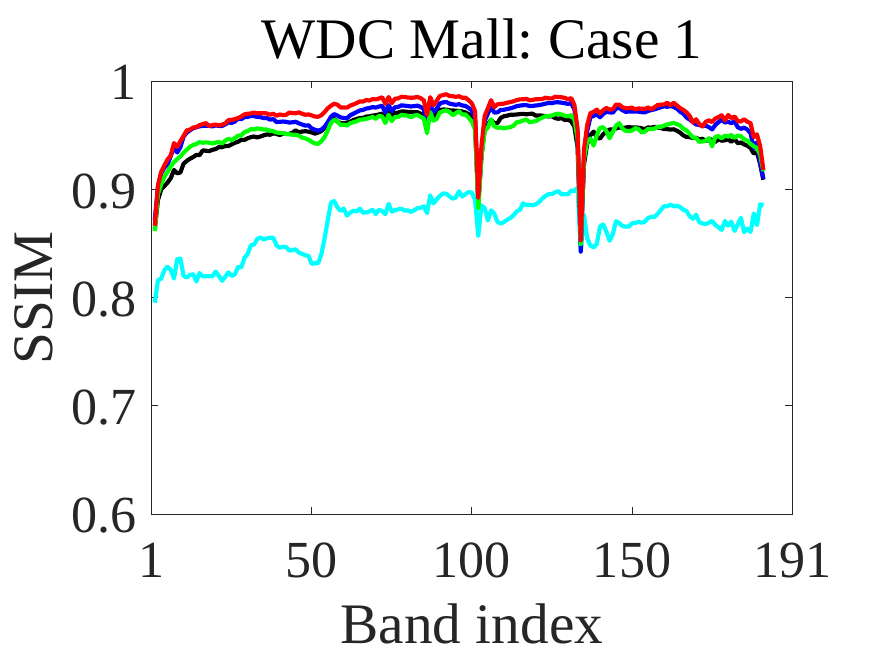}&
			\includegraphics[width=0.16\linewidth]{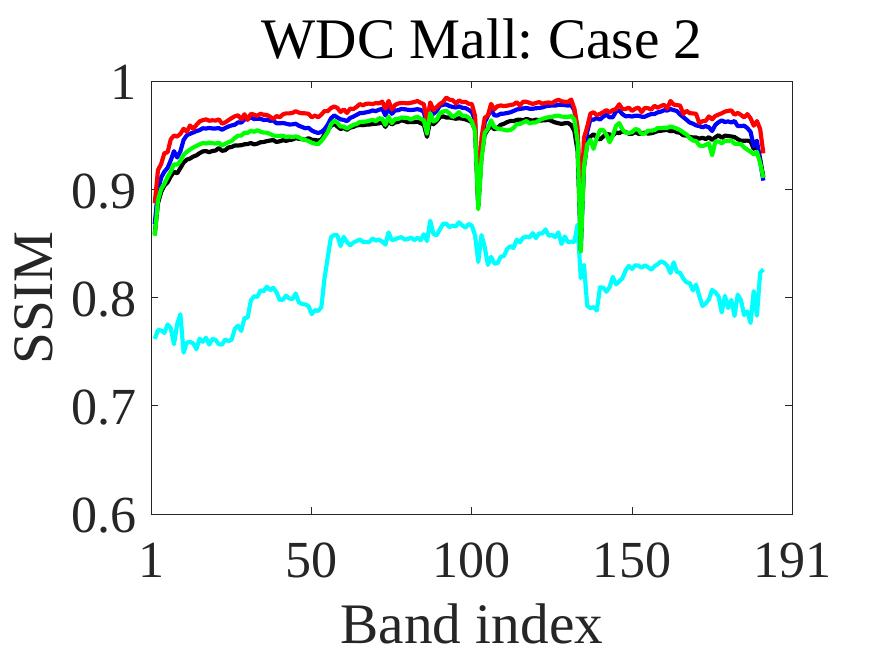}&
			\includegraphics[width=0.16\linewidth]{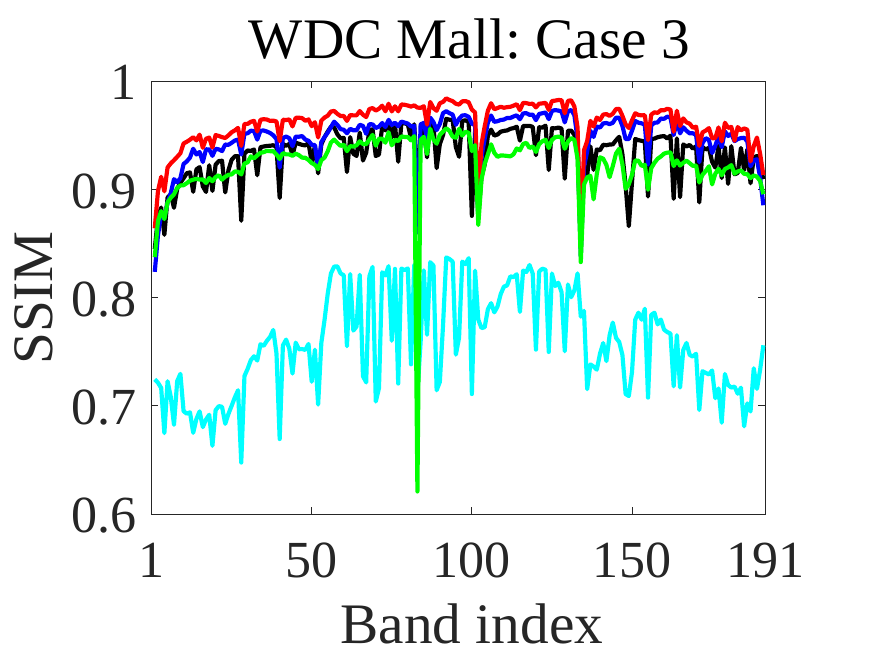}&
			\includegraphics[width=0.16\linewidth]{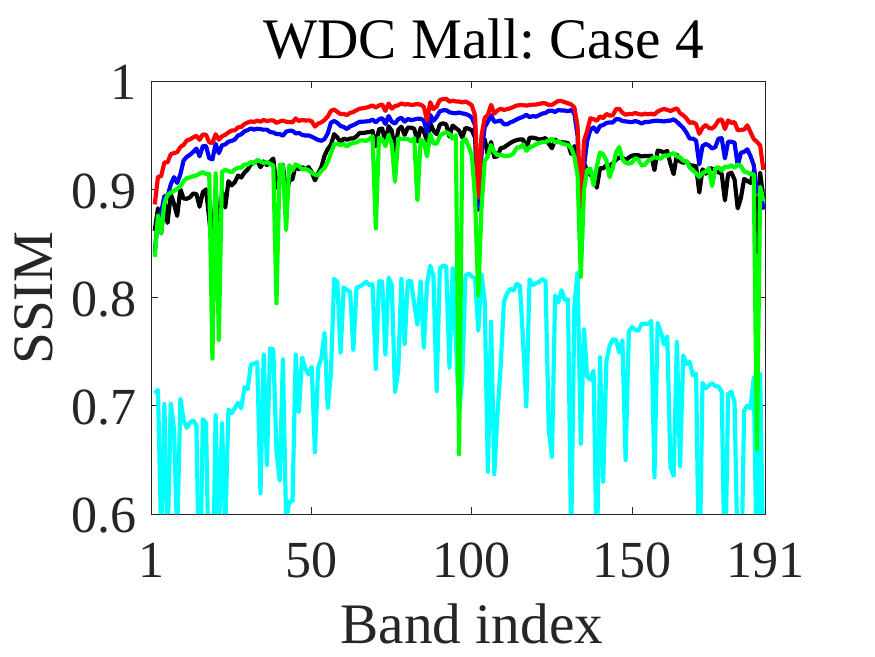}&				
			\includegraphics[width=0.16\linewidth]{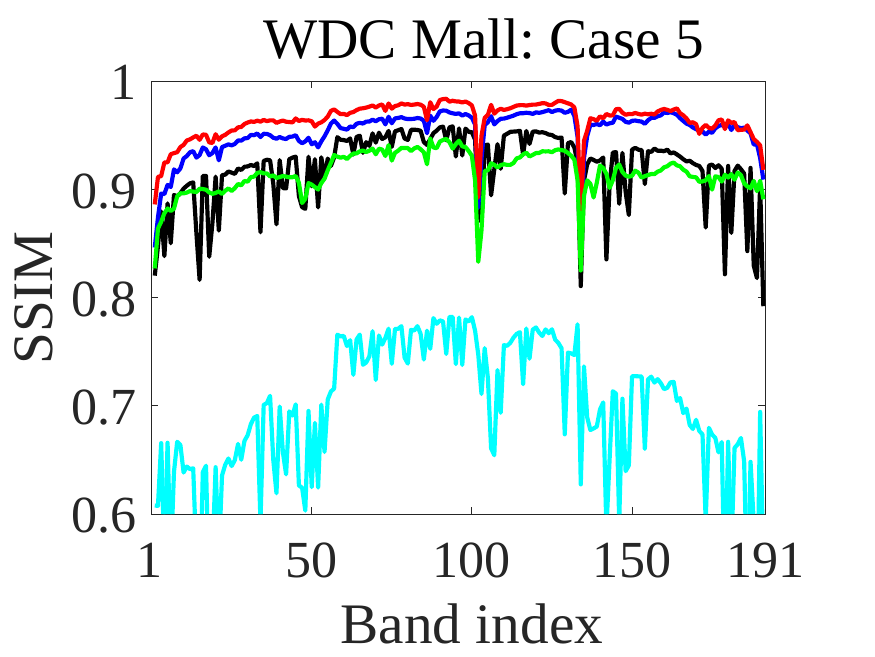}&
			\includegraphics[width=0.16\linewidth]{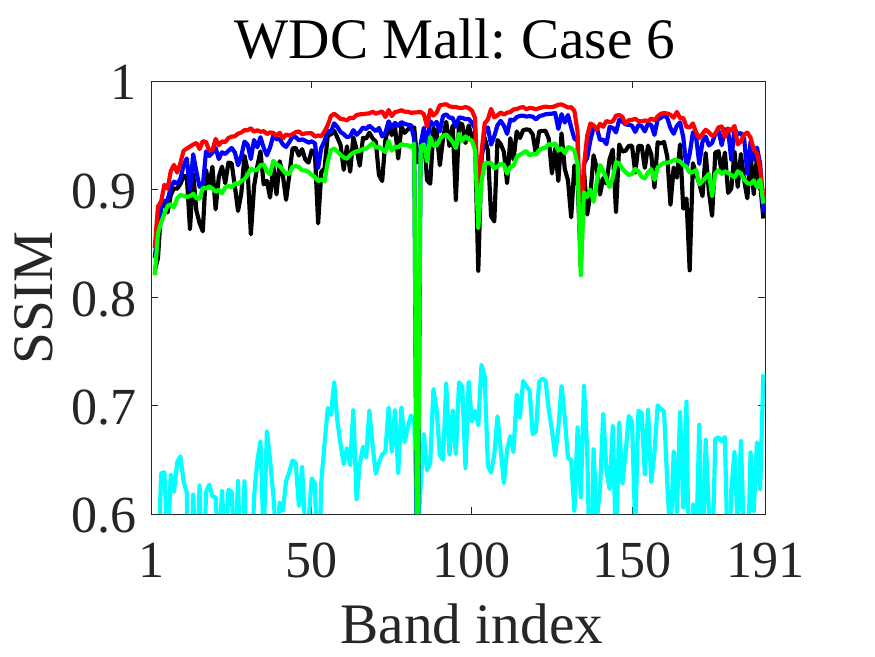}&\\								
			
		\end{tabular}
		\caption{PSNR and SSIM values of all bands obtained by different methods on HSI WDC Mall for Cases 1-6.}
		\label{every_bands}
	\end{figure*}
	
	\begin{figure*}[!htbp]
		\footnotesize
		\setlength{\tabcolsep}{1pt}
		\centering
		\begin{tabular}{ccccccc}
			Observed & DIP2D & LRMR & LRTDTV  & LRTFL0 & DS2DP & Ground truth\\
			
			\includegraphics[width=0.135\textwidth]{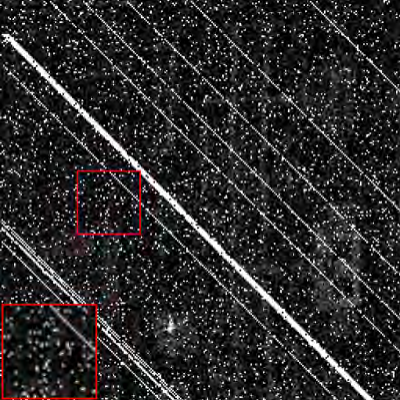}&
			\includegraphics[width=0.135\textwidth]{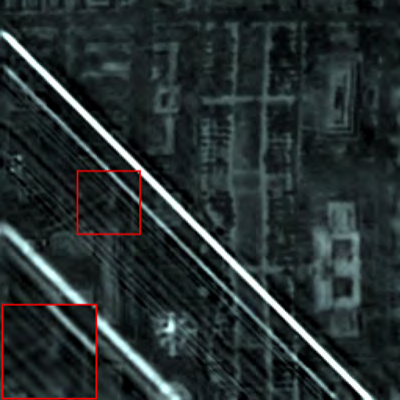}&
			\includegraphics[width=0.135\textwidth]{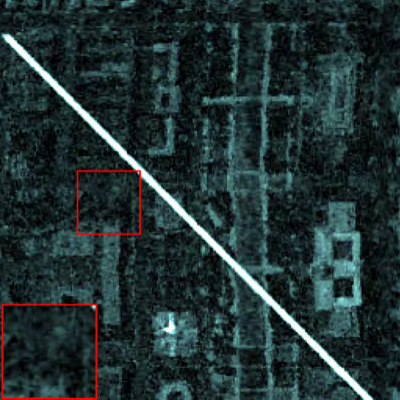}&
			\includegraphics[width=0.135\textwidth]{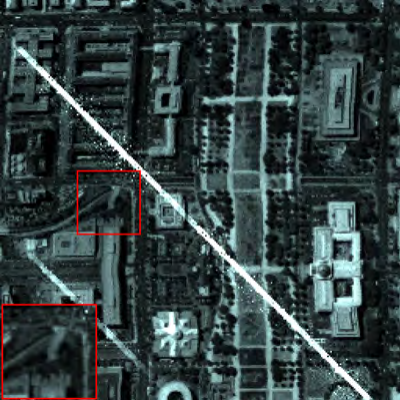}&
			\includegraphics[width=0.135\textwidth]{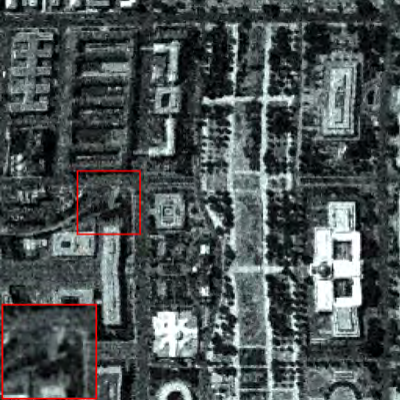}&
			\includegraphics[width=0.135\textwidth]{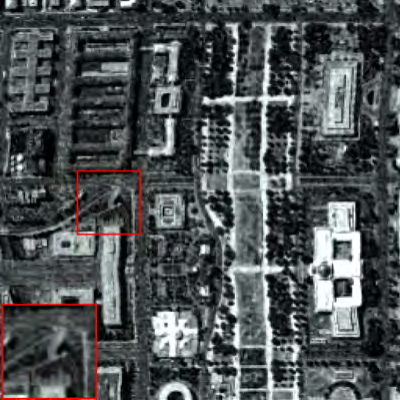}&
			\includegraphics[width=0.135\textwidth]{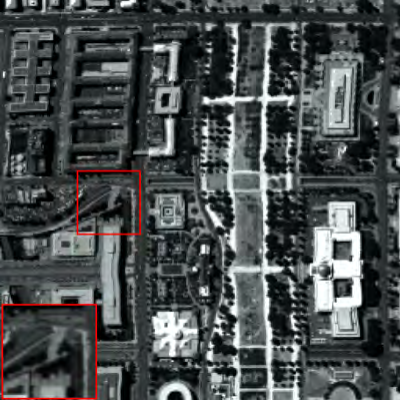}\\
			
			\includegraphics[width=0.135\textwidth]{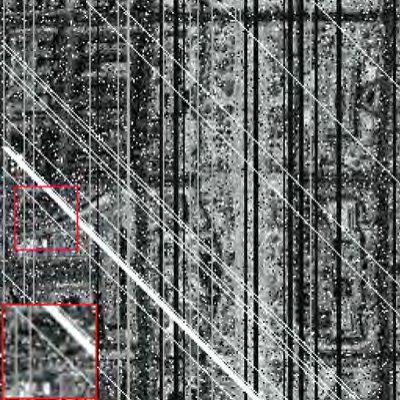}&
			\includegraphics[width=0.135\textwidth]{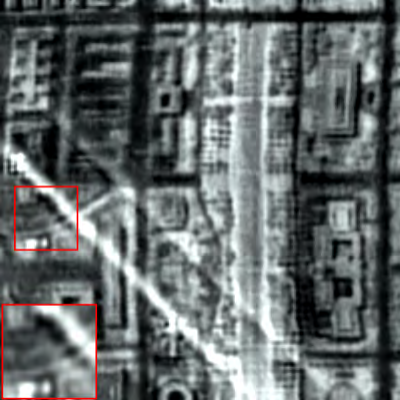}&
			\includegraphics[width=0.135\textwidth]{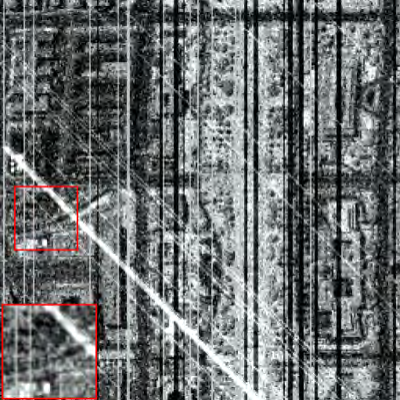}&
			\includegraphics[width=0.135\textwidth]{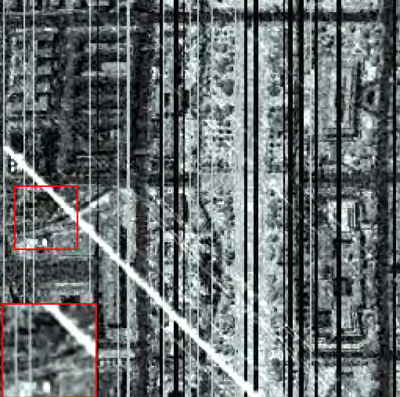}&
			\includegraphics[width=0.135\textwidth]{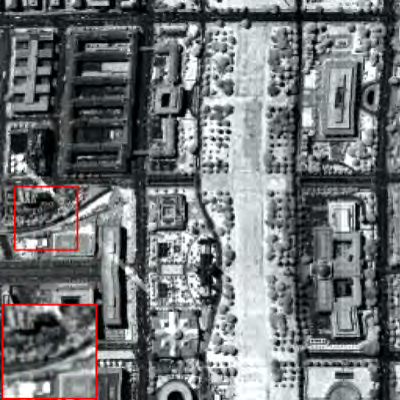}&
			\includegraphics[width=0.135\textwidth]{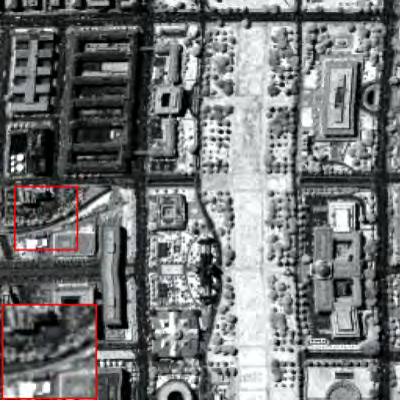}&
			\includegraphics[width=0.135\textwidth]{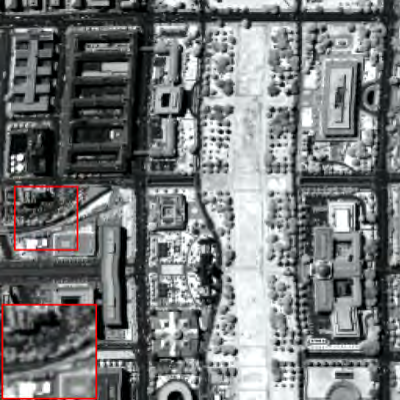}\\

			\includegraphics[width=0.135\textwidth]{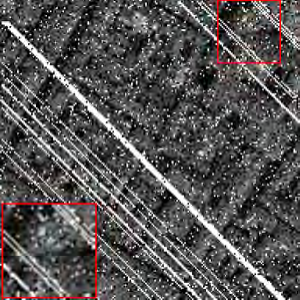}&
			\includegraphics[width=0.135\textwidth]{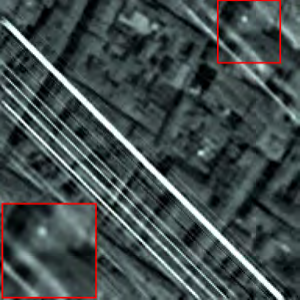}&
			\includegraphics[width=0.135\textwidth]{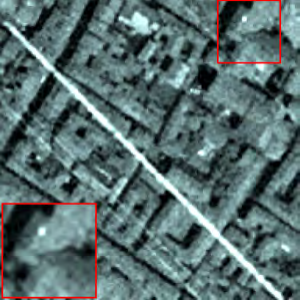}&
			\includegraphics[width=0.135\textwidth]{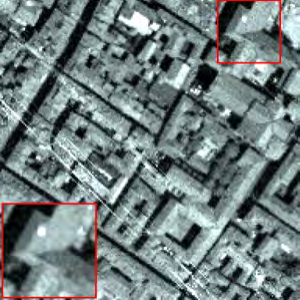}&
			\includegraphics[width=0.135\textwidth]{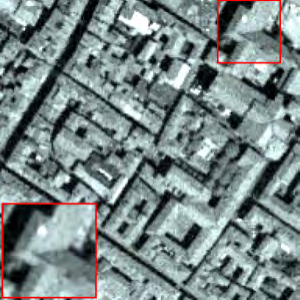}&
			\includegraphics[width=0.135\textwidth]{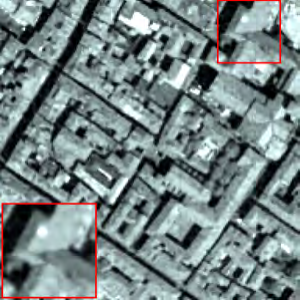}&
			\includegraphics[width=0.135\textwidth]{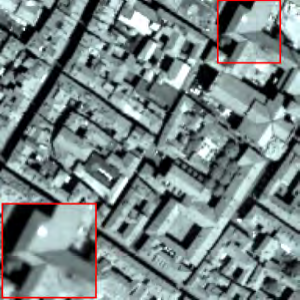}\\
			
			\includegraphics[width=0.135\textwidth]{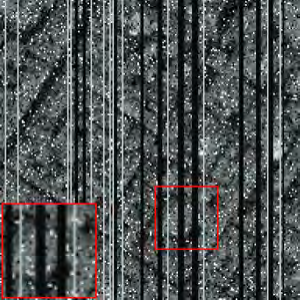}&
			\includegraphics[width=0.135\textwidth]{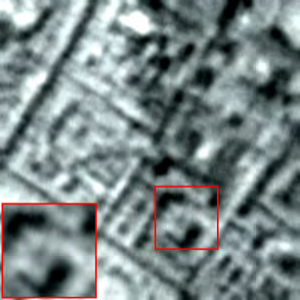}&
			\includegraphics[width=0.135\textwidth]{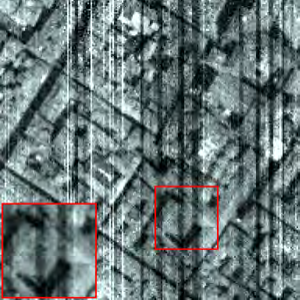}&
			\includegraphics[width=0.135\textwidth]{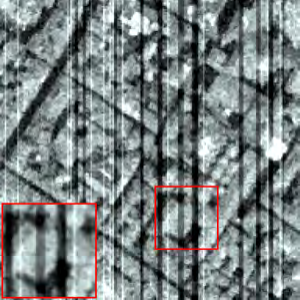}&
			\includegraphics[width=0.135\textwidth]{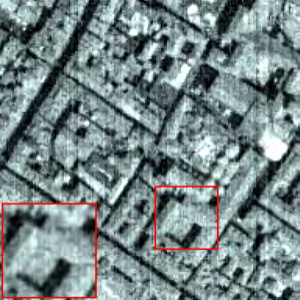}&
			\includegraphics[width=0.135\textwidth]{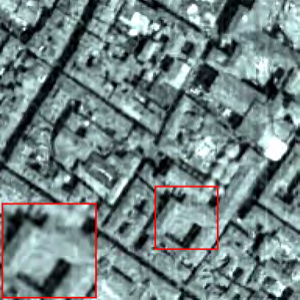}&
			\includegraphics[width=0.135\textwidth]{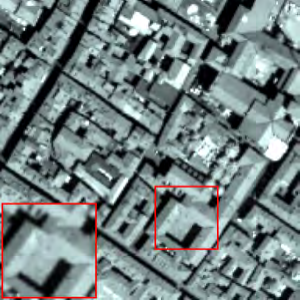}\\
			
			\includegraphics[width=0.135\textwidth]{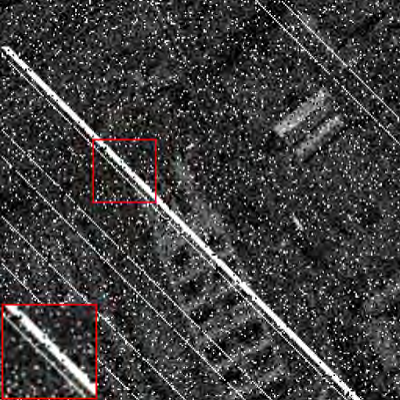}&
			\includegraphics[width=0.135\textwidth]{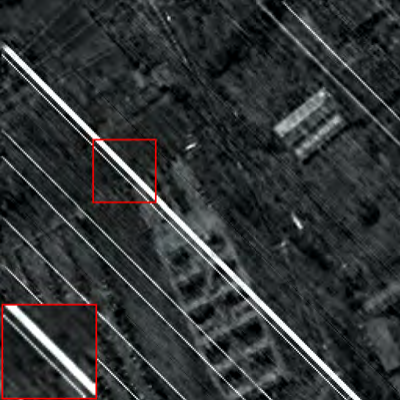}&
			\includegraphics[width=0.135\textwidth]{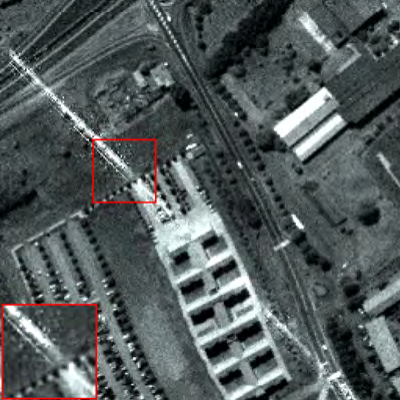}&
			\includegraphics[width=0.135\textwidth]{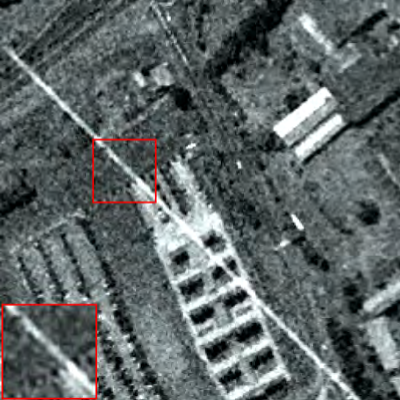}&
			\includegraphics[width=0.135\textwidth]{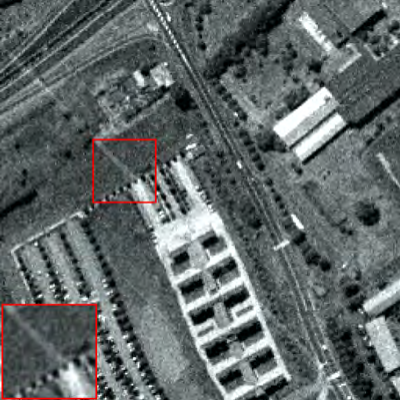}&
			\includegraphics[width=0.135\textwidth]{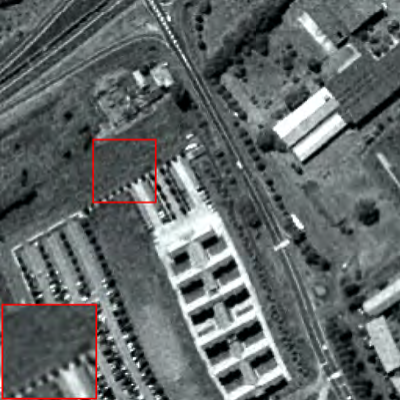}&			
			\includegraphics[width=0.135\textwidth]{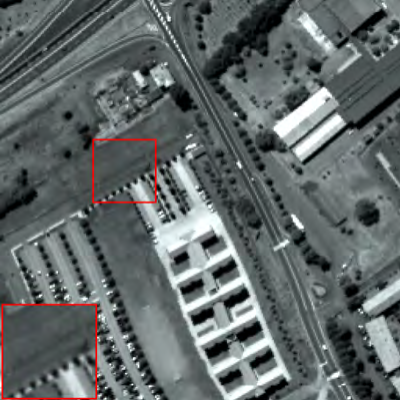}\\
			
			\includegraphics[width=0.135\textwidth]{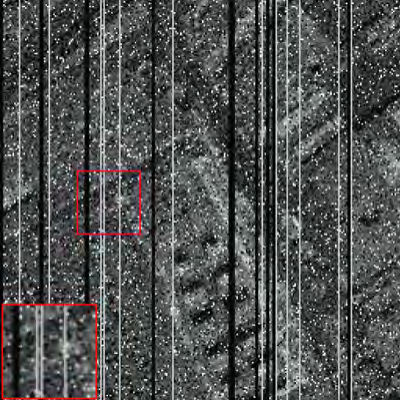}&
			\includegraphics[width=0.135\textwidth]{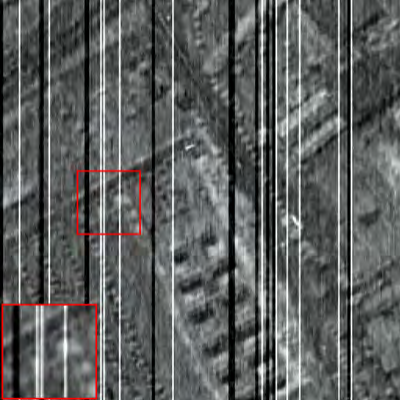}&
			\includegraphics[width=0.135\textwidth]{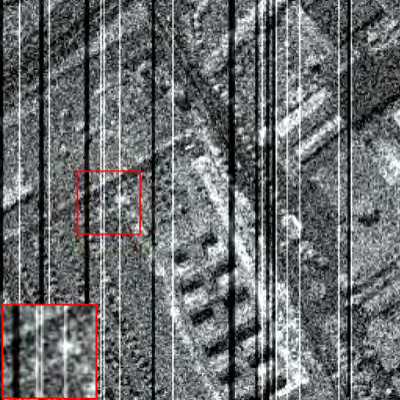}&
			\includegraphics[width=0.135\textwidth]{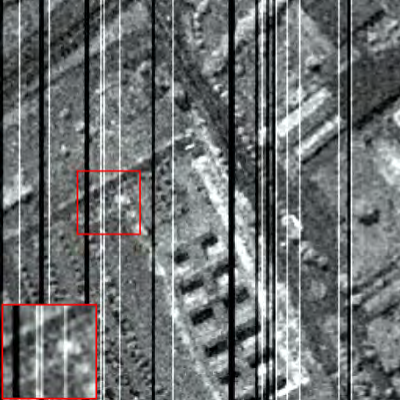}&
			\includegraphics[width=0.135\textwidth]{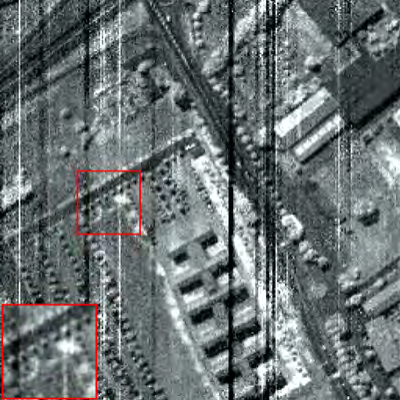}&
			\includegraphics[width=0.135\textwidth]{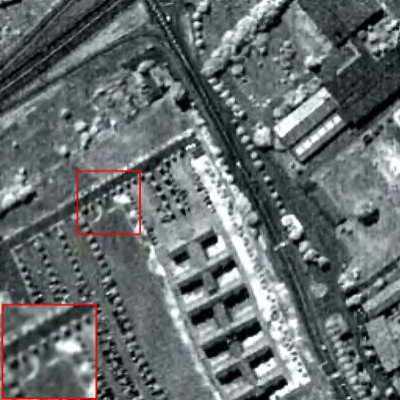}&			
			\includegraphics[width=0.135\textwidth]{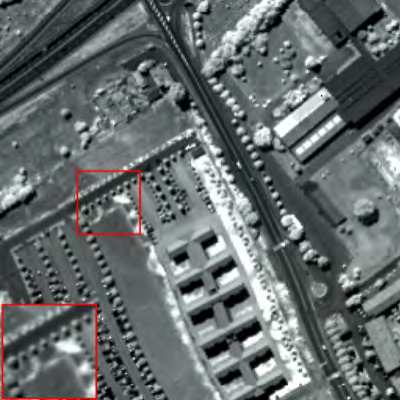}\\
			
		\end{tabular}
		\caption{Denoising results obtained by different methods. From left to right: the observed image, the denoising results by \texttt{DIP2D}, \texttt{LRMR}, \texttt{LRTDTV}, \texttt{LRTFL0}, \texttt{DS2DP} (proposed), and the ground truth, respectively. The first two rows are the denoising results on WDC Mall for Cases 4 and 6, respectively.  The second two rows are the denoising results on Pavia Centre for Cases 4 and 6, respectively. The last two rows are the denoising results on Pacia University for Cases 4 and 6, respectively.}
		
		\label{stimulate}
	\end{figure*}
	
	\begin{figure*}[!htbp]
		\footnotesize
		\setlength{\tabcolsep}{1pt}
		\centering
		\begin{tabular}{cccccccc}
			Observed & DIP2D & LRMR & LRTDTV & LRTFL0  & DS2DP & Ground truth\\
			
			\includegraphics[width=0.135\textwidth]{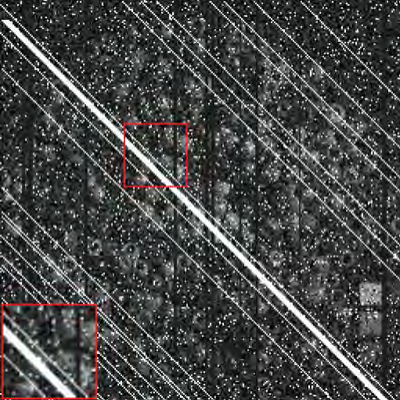}&
			\includegraphics[width=0.135\textwidth]{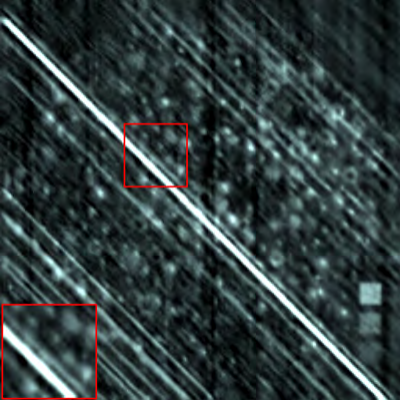}&
			\includegraphics[width=0.135\textwidth]{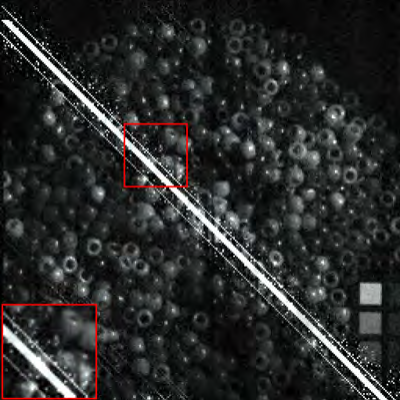}&
			\includegraphics[width=0.135\textwidth]{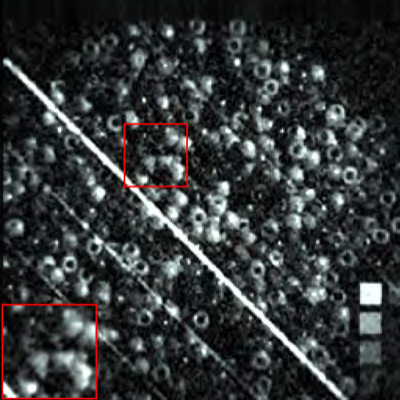}&
			\includegraphics[width=0.135\textwidth]{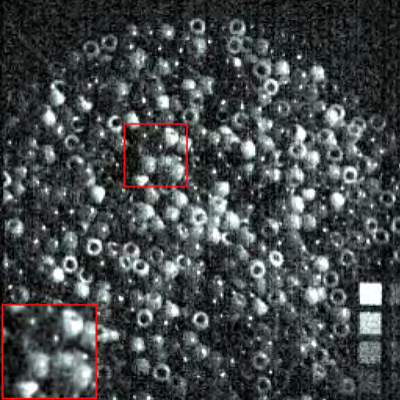}&
			\includegraphics[width=0.135\textwidth]{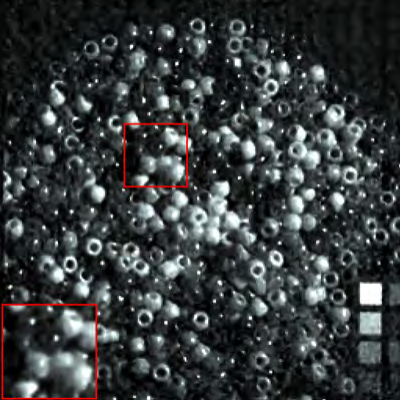}&
			\includegraphics[width=0.135\textwidth]{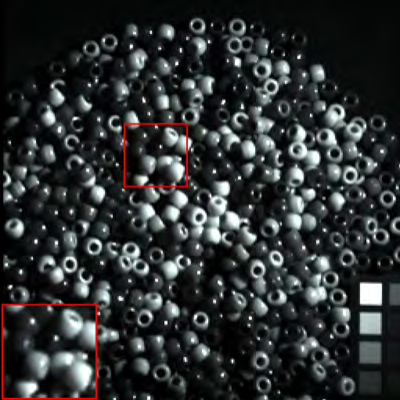}\\
			
			
			\includegraphics[width=0.135\textwidth]{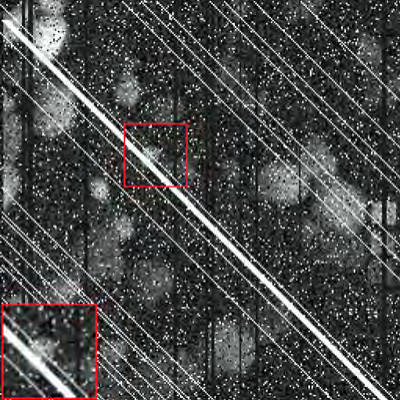}&		\includegraphics[width=0.135\textwidth]{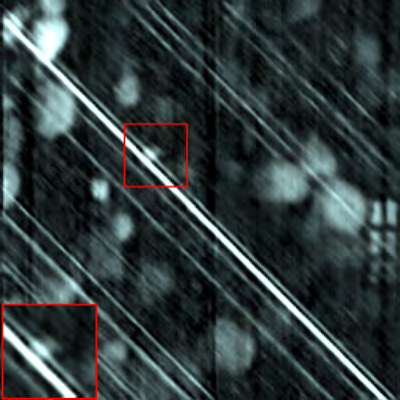}&
			\includegraphics[width=0.135\textwidth]{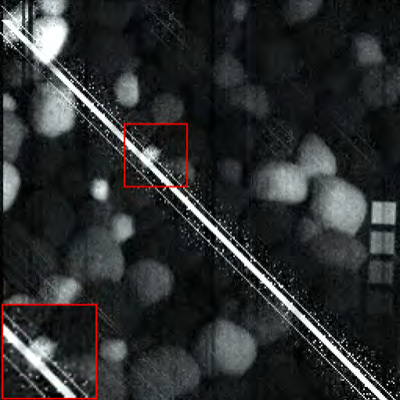}&
			\includegraphics[width=0.135\textwidth]{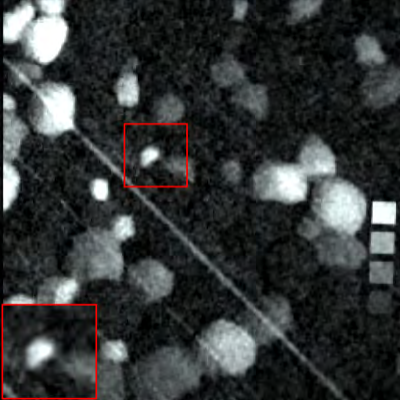}&
			\includegraphics[width=0.135\textwidth]{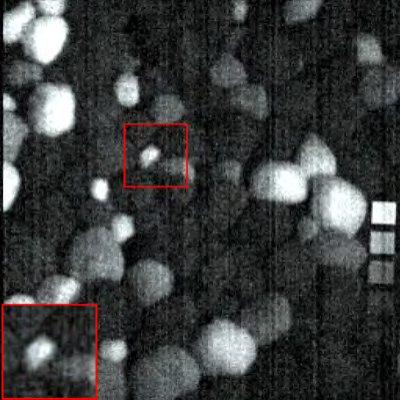}&
			\includegraphics[width=0.135\textwidth]{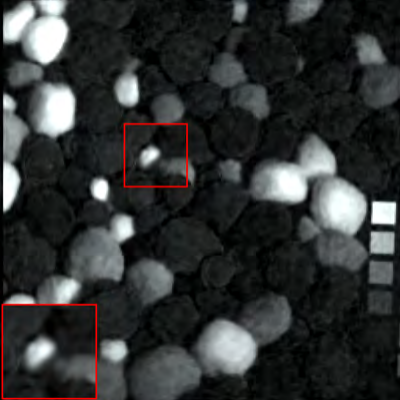}&
			\includegraphics[width=0.135\textwidth]{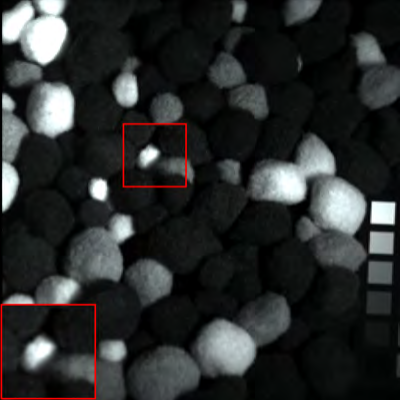}\\
			
			
			\includegraphics[width=0.135\textwidth]{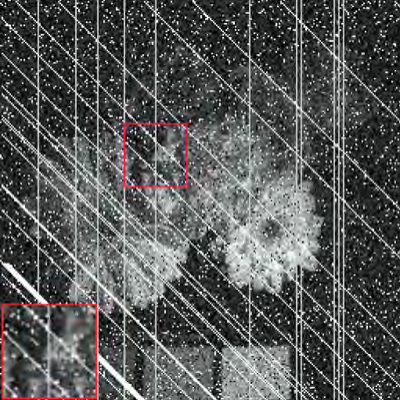}&
			\includegraphics[width=0.135\textwidth]{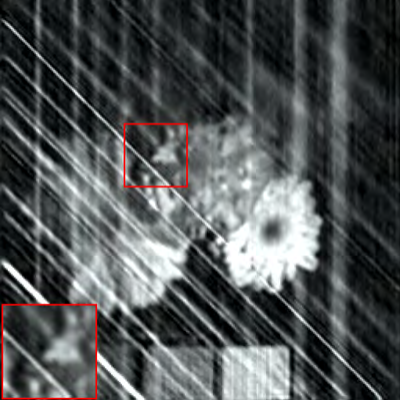}&
			\includegraphics[width=0.135\textwidth]{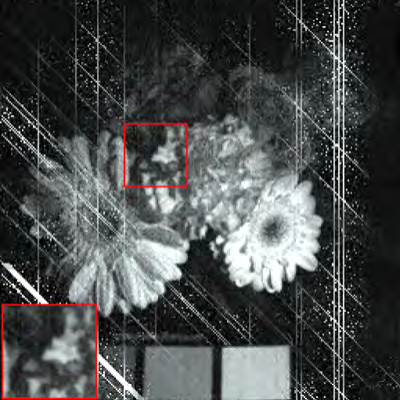}&
			\includegraphics[width=0.135\textwidth]{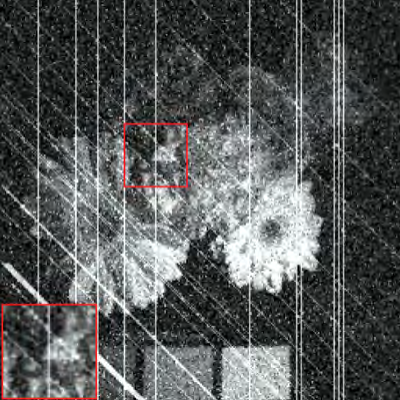}&
			\includegraphics[width=0.135\textwidth]{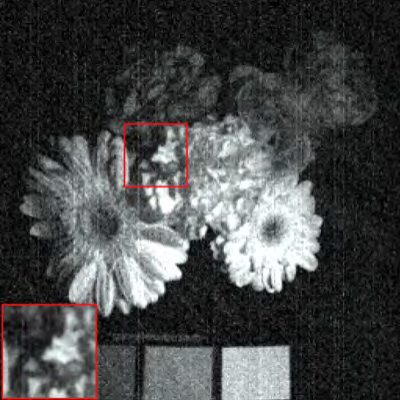}&
			\includegraphics[width=0.135\textwidth]{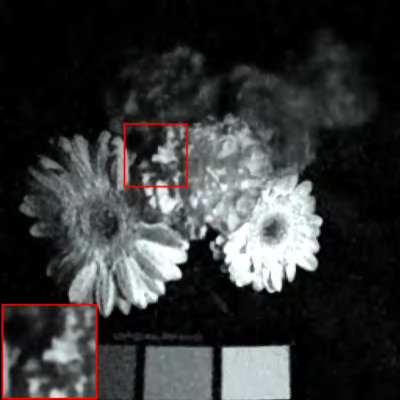}&
			\includegraphics[width=0.135\textwidth]{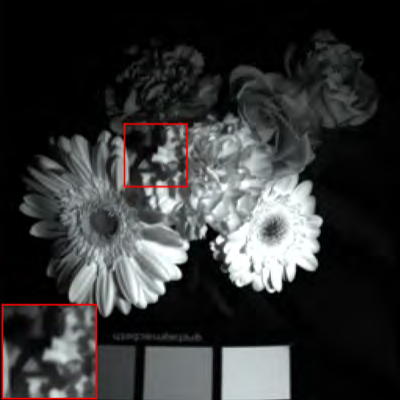}\\
			
		\end{tabular}
		\caption{Denoising results obtained by different methods for Case 6. From top to bottom: the band 4 of Beads, the band 4 of Pompoms, and the band 31 of Flowers, respectively. From left to right: the observed image, the denoising results by \texttt{DIP2D}, \texttt{LRMR}, \texttt{LRTDTV}, \texttt{LRTFL0}, \texttt{DS2DP}, and the ground truth, respectively.}
		\label{omsi}
	\end{figure*}
	
	\begin{figure*}[!htbp]
		\footnotesize
		\setlength{\tabcolsep}{1.2pt}
		\centering
		\begin{tabular}{ccccccc}
			DIP2D & DIP3D & LRMR & LRTDTV & LRTFL0  & DS2DP \\
			
			\includegraphics[width=0.16\textwidth]{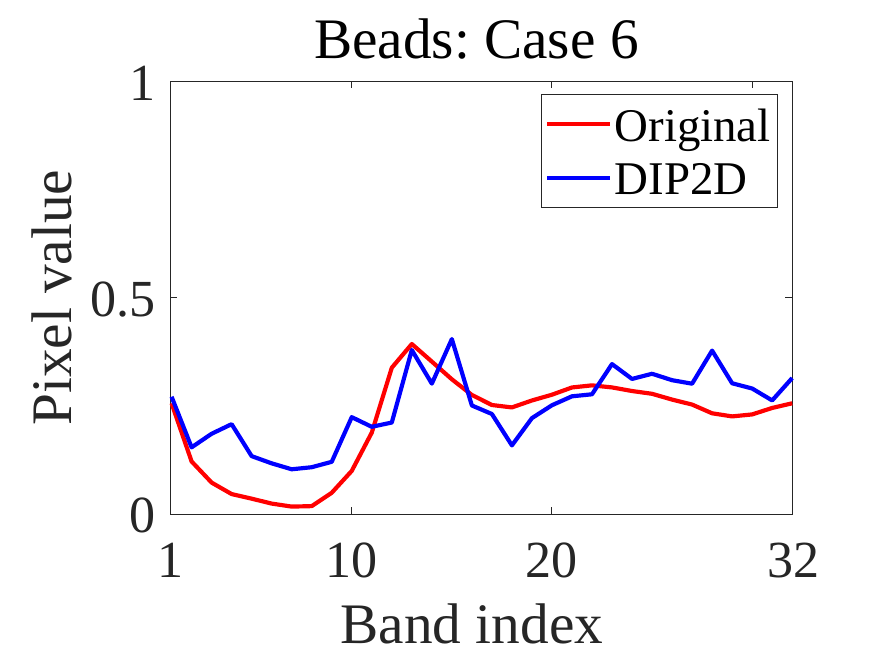}&
			\includegraphics[width=0.16\textwidth]{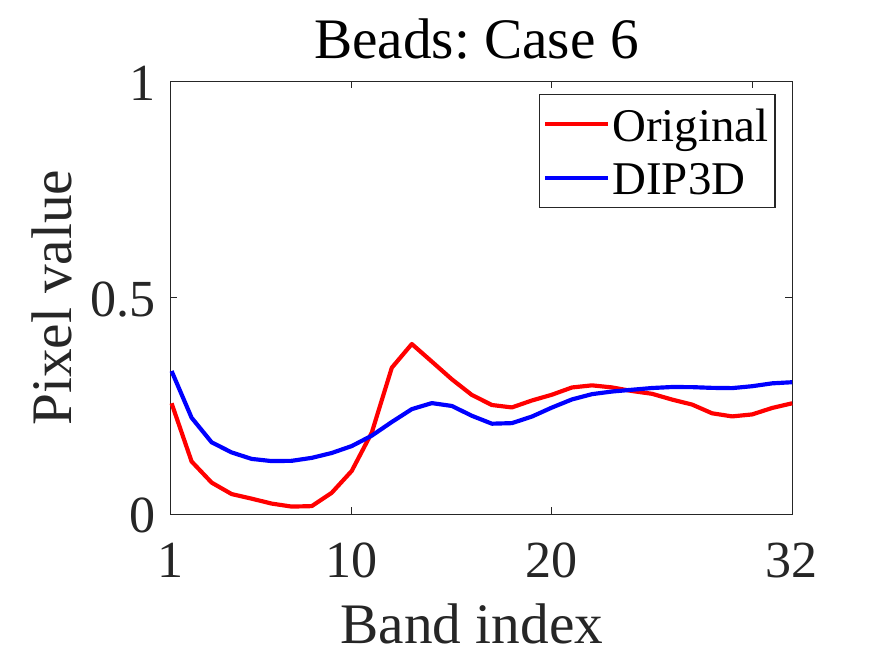}&
			\includegraphics[width=0.16\textwidth]{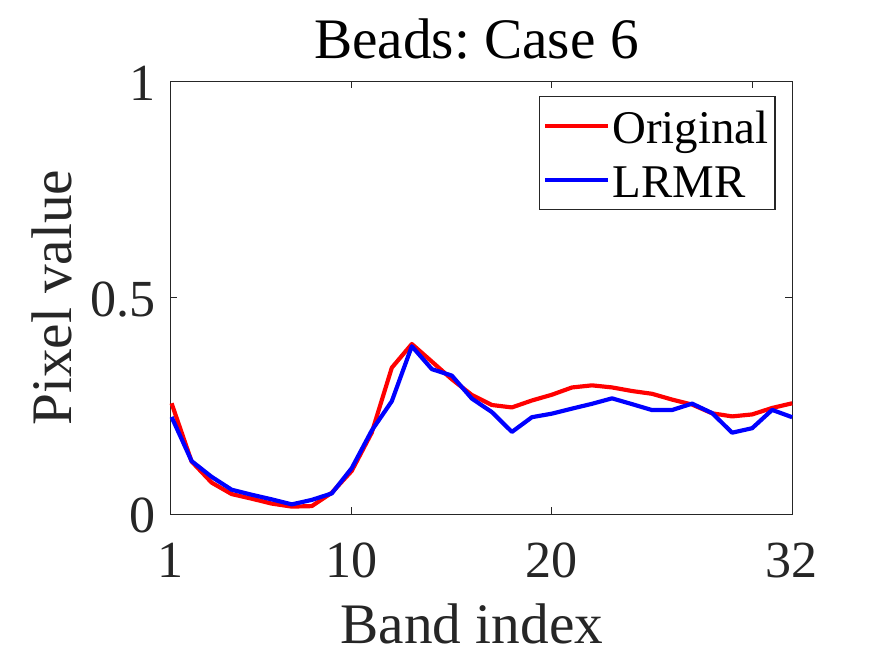}&
			\includegraphics[width=0.16\textwidth]{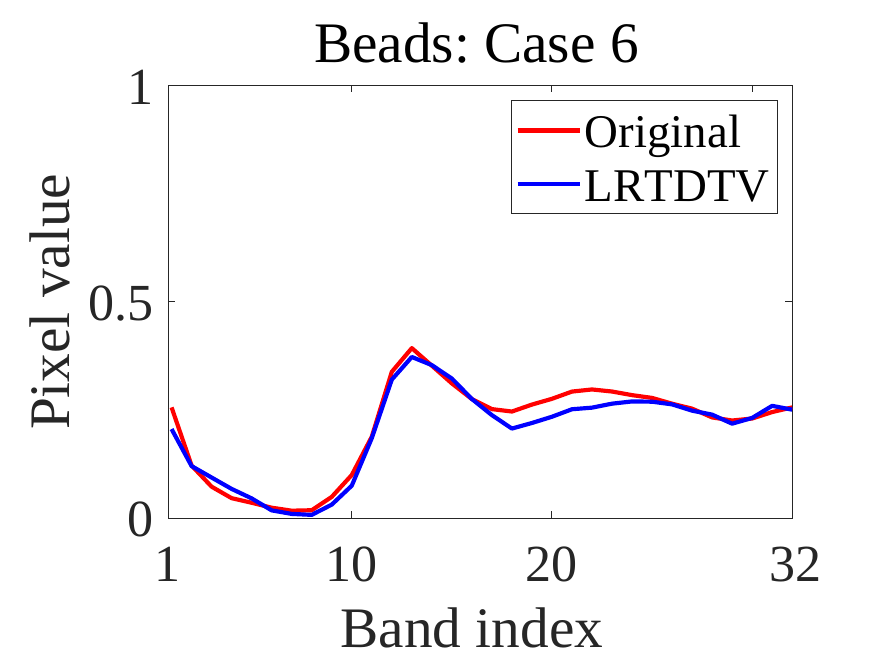}&
			\includegraphics[width=0.16\textwidth]{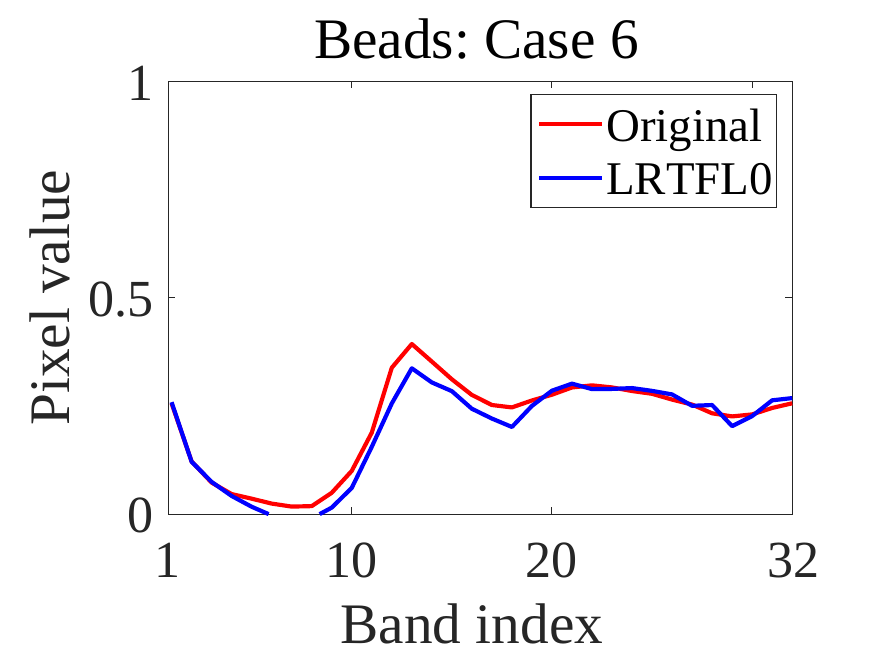}&
			\includegraphics[width=0.16\textwidth]{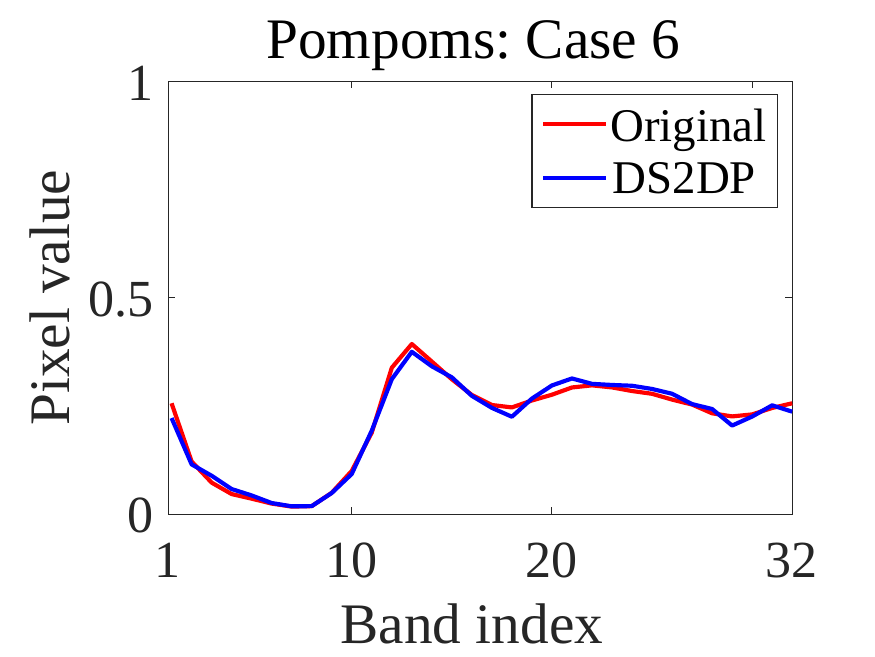}&\\
			\includegraphics[width=0.16\textwidth]{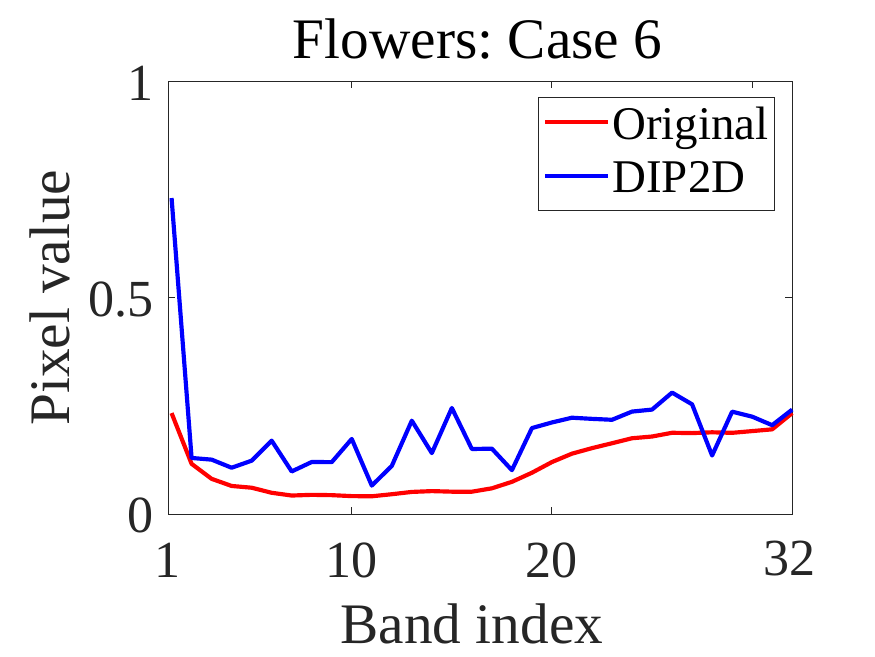}&
			\includegraphics[width=0.16\textwidth]{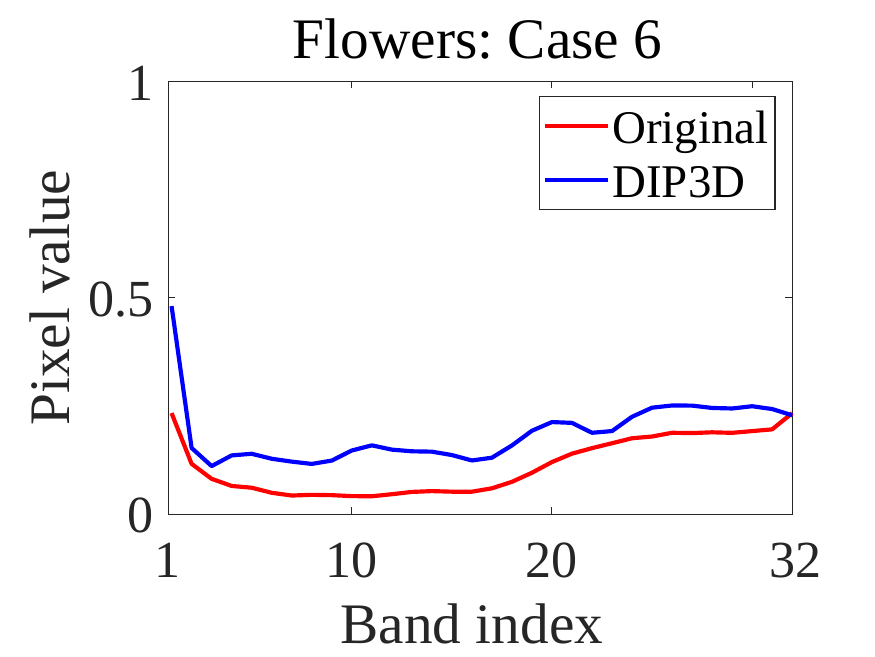}&
			\includegraphics[width=0.16\textwidth]{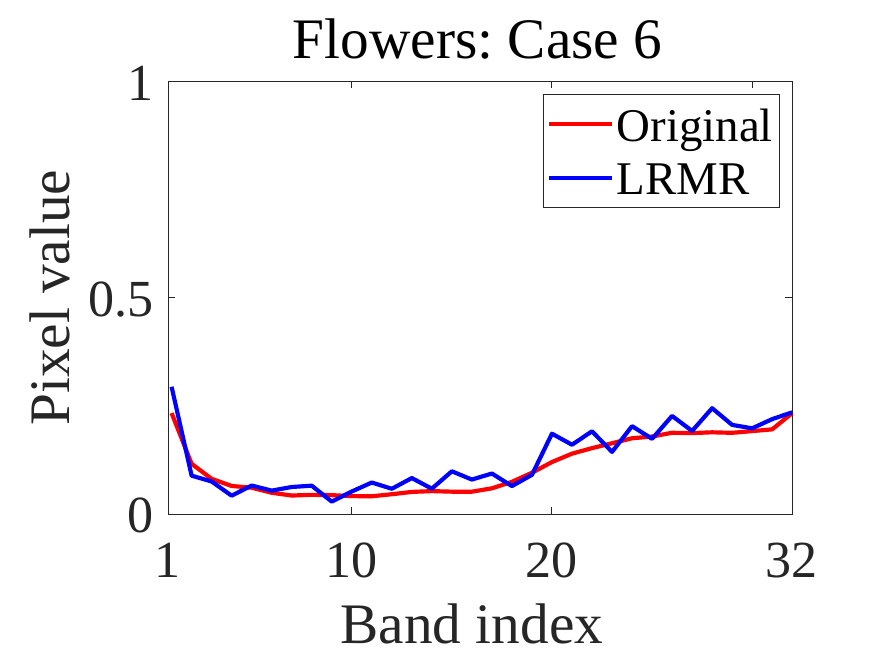}&
			\includegraphics[width=0.16\textwidth]{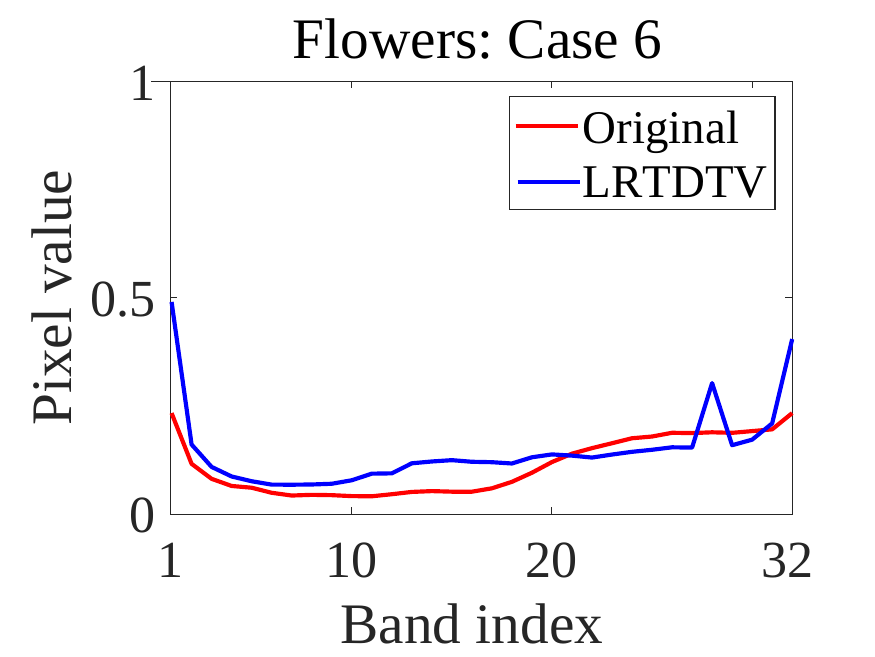}&
			\includegraphics[width=0.16\textwidth]{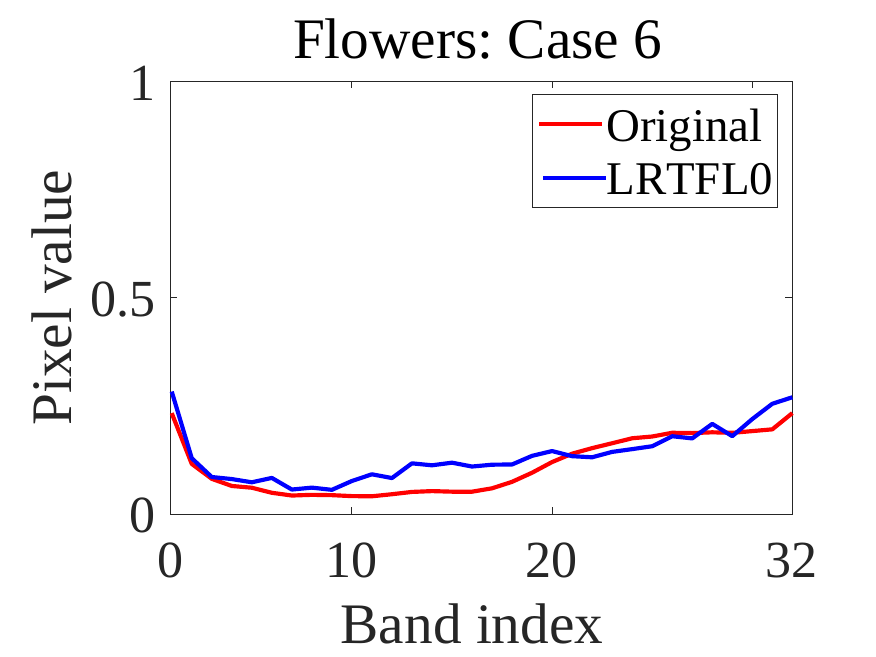}&
			\includegraphics[width=0.16\textwidth]{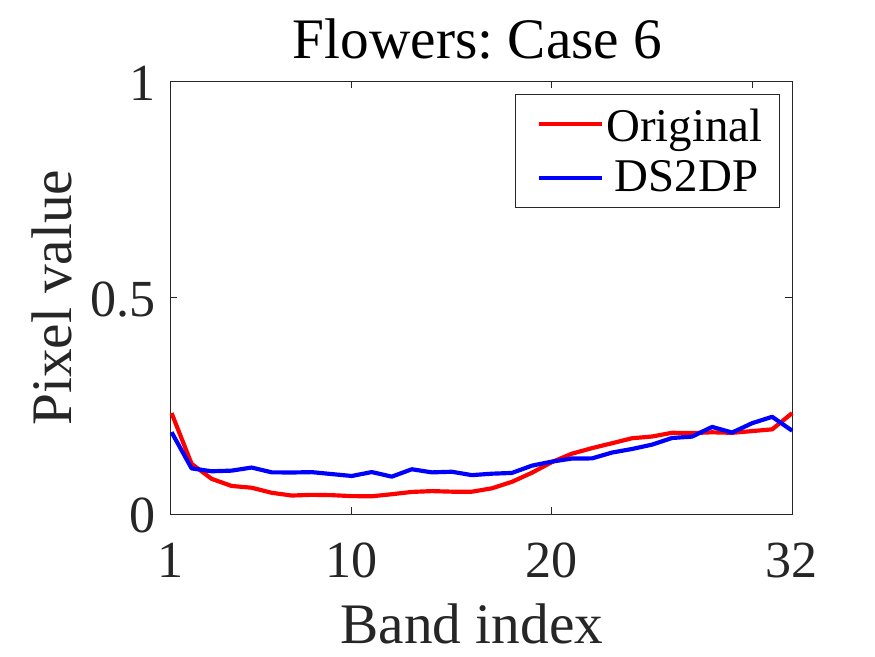}&\\
			\includegraphics[width=0.16\textwidth]{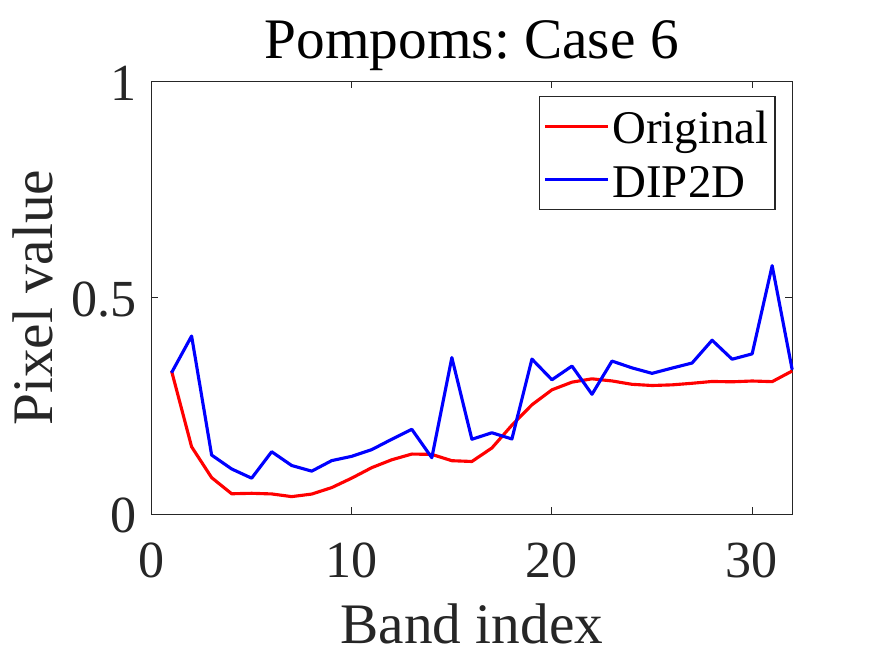}&
			\includegraphics[width=0.16\textwidth]{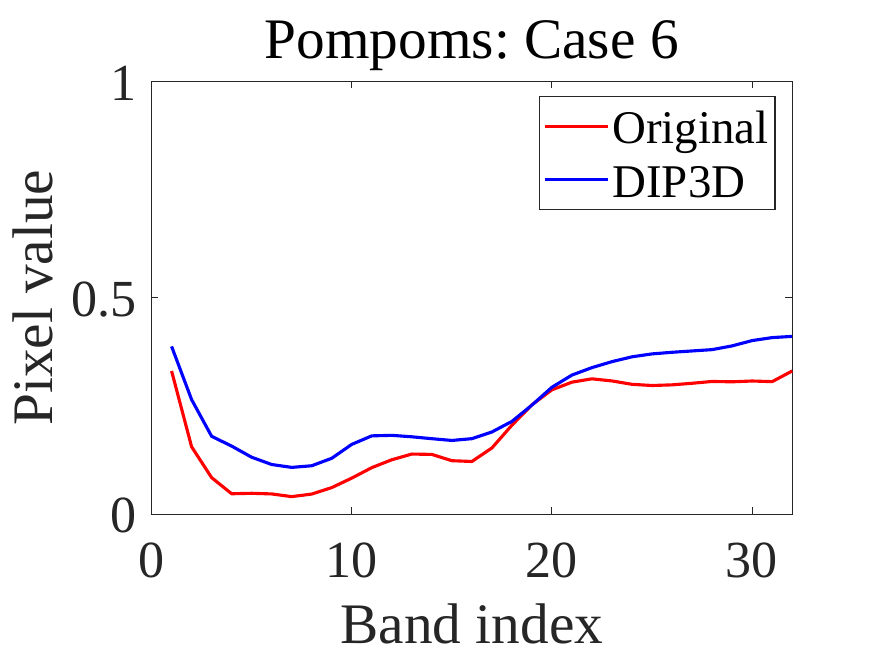}&
			\includegraphics[width=0.16\textwidth]{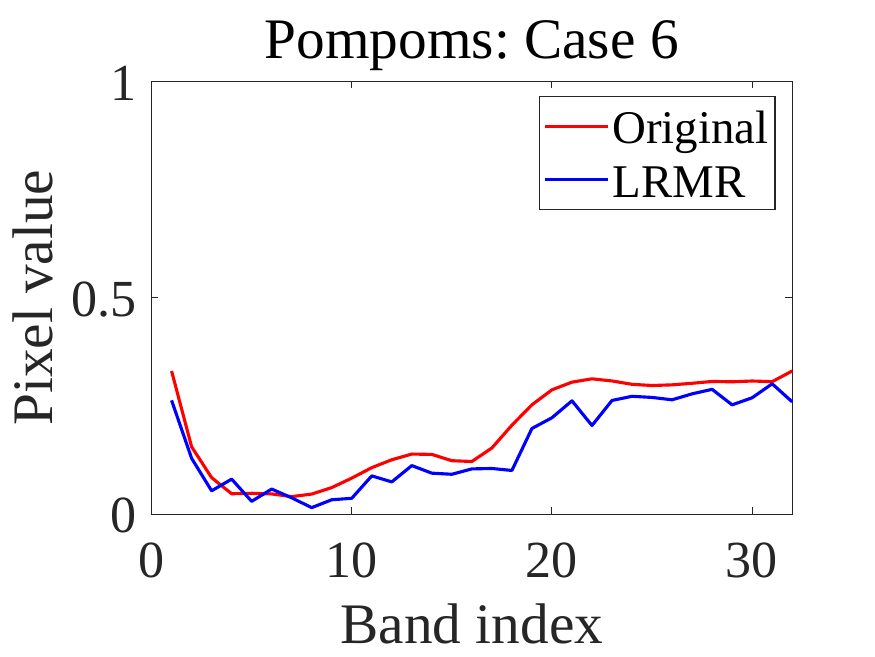}&
			\includegraphics[width=0.16\textwidth]{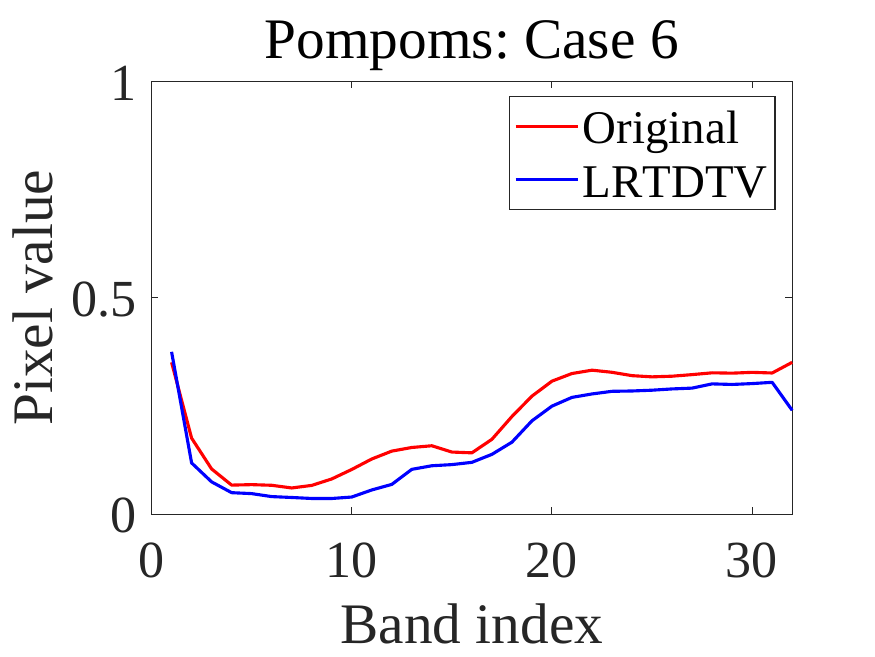}&
			\includegraphics[width=0.16\textwidth]{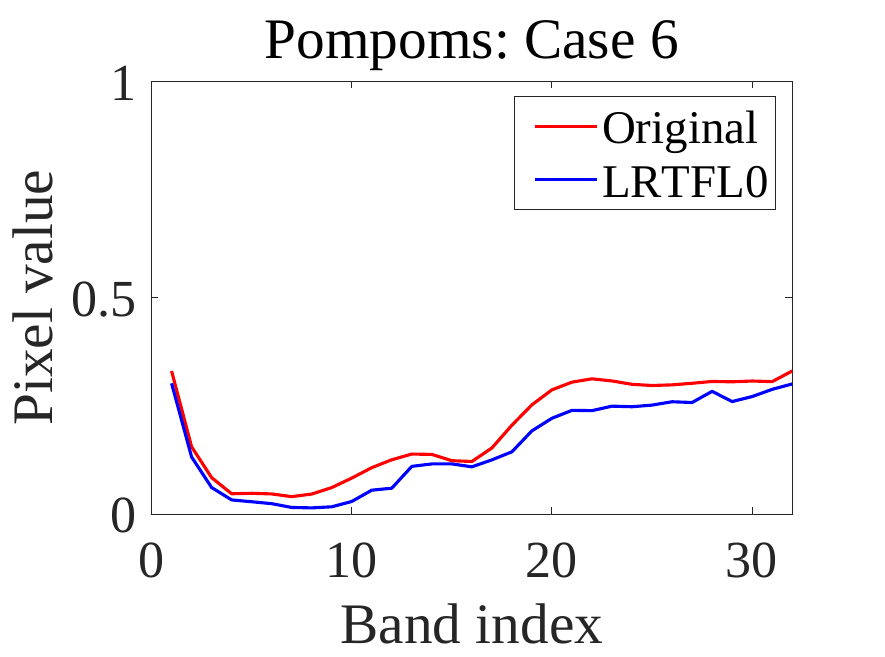}&
			\includegraphics[width=0.16\textwidth]{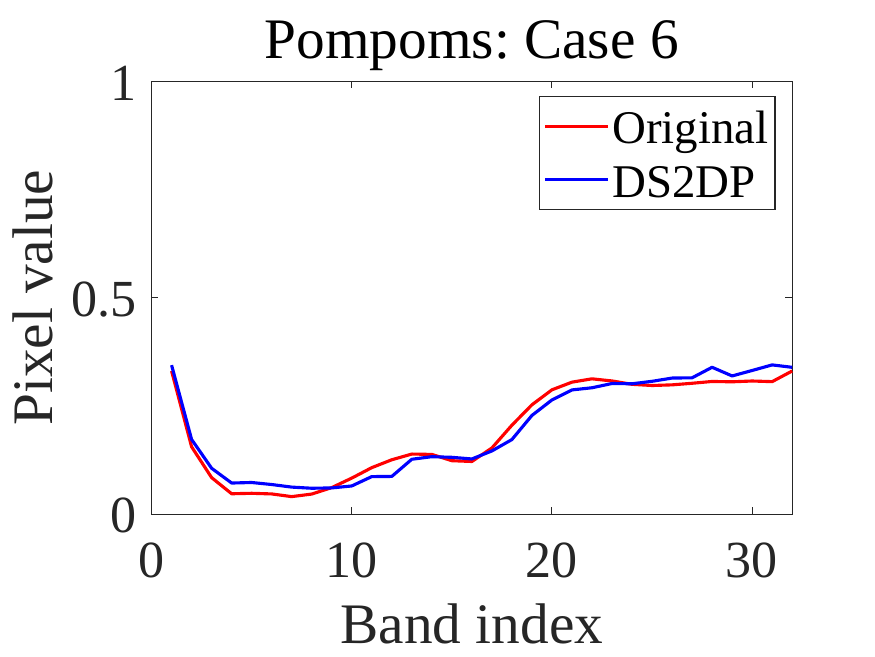}&\\
			
		\end{tabular}
		\caption{Spectral curves of the denoising results by different compared methods for Case 6. From left to right: the results by \texttt{DIP2D}, \texttt{DIP3D}, \texttt{LRMR}, \texttt{LRTDTV}, \texttt{LRTFL0}, and \texttt{DS2DP}, respectively. From top to bottom: the results at spatial location (55, 60) of MSI Beads, the results at spatial location (90, 45) of the MSI Flowers, and the results at spatial location (200, 200) of the MSI Pompoms, respectively. }
		\label{spectral_curve}
	\end{figure*}
	
	\begin{figure*}[]
		\footnotesize
		\setlength{\tabcolsep}{1pt}
		\centering
		\begin{tabular}{ccccccc}
			Observed & DIP2D & LRMR & LRTDTV  & LRTFL0  &DS2DP \\
			
			\includegraphics[width=0.16\textwidth]{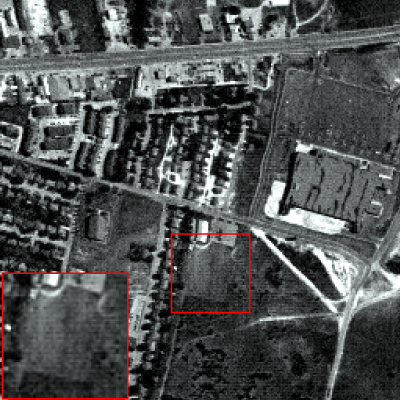}&
			\includegraphics[width=0.16\textwidth]{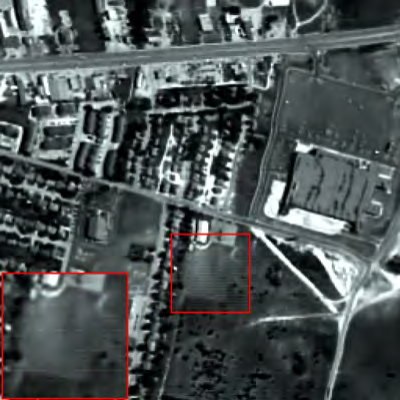}&
			\includegraphics[width=0.16\textwidth]{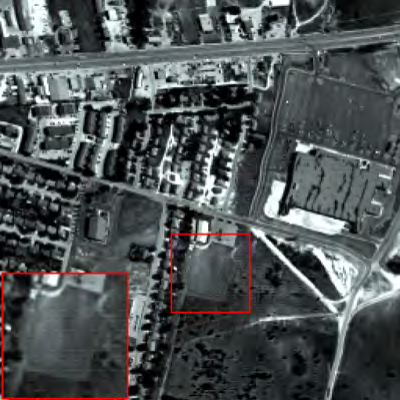}&
			\includegraphics[width=0.16\textwidth]{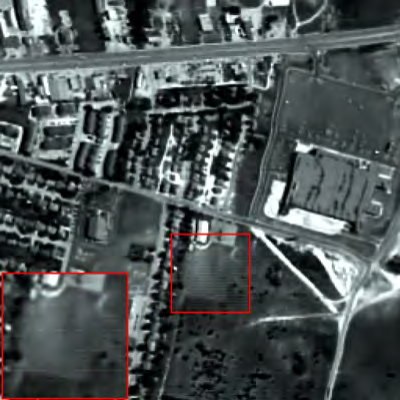}&
			\includegraphics[width=0.16\textwidth]{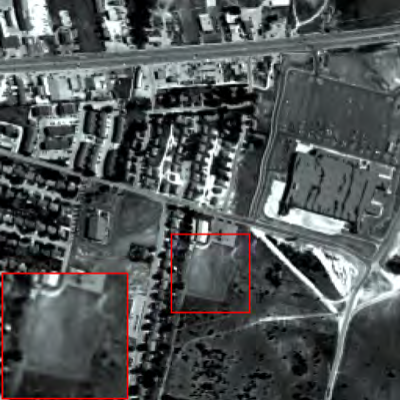}&
			\includegraphics[width=0.16\textwidth]{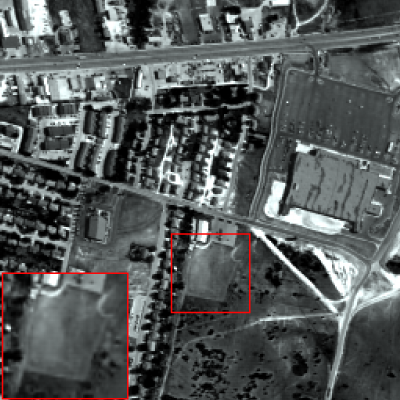}\\
			
			\includegraphics[width=0.16\textwidth]{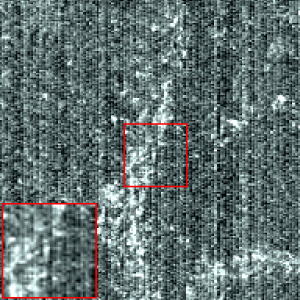}&
			\includegraphics[width=0.16\textwidth]{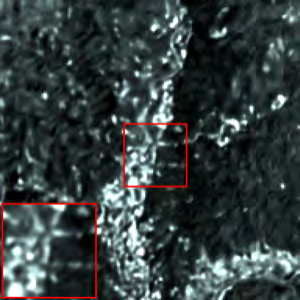}&
			\includegraphics[width=0.16\textwidth]{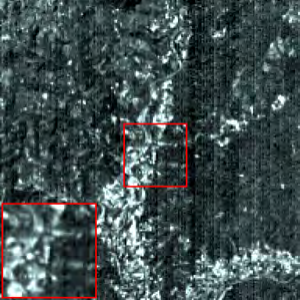}&
			\includegraphics[width=0.16\textwidth]{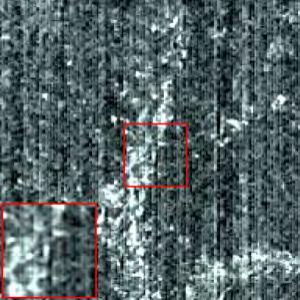}&
			\includegraphics[width=0.16\textwidth]{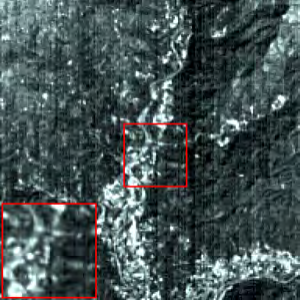}&
			\includegraphics[width=0.16\textwidth]{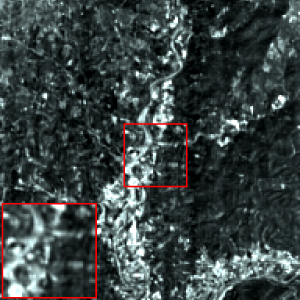}\\
			
			\includegraphics[width=0.16\textwidth]{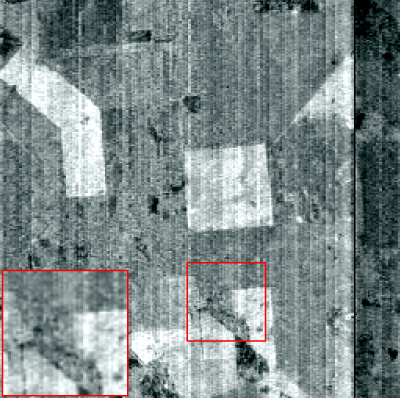}&
			\includegraphics[width=0.16\textwidth]{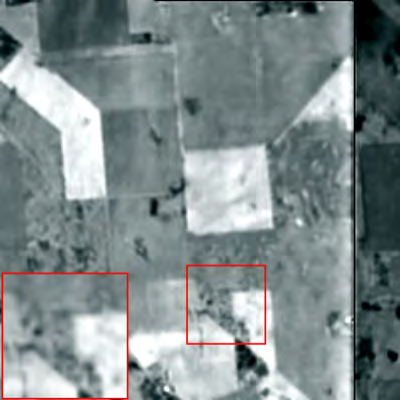}&
			\includegraphics[width=0.16\textwidth]{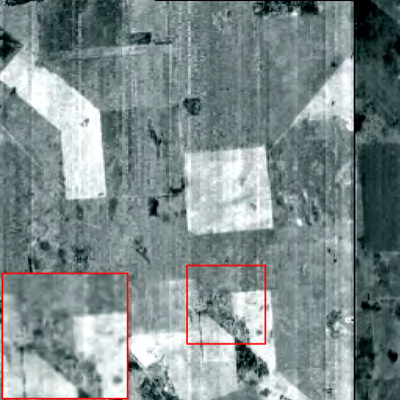}&
			\includegraphics[width=0.16\textwidth]{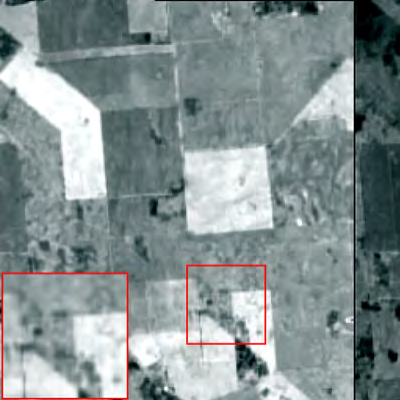}&
			\includegraphics[width=0.16\textwidth]{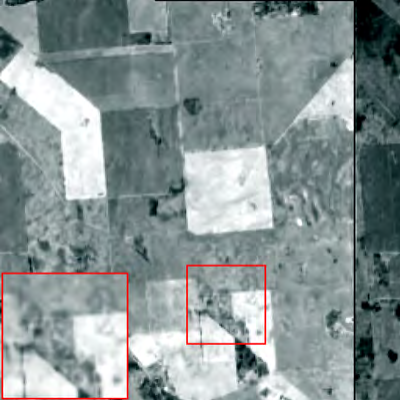}&
			\includegraphics[width=0.16\textwidth]{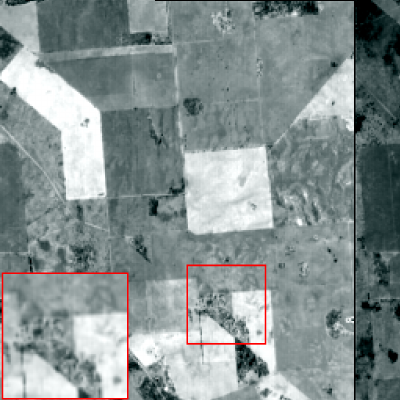}\\
			
			\includegraphics[width=0.16\textwidth]{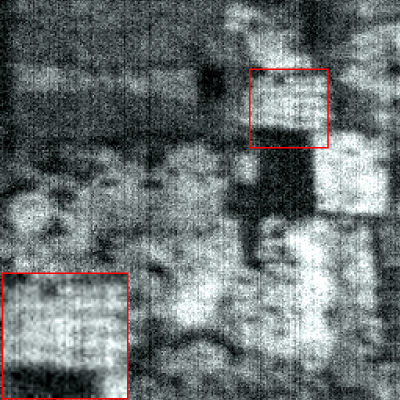}&
			\includegraphics[width=0.16\textwidth]{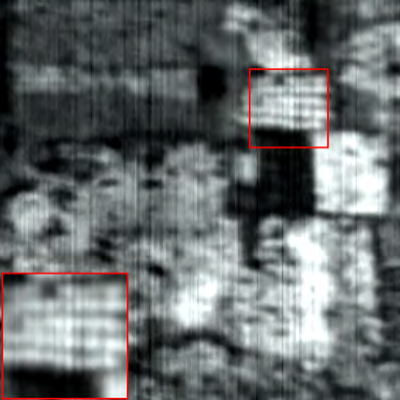}&
			\includegraphics[width=0.16\textwidth]{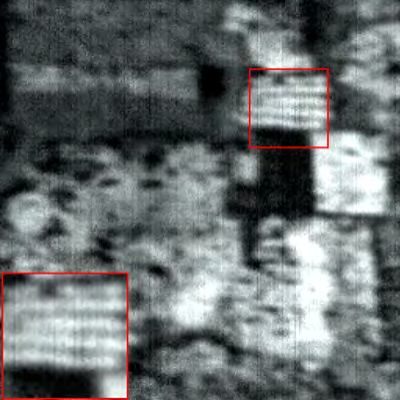}&
			\includegraphics[width=0.16\textwidth]{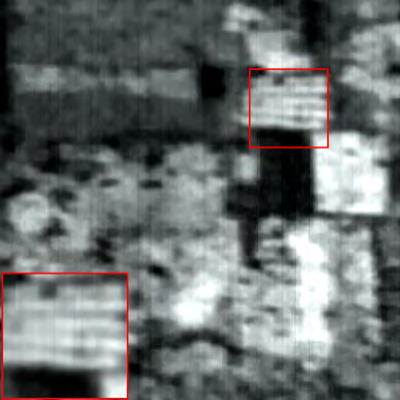}&
			\includegraphics[width=0.16\textwidth]{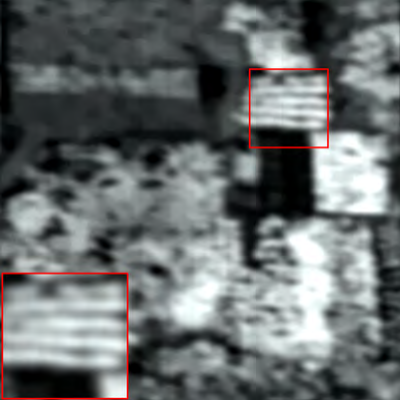}&
			\includegraphics[width=0.16\textwidth]{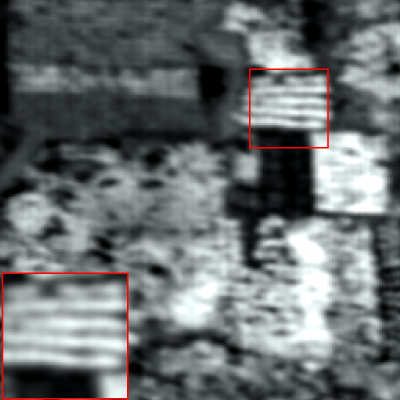}\\
			
			\includegraphics[width=0.16\textwidth]{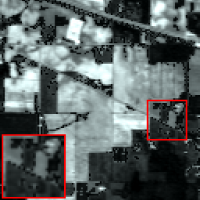}&
			\includegraphics[width=0.16\textwidth]{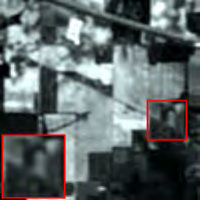}&
			\includegraphics[width=0.16\textwidth]{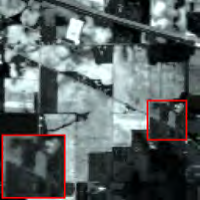}&
			\includegraphics[width=0.16\textwidth]{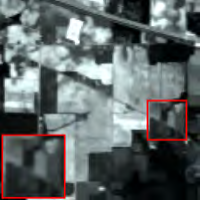}&
			\includegraphics[width=0.16\textwidth]{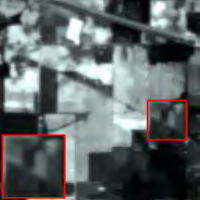}&
			\includegraphics[width=0.16\textwidth]{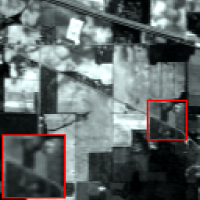}\\
		\end{tabular}
		\caption{Denoising results by different methods on Urban dataset, EO-1, Australian dataset, WHU HongHu dataset, and Indian Pines dataset. From top to bottom:  the band 203 of Urban dataset, the band 132 of EO-1 dataset, the band 87 of Australian dataset, the band 37 of WHU HongHu dataset, and the band 24 of the Indian Pines dataset, respectively. From left to right: the observed image, the results by \texttt{DIP2D}, \texttt{LRMR}, \texttt{LRTDTV}, \texttt{LRTFL0}, and \texttt{DS2DP}, respectively.}
		\label{real}
	\end{figure*}
	
	\section{Experiments}\label{4}
	In this section, we use simulated and real data to demonstrate the effectiveness of the proposed approach.
	
	\subsection{Baselines}

	To thoroughly evaluate the performance of \texttt{DS2DP}, we implemented five state-of-the-art methods as the baselines. These methods include two unsupervised methods, i.e., {\it deep image prior based on 2D convolution} (\texttt{DIP2D}) \cite{HSIDIP} and {\it deep image prior based on 3D convolution} (\texttt{DIP3D}) \cite{HSIDIP}, a matrix optimization-based method, i.e., {\it hyperspectral image restoration using low-rank matrix recovery} (\texttt{LRMR}) \cite{LRMR}, and two tensor optimization-based methods, i.e., {\it TV-regularized low-rank tensor decomposition} (\texttt{LRTDTV}) \cite{LRTDTV} and  {\it hyperspectral restoration via $ L_0 $ gradient regularized low-rank tensor factorization} (\texttt{LRTFL0}) \cite{LRTF_HR}. 
	
	For \texttt{DIP2D} and \texttt{DIP3D}, we set the maximum number of iterations to be 6,000 and report the best results during the iterations. For the proposed \texttt{DS2DP}, we set the maximum number of iterations to be 6,000 and report the results at the 6,000th iteration. For \texttt{LRMR}, \texttt{LRTDTV}, and \texttt{LRTFL0}, their parameters are set as suggested in  \cite{LRMR,LRTDTV,LRTF_HR}---with parameter fine-tuning effort to uplift its performance in some cases. The experiments of \texttt{DIP2D}, \texttt{DIP3D}, and \texttt{DS2DP} are executed using \texttt{Python} on a computer with a six-core Intel(R) Core(TM) i7-9750H CPU @ 2.60GHz, 32.0 GB of RAM, and an NVIDIA GeForce RTX 2070 GPU. The experiments of \texttt{LRMR}, \texttt{LRTDTV}, and \texttt{LRTFL0} are implemented in Matlab (2019a) on the same computer.
	
	\subsection{Simulated Data Experiments}
	\noindent
	\textbf{Evaluation Metrics.} We adopt three frequently used evaluation metrics, namely, peak signal-to-noise ratio (PSNR), structure similarity (SSIM), and spectral angle mapper (SAM) \cite{LRTF_HR}. Generally, better-restored denoising performance is reflected by higher PSNR and SSIM values and lower SAM values.
	
	\noindent
	\textbf{Simulated Data.} For simulated data, we use a number of HSIs to serve as our ground truth, which include Washington DC Mall (WDC Mall) \footnote{\url{http://lesun.weebly.com/hyperspectral-data-set.html}\label{HSI}} of size 256 $ \times $ 256 $ \times $ 191, Pavia Centre \footref{HSI} of size 200 $ \times $ 200 $ \times $ 80 that is clipped into 192 $ \times $ 192 $ \times $ 80, and Pavia University \footref{HSI} of size 256 $ \times $ 256 $ \times $ 87. The multispectral images (MSIs) in the CAVE dataset \footnote{\url{https://www.cs.columbia.edu/CAVE/databases/multispectral/}} of size 256 $\times$ 256 $\times$ 31 are also used to serve as our clean data $\underline{\bm  X }_\natural$.
	
	\noindent
	\textbf{Scenarios.}
	We consider a series of scenarios with various types of noise:
	
	\noindent  
	$ Case\ 1 $ \textit{(Gaussian Noise)}: In this basic scenario, the i.i.d. zero-mean Gaussian noise is added to all bands with the variance set to be 0.1. The signal-to-noise ratios (SNRs) (see definition in \cite{snr}) associated with different datasets can be found in Table \ref{tab1}. One can see the noise levels in different datasets are similar. Note that the HSIs with SNR being 6dB to 8dB are considered as severely corrupted data.
	
	\begin{table}[!htbp]
		\caption{The SNR of the degraded images for Case 1.}
		\centering
		\begin{tabular}{c|cccc}
			\Xhline{1.0pt}
			Data & WDC Mall & Pavia Centre & Pavia University & CAVE \bigstrut\\
			\hline
			SNR & 7.196 & 7.691 & 6.297 & 6.318\bigstrut\\
			\Xhline{1.0pt}
		\end{tabular}
		\label{tab1}
	\end{table}
	
	\noindent  
	$ Case\ 2 $ \textit{(Gaussian Noise + Impulse Noise)}: In this case, the Gaussian noise for Case 1 is kept. We also additionally consider impulse noise that often happens in real HSI analysis. The impulsive noise is also added to each band. Such noise is generated following the i.i.d. zero-mean Laplacian distribution with the density parameter being 0.1.
	
	\noindent  
	$ Case\ 3 $ \textit{(Gaussian Noise + Impulse Noise + Deadlines)}: To make the case more challenging, we include deadlines on top of Case 2; see Fig. \ref{stimulate} for illustration of deadlines. The deadlines are generated by nullifying some selected pixels and bands. We assume that the deadlines randomly affect 30\% of the bands. Moreover, for each selected band, the number of deadlines is randomly generated from 10 to 15, and the spatial width of the deadlines is randomly selected from 1 to 3 pixels.
	
	\noindent  
	$ Case\ 4 $ \textit{(Gaussian Noise + Impulse Noise + Diagonal Stripes)}: In this case, we replace the deadlines for Case 3 by diagonal stripes; see Fig. \ref{stimulate} for illustration. The elements of the diagonal stripes are all ones, which are used to simulate the constant brightness. As before, we assume that the stripes effect 30\% of the bands.  Moreover, for each selected band, the number of diagonal stripes is randomly generated from 15 to 30.
	
	\noindent  
	$ Case\ 5 $ \textit{(Gaussian Noise + Impulse Noise + Vertical Stripes)}: In this case, we use the setting as for Case 4, except that vertical (other than diagonal) stripes are added; see Fig. \ref{stimulate}. For each affected band, the number of vertical stripes is randomly generated from 10 to 15.  In this case, the elements of each vertical stripe are set to a certain value randomly generated from the range of [0.6, 0.8], to diversify our simulated scenarios.
	
	\noindent  
	$ Case\ 6 $ \textit{(Gaussian Noise + Impulse Noise + Deadlines + Diagonal Stripes + Vertical Stripes)}: To create an extra challenging case, Gaussian noise, impulse noise, and deadlines are added as for Case 3. Moreover, diagonal stripes and vertical stripes are added as for Case 4 and Case 5, respectively.
	
	\noindent
	\textbf{Parameter Setting.}
	In \texttt{DS2DP}, there are two parameters to be manually tuned, namely,  $\lambda$ and $ {R} $. For the parameter $ \lambda $, we generally set it as $ i \times 10^j\ (i =  2, \ 5,\ 8;\ j = -6, -5, -4, -3, -2)$ for Cases 1-6. Regarding the parameter $ R $, which is the number of endmembers in the HSI and can be determined by many existing algorithms, e.g., \cite{HySime, 7055246, Fu2016Unmixing}. 
	
	\noindent
	\textbf{Quantitative Comparison.}
	Table \ref{tab} lists the quantitative comparisons of the competing methods for Cases 1-6. 
	The symbol ``*'' in Table \ref{tab} means that the corresponding methods have exhausted the computational resources (memory or time) but still could not produce sensible results. For the CAVE dataset, we report the averaged evaluation results from 32 images. From Table \ref{tab}, it is easy to see that \texttt{DS2DP} outperforms the state-of-the-art approaches in most cases, in terms of PSNR, SSIM, and SAM. For example, for Case 1, \texttt{DS2DP} achieves around 1.4 dB gain in PSNR compared to the second-best method (\texttt{LRTFL0}) on Pavia Centre. For Case 5, when the clean image is corrupted by Gaussian noise, impulse noise, and vertical stripes, the proposed method also achieves around 1.2 dB gain in PSNR against the same second-best method (\texttt{LRTFL0}).
	
	To test our method's performance on every band, each band's PSNR and SSIM values on WDC Mall for Cases 1-6 are shown in Fig. \ref{every_bands}. As observed, \texttt{DS2DP} achieves the highest SSIM and PSNR values on most bands in all cases.
	
	\noindent
	\textbf{Visual Comparison.}
	Figs. \ref{stimulate} and \ref{omsi} show denoising results on HSIs and MSIs by different methods, respectively. The low-rank matrix model-based approach \texttt{LRMR} cannot effectively remove the stripes and deadlines. 
	Additionally, \texttt{LRTDTV} achieves noise removal in partial bands but fails to remove the stripes and deadlines in all bands. Besides, \texttt{LRTFL0} removes almost all of the noise but fails to capture the detailed information. Although there is some residual structured noise remaining in the result produced by \texttt{DS2DP}, the overall visual perception largely outperforms the baselines. We conjecture that such performance boost is mainly due to the deep spatial prior's ability to preserve the local spatial details---empowered by the expressiveness of appropriately crafted neural network structures. 
	
	Fig. \ref{spectral_curve} visualizes the denoising results by the algorithms in the spectral domain. One can see that, among all algorithms, the \texttt{DS2DP}-produced spectral signatures (on randomly selected pixel) also exhibit the highest visual similarity with those from the ground-truth image. This is consistent with its good performance in the spatial domain.
	
	Compared with hand-crafted prior and deep image prior, the promising results of the proposed  DS2DP can be attributed to that unsupervised disentangled spatio-spectral deep priors can characterize the complex scenes finely, which is beneficial to stripe removal.

	\subsection{Real Data Experiments}
	For real-data experiments, we choose five real-world HSI datasets to test the real noise removal, i.e., Urban dataset\footnote{\url{https://sites.google.com/site/feiyunzhuhomepage/datasets-ground-truths}}, Earth Observing-1
	(EO-1) Hyperion dataset\footnote{\url{http://www.lmars.whu.edu.cn/prof web/zhanghongyan/resource/noiseEOI.zip}}, Australian dataset\footnote{\url{http://remote-sensing.nci.org.au/u39/public/html/index.shtml}}, WHU HongHu dataset\footnote{\url{http://rsidea.whu.edu.cn/resource_WHUHi_sharing.htm}}, and Indian Pines dataset\footnote{\url{http://www.ehu.eus/ccwintco/index.php/Hyperspectral_Remote_Sensing_Scenes}}
	More precisely, the size of Urban dataset is 256$ \times $256$ \times $210, the size of EO-1 dataset is 192$ \times $192$ \times $166, the size of Australian dataset is 256$ \times $256$ \times $128, the size of WHU HongHu dataset is 256$\times$256$\times$64, and the size of Indian Pines dataset is 128$\times$128$\times$220. Regarding the proposed \texttt{DS2SP}, the parameters $ {R} $ is set as as 3, 2, 6, 2, and 5 for Urban, EO-1, Australian, WHU HongHu, and Indian Pines respectively. $ \lambda $ is set as 0.01, 0.01, 0.001, 0.1, and 0.000001 for Urban, EO-1, Australian, WHU HongHu, and Indian Pines, respectively. 
	
	The denoising results on these real-world datasets are shown in Fig. \ref{real}.
	One can see that all algorithms offer reasonable results on Urban dataset, perhaps because the data is not severely corrupted. Nevertheless, the proposed method produces the visually sharpest results. In particular, in the zoomed-in area, one can see that the proposed method's result does not have horizontal stripes, while such stripes still appear in results given by most of the baselines. 
	For EO-1 dataset, since the selected band was severely damaged by sparse noise, the denoising task is particularly challenging. 
	One can see that traditional methods can hardly produce satisfactory results. Nonetheless, \texttt{DS2DP} removes almost all of the noise---with the price of blurring the image to a certain extent---and offers the most visually pleasing result. For Australian, WHU HongHu, and Indian Pines datasets, we can see that there exists visible sparse noise in the observed images. The proposed \texttt{DS2DP} preserves more details while removing such noise, compared with other competing methods.
	
	\section{Further Discussions}\label{5}
	
	\subsection{Analysis of Algorithm Complexity}
	In this part, we analyze the algorithm complexity of the proposed method on HSI WDC Mall and MSI Superballs for Case 6. \texttt{DIP2D} and  \texttt{DIP3D} are selected as the baseline models since they stand for the unsupervised HSI denoising models. For \texttt{DIP2D} and  \texttt{DIP3D}, we select the network structure with the best performance according to the original implementation. 
	
	For a fair comparison, the network structure utilized in \texttt{DS2DP}, which is expected to capture the spatial prior information, is simply designed as U-Net-like ``hourglass" architecture. Moreover, we do not focus on meticulous designs on reducing the model scale in this work, i.e., depth-wise separable convolution, model pruning, and model compression \cite{compress}. These techniques may be used to reduce the network complexity of all methods (including ours), but this is beyond the scope of this work.
	Table \ref{tab2} lists the scale of parameters of different methods on HSI WDC Mall and MSI Superballs. Moreover, the corresponding values of PSNR, SSIM, and the execution time (in minutes) are also reported in Table \ref{tab2}.
	
	\begin{table}[!htbp]
		\centering
		\selectfont
		\setlength{\tabcolsep}{5pt}
		\renewcommand\arraystretch{1.2}
		\caption{The relevant indicators of DIP3D, DIP2D, and DS2DP on HSI WDC Mall and MSI Superballs for Case 6.  The \textbf{best} and \underline{second-best} values are highlighted in bold and underlined, respectively.}
		\centering
		\vspace{-0.2cm}
		\begin{tabular}{clccccccc}
			\Xhline{1.0pt}
			Data & Methods & Params & PSNR & SSIM & Time  \\
			\hline
			\multirow{4}[1]{*}{\begin{tabular}[c]{@{}c@{}}\\ HSI: WDC Mall\\ (256 $\times$ 256 $\times$ 191)\end{tabular}}
			& DIP3D  &  6.275M & * & * & *   \\
			& DIP2D  &  2.342M & 21.759 & 0.594 & \underline{15.625}  \\
			& DS2DP  &  \underline{2.150M} & \textbf{34.352} & \textbf{0.967} & 19.544 \\
			& DS2DP*  &  \textbf{0.574M} & \underline{32.710} & \underline{0.942} &  \textbf{12.353} \\

			\hline
			\multirow{4}[1]{*}{\begin{tabular}[c]{@{}c@{}}\\ MSI: Superballs\\ (256 $\times$ 256 $\times$ 32)\end{tabular}}  
			& DIP3D  &  6.275M & 20.705 & 0.399 &  16.091 \\
			& DIP2D  & \underline{2.138M} & 20.901 & 0.408 &  \underline{3.314}  \\
			& DS2DP  &  2.150M & \textbf{35.037} & \textbf{0.881}  & 4.677  \\
			& DS2DP*  &  \textbf{0.574M} & \underline{34.779} & \underline{0.871} & \textbf{2.181}  \\
			\Xhline{1.0pt}
		\end{tabular}
		\label{tab2}
	\end{table}

	As shown in Table \ref{tab2}, the proposed \texttt{DS2DP} achieves significantly better performance with roughly equal parameters and slightly longer execution time compared with the baseline models. More precisely, \texttt{DS2DP} outperforms \texttt{DIP2D} by 12.593 dB and 14.136 dB in terms of PSNR on HSI WDC Mall and MSI Superballs, respectively. \texttt{DS2DP} achieves performance gains over \texttt{DIP3D} with about 14 dB on MSI Superballs.  
	
	In our original implementation, to push \texttt{DS2DP} to attain the (empirically) achievable ``best'' performance, we use several parallel networks with the same architecture to generate the abundance maps. 
	To reduce the number of the parameters, we let the parameters be shared between several parallel networks.  This method is denoted by \texttt{DS2DP}* and its performance is also shown in Table \ref{tab2}. This way, the parameter amount reduces by 3/4 and the execution time reduces by 2/5,  while the PSNR is essentially unaffected.

	\subsection{Effectiveness of the Deep Spectral and Spatial Priors}
\begin{figure}[!htbp]
		\footnotesize
		\setlength{\tabcolsep}{0pt}
		\centering
		\begin{tabular}{ccc}
			\includegraphics[width=0.16\textwidth]{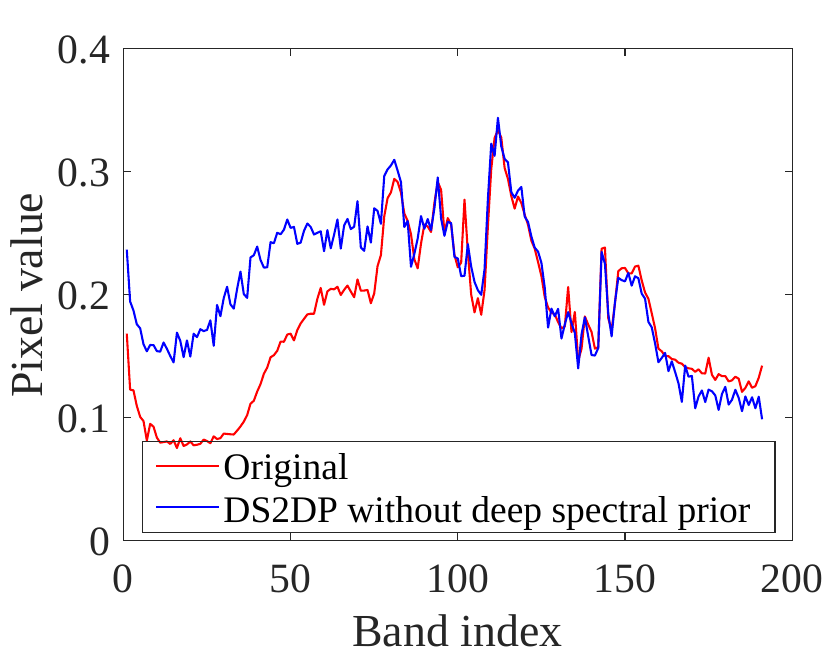}&
			\includegraphics[width=0.16\textwidth]{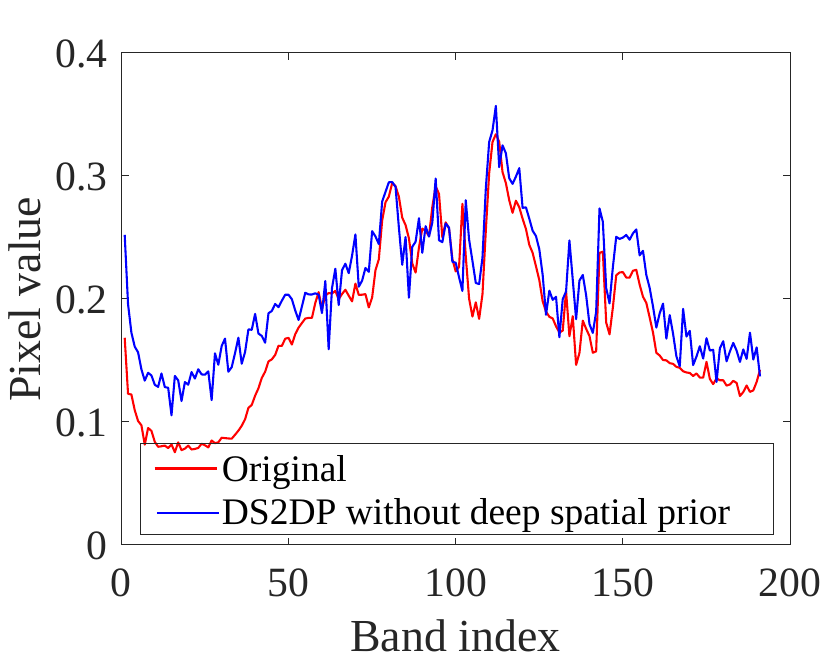}&
			\includegraphics[width=0.16\textwidth]{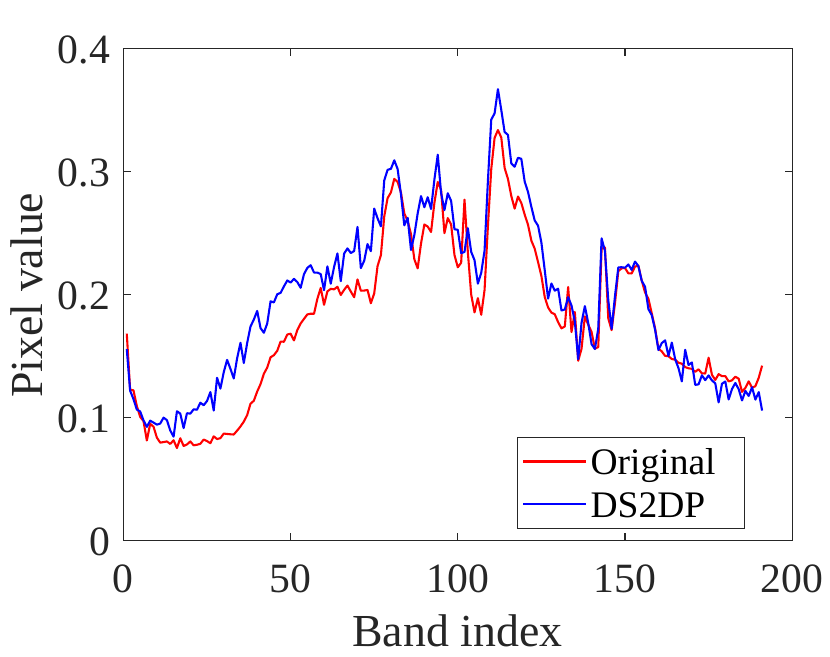}\\
			(a)&(b)&(c)\\
		\end{tabular}
		\caption{Effectiveness of the deep priors in the spectral domain. The red curve is the ground truth of a selected pixel at spatial location (120, 120) for illustration. The blue curves correspond to: (a) the estimated spectrum by \texttt{DS2DP} without deep spectral prior; (b) the estimated spectrum by \texttt{DS2DP} without deep spatial prior; and (c) the estimated spectrum by the proposed \texttt{DS2DP}.}
		\label{discussion1}
	\end{figure}

\begin{figure}[!htbp]
		\footnotesize
		\setlength{\tabcolsep}{1pt}
		\centering
		\begin{tabular}{cccc}	
			\includegraphics[width=0.11\textwidth]{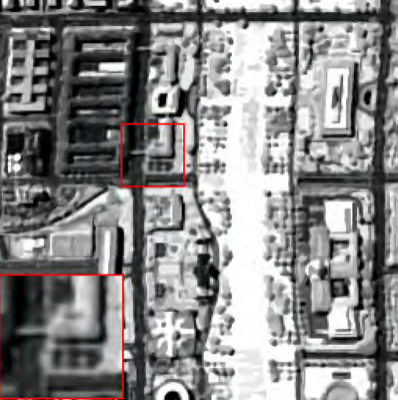}&
			\includegraphics[width=0.11\textwidth]{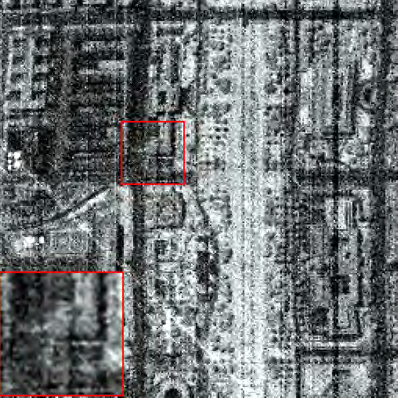}&
			\includegraphics[width=0.11\textwidth]{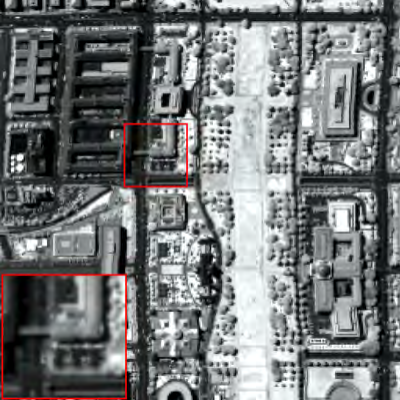}&
			\includegraphics[width=0.11\textwidth]{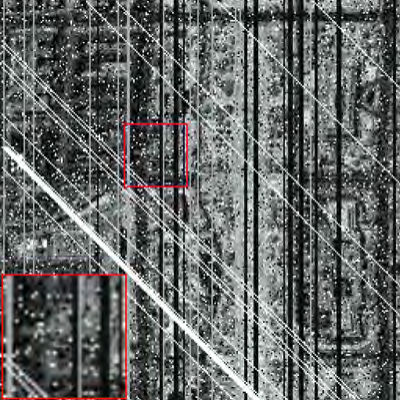}\\
			(a)&(b)&(c)&(d)\\
		\end{tabular}
		\caption{Effectiveness of the deep priors in the spatial domain. The four figures correspond to: (a) the denoising result by \texttt{DS2DP} without deep spectral prior; (b) the denoising result by \texttt{DS2DP} without deep spatial prior; (c) the denoising result by \texttt{DS2DP}; and (d) the observed image.}
		\label{discussion2}
	\end{figure}

	In this part, we take a deeper look at the deep spectral and spatial priors in \texttt{DS2DP}. To verify these two priors' effectiveness, we conduct ablation studies for Case 6 using the WDC Mall data. The impacts of our designed priors in spectral and spatial domains are shown in Fig. \ref{discussion1} and Fig. \ref{discussion2}, respectively.
	
	Fig. \ref{discussion1} (a) shows that when only employing deep spatial prior in \texttt{DS2DP} without the deep spectral prior, the estimated spectrum of the selected pixel is not accurate. In contrast, when considering both types of priors in \texttt{DS2DP}, the results become much more promising; see (c). Besides, \texttt{DS2DP} without the deep spatial prior and the complete \texttt{DS2DP} both achieve satisfactory performance on most bands. This supports our idea for disentangling the spatial and spectral information and modeling them individually.
	
	Fig. \ref{discussion2} shows similar effects in the spatial domain. One can see that there is obviously visible noise in the results when only employing the deep spectral prior. However, when considering the two priors, the performance is clearly much more visually pleasing. In addition,  Fig. \ref{discussion2} (c) also clearly demonstrates the disentanglement between the spatial and spectral effects.
	
	Moreover, the quantitative comparisons of the denoising results by \texttt{DS2DP} without deep spectral prior, \texttt{DS2DP} without deep spatial prior, and \texttt{DS2DP} are shown in Table \ref{tab4}.
	
	\begin{table}[htbp!]
		\centering
		\caption{Quantitative comparison of \texttt{DS2DP} without deep spectral prior, \texttt{DS2DP} without deep spatial prior, and \texttt{DS2DP}. The \textbf{best} and \underline{second-best} values are highlighted in bold and underlined, respectively.} 
		\begin{tabular}{c|ccc}
			\Xhline{1.0pt}
			Method & PSNR & SSIM & SAM \bigstrut\\
			\hline
			\texttt{DS2DP} without deep spectral prior & 27.316 & 0.837 & \underline{0.167}\bigstrut\\
			\hline
			\texttt{DS2DP} without deep spatial prior & \underline{31.927} & \underline{0.903} & 0.181\bigstrut\\
			\hline
			\texttt{DS2DP} & \textbf{34.352} & \textbf{0.967} & \textbf{0.116}\bigstrut\\
			\Xhline{1.0pt}
		\end{tabular}
		\label{tab4}
	\end{table}
	\subsection{Effectiveness of the Sparsity Regularization}

	\begin{figure}[!htbp] 
		\centering
		\includegraphics[scale=0.62]{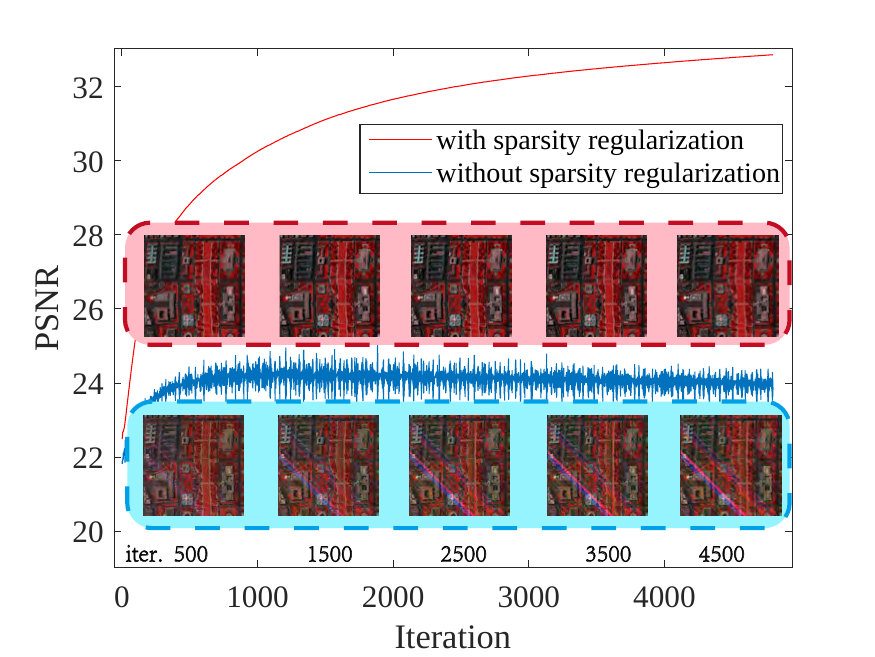}
		\caption{The history of PSNR values and the corresponding denoising results by \texttt{DS2DP} with and without sparsity regularization. } \label{discuss1}
	\end{figure}
	
	To verify the sparsity regularization's effect, we design a comparative experiment, also using Case 6 and the WDC Mall data. The result is shown in Fig. \ref{discuss1}.
	One can see that when the sparsity regularization is not applied, the PSNR first increases and then declines slowly as the number of iterations increases. In contrast, when sparsity regularization is employed, the PSNR maintains an upward trend during the iterations---and eventually exhibits a big PSNR improvement relative to the former case.
	
	Fig. \ref{discuss1} also shows the visualization of the algorithm with and without the sparsity regularization in the 1,000th iteration. One can see that the proposed method produces a relatively clean image, which clearly shows an advantage over the case without the $L_1$ term. 
	
	Moreover, to further support the effectiveness of the sparsity regularization, we list the quantitative indexes (i.e., PSNR, SSIM, and SAM) of the denoising results by \texttt{DS2DP} without sparsity regularization and \texttt{DS2DP} in Table \ref{tab5}.
	
	\begin{table}[htbp!]
		\centering
			\caption{The quantitative indexes (i.e., PSNR, SSIM, and SAM) of the denoising results by \texttt{DS2DP} without sparsity regularization and \texttt{DS2DP}. The \textbf{best} is highlighted in bold.} 
			\begin{tabular}{c|ccc}
				\Xhline{1.0pt}
				Method & PSNR & SSIM & SAM \bigstrut\\
				\hline
				\texttt{DS2DP} without sparsity regularization & 23.762 & 0.668 & 0.253\bigstrut\\
				\hline
				\texttt{DS2DP} & \textbf{34.352} & \textbf{0.967} & \textbf{0.116}\bigstrut\\
				\Xhline{1.0pt}
			\end{tabular}
			\label{tab5}
	\end{table}

		\subsection{Sensitivity Analysis and Selection of the Parameters ${R}$ and $\lambda$}
		\subsubsection{Parameter $R$}
		In this part, we study the parameter sensitivity of the number of materials $R$ on real HSI Indian Pines---which contains 16 materials. Then, we further discuss the selection of the parameter $R$ in real scenarios.
		
\begin{figure}[!htbp]
			\scriptsize
			\setlength{\tabcolsep}{1pt}
			\centering
			\begin{tabular}{ccccc}	
				\includegraphics[width=0.11\textwidth]{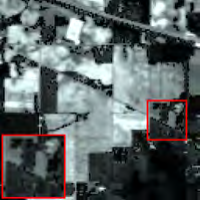}&
				\includegraphics[width=0.11\textwidth]{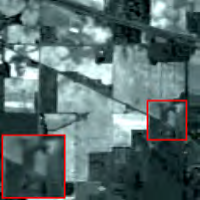}&
				\includegraphics[width=0.11\textwidth]{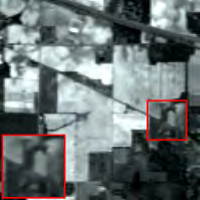}&
				\includegraphics[width=0.11\textwidth]{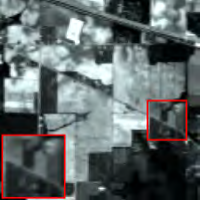}\\
				Observed & $ R $ = 2 (1.150 M) & $ R $ = 4 (2.298 M) & $ R $ = 6 (3.449 M)\\
				\includegraphics[width=0.11\textwidth]{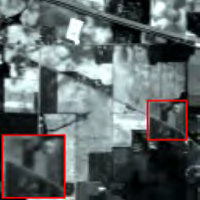}&
				\includegraphics[width=0.11\textwidth]{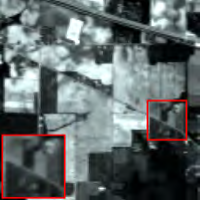}&
				\includegraphics[width=0.11\textwidth]{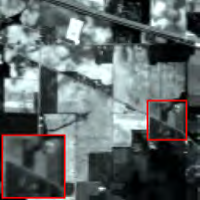}&
				\includegraphics[width=0.11\textwidth]{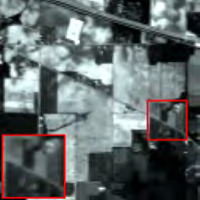}\\
				$ R $ = 8 (4.597 M)& $ R $ = 10 (5.747 M) & $ R $ = 12 (6.896 M) & $ R $ = 14 (8.046 M) \\
				\includegraphics[width=0.11\textwidth]{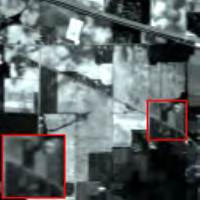}&
				\includegraphics[width=0.11\textwidth]{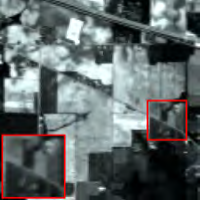}&
				\includegraphics[width=0.11\textwidth]{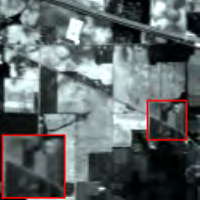}&
				\includegraphics[width=0.11\textwidth]{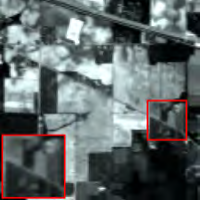}\\
				$ R $ = 16 (9.197 M)&$ R $ = 18 (10.3462 M) & $ R $ = 20 (11.495 M) & $ R $ = 22 (12.644 M)\\
			\end{tabular}
			\caption{The band 24 of the denoising results by \texttt{DS2DP} on real-world HSI Indian Pines 
					with the corresponding number of endmembers $R$ and network parameters.}
			\label{RR}
		\end{figure}

		Due to the absence of the ground truth, we display the denoising images with the corresponding number of endmembers $ R $ and network parameters in Fig. \ref{RR}. We can observe that when $R$ is smaller than 3, the visual effect is not satisfactory. When $R$ is larger than 3, the denoising results are visually the same. This observation demonstrates that the proposed \texttt{DS2DP} is robust with respect to the parameter $ R $. Such robustness against underestimated $R$ is a bit surprising at first glance, yet understandable---under the linear mixture model, the range space spanned by the first several principal components may contain most of the energy. Hence, underestimating $R$ may not be very detrimental in many cases. Note that when selecting $R$, the number of network parameters should also be considered and balanced, since the number of the network parameters increases {\it substantially} with the increasing value of $R$. Thus, we select the value of $R$ from a starting number 2 with the increments 1 in practice, which balances between the visual effect and the number of the parameters.
		
		Other than using visual validation, the parameter $R$ could also be selected by the existing effective number of endmember estimation algorithms, e.g.,  \cite{HySime, 7055246}.
		
		\subsubsection{Parameter $\lambda$}
		In this subsection, we conduct an empirical sensitivity analysis of the parameter $\lambda$, using the real-world HSI Australian, Urban, and WHU HongHu with different corruption levels. Then, we further discuss the selection of the parameter $\lambda$ in real scenarios.
		
		\begin{figure}[!htbp]
			\footnotesize
			\setlength{\tabcolsep}{1pt}
			\centering
			\begin{tabular}{cccc}	
				Observed & $\lambda$ = 0.001 & $\lambda$ = 0.01 & $\lambda$ = 0.1 \\
				\includegraphics[width=0.11\textwidth]{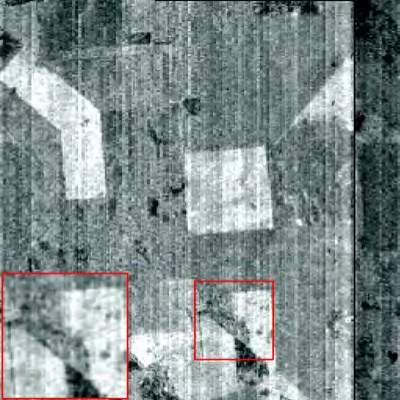}&
				\includegraphics[width=0.11\textwidth]{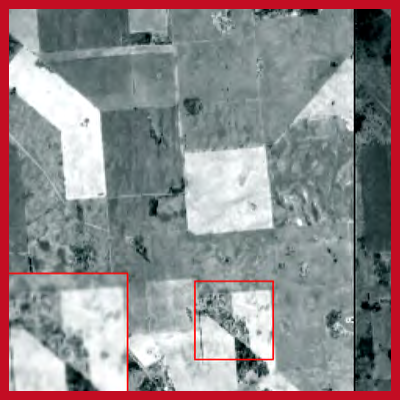}&
				\includegraphics[width=0.11\textwidth]{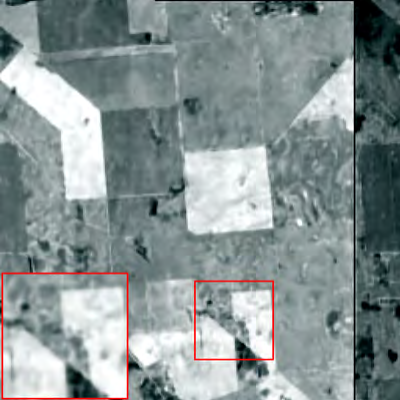}&
				\includegraphics[width=0.11\textwidth]{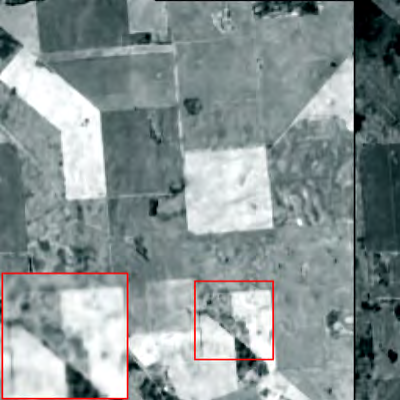}\\
				\includegraphics[width=0.11\textwidth]{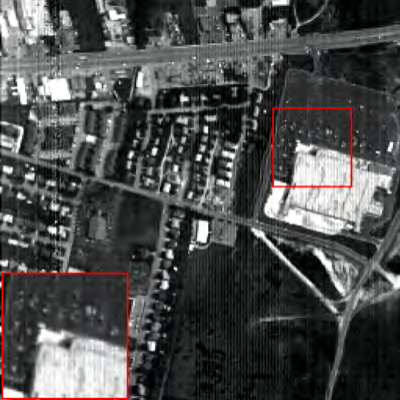}&
				\includegraphics[width=0.11\textwidth]{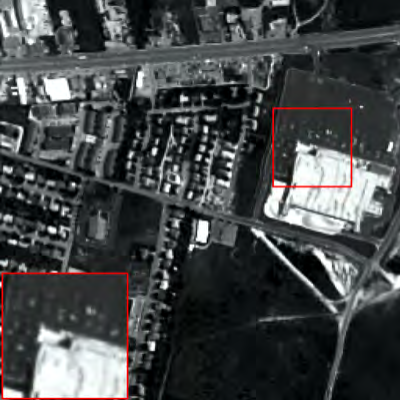}&
				\includegraphics[width=0.11\textwidth]{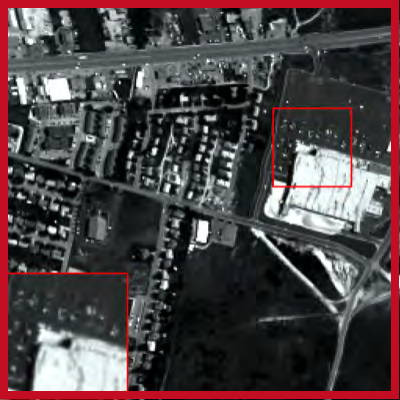}&
				\includegraphics[width=0.11\textwidth]{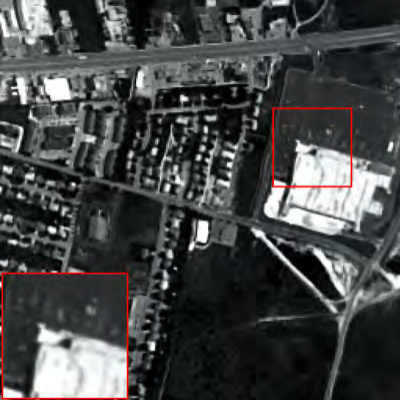}\\
				\includegraphics[width=0.11\textwidth]{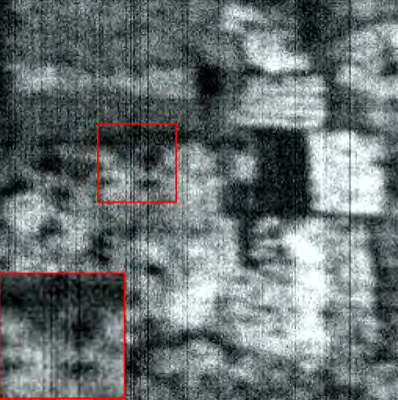}&
				\includegraphics[width=0.11\textwidth]{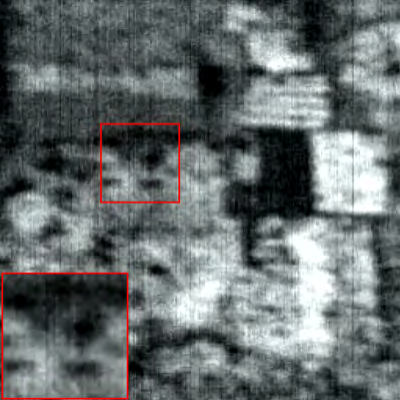}&
				\includegraphics[width=0.11\textwidth]{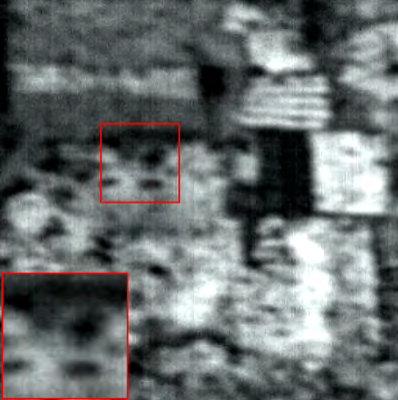}&
				\includegraphics[width=0.11\textwidth]{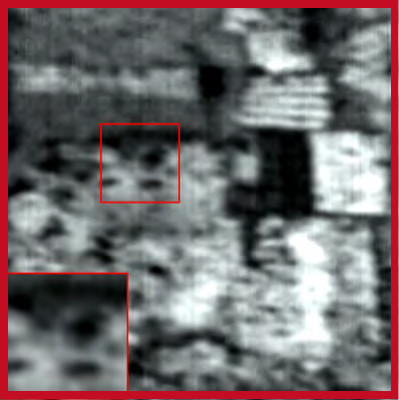}\\
			\end{tabular}
			\caption{Denoising results by \texttt{DS2DP} with different $\lambda$ on real-world HSI Australian, Urban, and WHU HongHu datasets. From top to bottom: the band 87 of Australian dataset, the band 1 of Urban dataset, and the band 37 of WHU HongHu dataset, respectively. From left to right: the observed image, the results with $\lambda$ = 0.001, 0.01, and 0.1, respectively.}
			\label{lambda}
		\end{figure}
		
		Fig. \ref{lambda} shows the denoising results by \texttt{DS2DP} with different $\lambda$ on these three real HSIs. We can observe that the proposed \texttt{DS2DP} attains the best visual effect when $\lambda$ = 0.001, 0.01, and 0.1 for real HSIs Australian, Urban, and WHU HongHu, respectively. This observation reflects that the regularization parameter $\lambda$ is sensitive to the weight of the sparse noise. Thus, we apply (rough) grid search from a range of parameters to select a relatively ``good'' one in terms of visual effect. For example, we select $\lambda$ from the candidate set \{0.001, 0.01, 0.1\} and visually determine which parameter is more plausible.

		Additionally, note that for images that are more severely contaminated by sparse noise, $\lambda$ should be larger. Hence, in addition to visual validation, $\lambda$ could also be selected from a collection of candidates (e.g., \{0.001, 0.01, 0.1\}) by estimating the corruption level. The level of corruption/noise can be roughly estimated by some hyperspectral noise estimation algorithms, e.g., those in \cite{LBSR_HR, 8760540}. These noise/outlier estimation methods often use less powerful models relative to our deep prior-based one in terms of expressiveness, but may run fairly fast to yield some initial estimations for the noise level, which can serve our parameter selection purpose.
	
	\subsection{Impact of the Random Input to \texttt{DS2DP}}
	As illustrated previously, the input of the proposed \texttt{DS2DP} is random but known noise sampled from a uniform distribution. One may wonder if the input $\bm z_r$ has a significant impact on results? The answer is negative. We show this by calculating the means and standard deviations of the algorithm outputs' PSNR for Cases 1-6 on WDC Mall. For each case, we run ten trials with different $\bm z_r$'s that are randomly generated from \textit{U}(-0.05, 0.05), where \textit{U} stands for uniform distribution. The results are shown in Table \ref{tab3}. One can see that, perhaps a bit surprisingly, the standard deviations of the results are fairly small---which means the method is essentially not affected by the random input to a good extent.
	
	\begin{table}[!htbp]
		\selectfont
		\setlength{\tabcolsep}{10pt}
		\renewcommand\arraystretch{1.2}
		\caption{The denoising results' PSNR values (Mean$\pm$Std.Dev) for Cases 1-6 on WDC Mall}
		\centering
		\begin{tabular}{c|ccc}
			\Xhline{1.0pt}
			Case & Case 1 & Case 2 & Case 3\\  
			PSNR  & 36.213$\pm$0.254       & 35.636$\pm$0.358      & 34.297$\pm$0.276\\ \hline
			Case & Case 4 & Case 5 & Case 6\\
			PSNR & 35.511$\pm$0.344 & 34.802$\pm$0.297 & 34.173$\pm$0.221\\ 
			\Xhline{1.0pt}
		\end{tabular}
		\label{tab3}
	\end{table}
	
	\subsection{Impact of the Distribution Parameter of the Random Input to \texttt{DS2DP}}
\begin{figure}[!htb]
		\centering\scriptsize
			\renewcommand\arraystretch{1}
			\setlength{\tabcolsep}{1pt}
			\begin{tabular}{cccccccc}
				\includegraphics[width=0.7\linewidth]{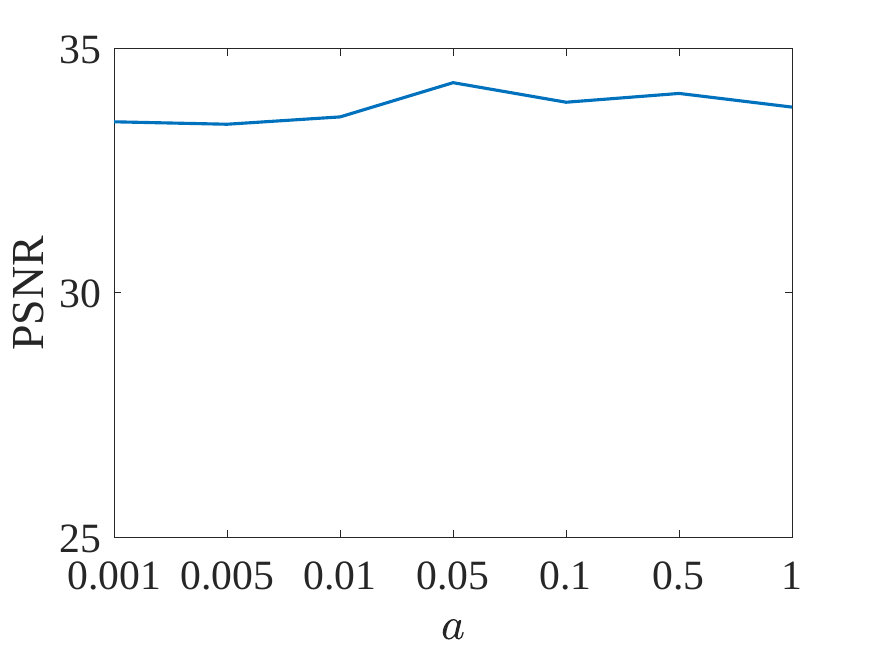}\\
			\end{tabular}
			\caption{The influence of $ a $ on HSI WDC Mall for Case 6}
			\label{a}
	\end{figure}

	We denote the uniform distribution as \textit{U}(-$ a $, $ a $), where -$ a $ and $ a $ are the lower boundary and upper boundary, respectively. In our experiment, the parameter $ a $ is set as 0.05 following \cite{DIP}. In this part, we have conducted an empirical sensitivity analysis of parameter $ a $. Fig. \ref{a} presents the PSNR values by the proposed method with different $ a $ for Case 6. We can see that, the PSNR value does not fluctuate greatly with different $ a $. Therefore, the denoising result is not sensitive to the value of $ a $.\\

	\subsection{HSI Classification Validation}
	
	\begin{table}[htbp!]
		\centering
			\scriptsize
			\setlength{\tabcolsep}{7pt}

			\caption{Quantitative comparisons of classification accuracy of the denoised HSI by the competing methods.  The \textbf{best} and \underline{second-best} values are highlighted in bold and underlined, respectively.}
			\begin{tabular}{c|cccccc}
				\Xhline{1.0pt}
				Metrics & Original & LRMR & LRTDTV & LRTFL0 & DS2DP \bigstrut\\
				\hline
				Overall accuracy & 76.5\% & 83.6\% & \underline{83.8\%} & 81.7\% & \textbf{84.9\%}  \bigstrut\\
				\hline
				Kappa coefficient & 0.733 & 0.814 & \underline{0.816} & 0.796 & \textbf{0.828} \bigstrut\\
				\Xhline{1.0pt}
			\end{tabular}
			\label{tab6}
	\end{table}
	
	To further evaluate the effectiveness of the proposed model, we considered a hyperspectral classification task on the Indian Pine  data\footnote{\url{http://www.ehu.eus/ccwintco/index.php/Hyperspectral_Remote_Sensing_Scenes}\label{Indian}}. We use the observed data and the outputs of different denoising algorithms as the inputs to the classification task---which is performed by a support vector machine (SVM) classifier. For each class, the number of training samples is 40, and the number of test samples ranges from 6 to 2415 for various classes.
	
	Table \ref{tab6} reports the classification performance.  The performance is measured by two commonly used metrics for classification tasks, namely, the overall accuracy \cite{8413118} and the Kappa coefficient \cite{8413118}. One can see that the denoising methods can improve the classification performance by simply using the observed data. In addition, one can observe that \texttt{DS2DP} achieves around 1.1\% and 0.012 gain in terms of the overall accuracy and the Kappa coefficient, respectively, compared to the second-best method (\texttt{LRTDTV}). \\

	\section{Conclusions}
	We proposed an unsupervised deep prior-based HSI denoising framework. Unlike existing methods that directly learn deep generative networks for the entire HSI, our method leverages the classic LMM to disentangle the spatial and spectral information, and learns two types of deep priors for the abundance maps and the spectral signatures of the endmembers, respectively.
	Our design is driven by the challenges that network structures used in deep priors for different types of images (in particular, HSIs) may be hard to search. 
	Using our information-disentangled framework, empirically validated unsupervised deep image prior structures for natural images can be easily incorporated for HSI denoising. Besides, the network complexity can be substantially reduced with proper parameter sharing, making the learning process more affordable than existing approaches. We also proposed a structured noise-robust optimization criterion that is tailored for HSI denoising. We tested our method using extensive experiments with various cases and ablation studies.
	The numerical results demonstrated promising HSI denoising performance of the proposed approach.
	A note is that, despite its promising performance, unsupervised DIP research has still been largely empirical---Rigorous analysis has been elusive. Interesting future directions include enhanced theoretical understanding to unsupervised DIP (e.g., in terms of sample complexity and generalization error analysis), which may be of broader interest beyond HSI denoising.
	\normalem
	\bibliography{referrence}

\begin{thebibliography}{10}

\bibitem{Bioucas2012HUOverview}
J.~M. {Bioucas-Dias}, A.~{Plaza}, N.~{Dobigeon}, M.~{Parente}, Q.~{Du},
  P.~{Gader}, and J.~{Chanussot}, ``Hyperspectral unmixing overview:
  Geometrical, statistical, and sparse regression-based approaches,'' {\em IEEE
  J. Sel. Topics Appl. Earth Observ. Remote Sens.}, vol.~5, no.~2,
  pp.~354--379, 2012.

\bibitem{MF}
W.~{He}, H.~{Zhang}, L.~{Zhang}, and H.~{Shen}, ``Total-variation-regularized
  low-rank matrix factorization for hyperspectral image restoration,'' {\em
  IEEE Trans. Geosci. Remote Sens.}, vol.~54, no.~1, pp.~178--188, 2016.

\bibitem{TF}
Y.~{Wang}, J.~{Peng}, Q.~{Zhao}, Y.~{Leung}, X.~{Zhao}, and D.~{Meng},
  ``Hyperspectral image restoration via total variation regularized low-rank
  tensor decomposition,'' {\em IEEE J. Sel. Top. Appl. Earth Observ. Remote
  Sens.}, vol.~11, no.~4, pp.~1227--1243, 2018.

\bibitem{BTD}
F.~Xiong, J.~Zhou, and Y.~Qian, ``Hyperspectral restoration via $ l\_0 $
  gradient regularized low-rank tensor factorization,'' {\em IEEE Trans.
  Geosci. Remote Sens.}, vol.~57, no.~12, pp.~10410--10425, 2019.

\bibitem{LBSR_HR}
L.~{Zhuang} and M.~K. {Ng}, ``Hyperspectral mixed noise removal by $\ell
  _1$-norm-based subspace representation,'' {\em IEEE J. Sel. Top. Appl. Earth
  Observ. Remote Sens.}, vol.~13, pp.~1143--1157, 2020.

\bibitem{BM3D}
K.~{Dabov}, A.~{Foi}, V.~{Katkovnik}, and K.~{Egiazarian}, ``Image denoising by
  sparse 3-{D} transform-domain collaborative filtering,'' {\em IEEE Trans.
  Image Process.}, vol.~16, no.~8, pp.~2080--2095, 2007.

\bibitem{S1}
J.~Mairal, F.~Bach, J.~Ponce, G.~Sapiro, and A.~Zisserman, ``Non-local sparse
  models for image restoration,'' in {\em Proc. IEEE Int. Conf. Comput. Vis.},
  pp.~2272--2279, 2009.

\bibitem{S2}
W.~Dong, L.~Zhang, G.~Shi, and X.~Li, ``Nonlocally centralized sparse
  representation for image restoration,'' {\em IEEE Trans. Image Process.},
  vol.~22, no.~4, pp.~1620--1630, 2012.

\bibitem{chen1}
Y.~Chen, J.~Li, and Y.~Zhou, ``Hyperspectral image denoising by total
  variation-regularized bilinear factorization,'' {\em Signal Process.},
  vol.~174, p.~107645, 2020.

\bibitem{PCA}
G.~Chen and S.~Qian, ``Denoising of hyperspectral imagery using principal
  component analysis and wavelet shrinkage,'' {\em IEEE Trans. Geosci. Remote
  Sens.}, vol.~49, no.~3, pp.~973--980, 2011.

\bibitem{MSBCRF}
P.~Zhong and R.~Wang, ``Multiple-spectral-band {C}{R}{F}s for denoising junk
  bands of hyperspectral imagery,'' {\em IEEE Trans. Geosci. Remote Sens.},
  vol.~51, no.~4, pp.~2260--2275, 2013.

\bibitem{BM4D}
M.~Maggioni, V.~Katkovnik, K.~Egiazarian, and A.~Foi, ``Nonlocal
  transform-domain filter for volumetric data denoising and reconstruction,''
  {\em IEEE Trans. Image Process.}, vol.~22, no.~1, pp.~119--133, 2013.

\bibitem{Meng2018}
Y.~{Chen}, X.~{Cao}, Q.~{Zhao}, D.~{Meng}, and Z.~{Xu}, ``Denoising
  hyperspectral image with non-i.i.d. noise structure,'' {\em IEEE Trans.
  Cybern.}, vol.~48, no.~3, pp.~1054--1066, 2018.

\bibitem{chang1}
Y.~{Chang}, L.~{Yan}, X.~L. {Zhao}, H.~{Fang}, Z.~{Zhang}, and S.~{Zhong},
  ``Weighted low-rank tensor recovery for hyperspectral image restoration,''
  {\em IEEE Trans. Cybern.}, vol.~50, no.~11, pp.~4558--4572, 2020.

\bibitem{chen2}
Y.~{Chen}, Y.~{Guo}, Y.~{Wang}, D.~{Wang}, C.~{Peng}, and G.~{He}, ``Denoising
  of hyperspectral images using nonconvex low rank matrix approximation,'' {\em
  IEEE Trans. Geosci. Remote Sens.}, vol.~55, no.~9, pp.~5366--5380, 2017.

\bibitem{chen3}
F.~{Xu}, Y.~{Chen}, C.~{Peng}, Y.~{Wang}, X.~{Liu}, and G.~{He}, ``Denoising of
  hyperspectral image using low-rank matrix factorization,'' {\em IEEE Geosci.
  Remote Sens. Lett.}, vol.~14, no.~7, pp.~1141--1145, 2017.

\bibitem{zhang1}
H.~{Zhang}, W.~{He}, L.~{Zhang}, H.~{Shen}, and Q.~{Yuan}, ``Hyperspectral
  image restoration using low-rank matrix recovery,'' {\em IEEE Trans. Geosci.
  Remote Sens.}, vol.~52, no.~8, pp.~4729--4743, 2014.

\bibitem{zhuang1}
L.~{Zhuang} and J.~M. {Bioucas-Dias}, ``Hyperspectral image denoising based on
  global and non-local low-rank factorizations,'' in {\em Proc. Int. Conf.
  Image Process.}, pp.~1900--1904, 2017.

\bibitem{Deep}
K.~He, X.~Zhang, S.~Ren, and J.~Sun, ``Deep residual learning for image
  recognition,'' in {\em Proc. IEEE Conf. Comput. Vis. Pattern Recog.}, 2016.

\bibitem{VAE}
D.~P. Kingma and M.~Welling, ``{Auto-Encoding Variational Bayes},'' in {\em
  Proc. Int. Conf. Learn. Representations}, 2014.

\bibitem{AE}
W.~{Wang}, Y.~{Huang}, Y.~{Wang}, and L.~{Wang}, ``Generalized autoencoder: A
  neural network framework for dimensionality reduction,'' in {\em Proc. IEEE
  Conf. Comput. Vis. Pattern Recog.}, pp.~496--503, 2014.

\bibitem{GAN}
Z.~Wang, Q.~She, and T.~E. Ward, ``Generative adversarial networks in computer
  vision: A survey and taxonomy,'' {\em arXiv preprint arXiv:1906.01529}, 2019.

\bibitem{yuan1}
Q.~{Yuan}, Q.~{Zhang}, J.~{Li}, H.~{Shen}, and L.~{Zhang}, ``Hyperspectral
  image denoising employing a spatial–-spectral deep residual convolutional
  neural network,'' {\em IEEE Trans. Geosci. Remote Sens.}, vol.~57, no.~2,
  pp.~1205--1218, 2019.

\bibitem{Dong}
W.~{Dong}, H.~{Wang}, F.~{Wu}, G.~{Shi}, and X.~{Li}, ``Deep spatial–spectral
  representation learning for hyperspectral image denoising,'' {\em IEEE Trans.
  Comput. Imag.}, vol.~5, no.~4, pp.~635--648, 2019.

\bibitem{yuan2}
Q.~{Yuan}, Y.~{Wei}, X.~{Meng}, H.~{Shen}, and L.~{Zhang}, ``A multiscale and
  multidepth convolutional neural network for remote sensing imagery
  pan-sharpening,'' {\em IEEE J. Sel. Topics Appl. Earth Observ. Remote Sens.},
  vol.~11, no.~3, pp.~978--989, 2018.

\bibitem{yuan3}
Q.~{Zhang}, Q.~{Yuan}, J.~{Li}, X.~{Liu}, H.~{Shen}, and L.~{Zhang}, ``Hybrid
  noise removal in hyperspectral imagery with a spatial–spectral gradient
  network,'' {\em IEEE Trans. Geosci. Remote Sens.}, vol.~57, no.~10,
  pp.~7317--7329, 2019.

\bibitem{chang2}
Y.~{Chang}, M.~{Chen}, L.~{Yan}, X.~{Zhao}, Y.~{Li}, and S.~{Zhong}, ``Toward
  universal stripe removal via wavelet-based deep convolutional neural
  network,'' {\em IEEE Trans. Geosci. Remote Sens.}, vol.~58, no.~4,
  pp.~2880--2897, 2020.

\bibitem{8693549}
W.~{Dong}, H.~{Wang}, F.~{Wu}, G.~{Shi}, and X.~{Li}, ``Deep spatial–spectral
  representation learning for hyperspectral image denoising,'' {\em IEEE Trans.
  Comput. Imag.}, vol.~5, no.~4, pp.~635--648, 2019.

\bibitem{DIP}
V.~{Lempitsky}, A.~{Vedaldi}, and D.~{Ulyanov}, ``Deep image prior,'' in {\em
  Proc. IEEE Conf. Comput. Vis. Pattern Recog.}, pp.~9446--9454, 2018.

\bibitem{HSIDIP}
O.~{Sidorov} and J.~Y. {Hardeberg}, ``Deep hyperspectral prior: Single-image
  denoising, inpainting, super-resolution,'' in {\em Proc. IEEE Int. Conf.
  Comput. Vis.}, pp.~3844--3851, 2019.

\bibitem{Ma2014unmixing}
W.-K. {Ma}, J.~M. {Bioucas-Dias}, T.-H. {Chan}, N.~{Gillis}, P.~{Gader}, A.~J.
  {Plaza}, A.~{Ambikapathi}, and C.-Y. {Chi}, ``A signal processing perspective
  on hyperspectral unmixing: Insights from remote sensing,'' {\em IEEE Signal
  Process. Mag.}, vol.~31, no.~1, pp.~67--81, 2014.

\bibitem{Fu2016Unmixing}
X.~{Fu}, W.-K. {Ma}, J.~M. {Bioucas-Dias}, and T.-H. {Chan}, ``Semiblind
  hyperspectral unmixing in the presence of spectral library mismatches,'' {\em
  IEEE Trans. Geosci. Remote Sens.}, vol.~54, no.~9, pp.~5171--5184, 2016.

\bibitem{Kanatsoulis2018HSR}
C.~I. {Kanatsoulis}, X.~{Fu}, N.~D. {Sidiropoulos}, and W.-K. {Ma},
  ``Hyperspectral super-resolution: A coupled tensor factorization approach,''
  {\em IEEE Trans. Signal Process.}, vol.~66, no.~24, pp.~6503--6517, 2018.

\bibitem{sidiropoulos2017tensor}
N.~D. Sidiropoulos, L.~De~Lathauwer, X.~Fu, K.~Huang, E.~E. Papalexakis, and
  C.~Faloutsos, ``Tensor decomposition for signal processing and machine
  learning,'' {\em IEEE Trans. Signal Process.}, vol.~65, no.~13,
  pp.~3551--3582, 2017.

\bibitem{TV5}
J.~{Liu}, Y.~{Sun}, X.~{Xu}, and U.~S. {Kamilov}, ``Image restoration using
  total variation regularized deep image prior,'' in {\em Proc. IEEE Int. Conf.
  Acoust. Speech Signal Process.}, pp.~7715--7719, 2019.

\bibitem{ye2014multitask}
M.~Ye, Y.~Qian, and J.~Zhou, ``Multitask sparse nonnegative matrix
  factorization for joint spectral--spatial hyperspectral imagery denoising,''
  {\em IEEE Trans. Geosci. Remote Sens.}, vol.~53, no.~5, pp.~2621--2639, 2014.

\bibitem{veganzones2015nonnegative}
M.~A. Veganzones, J.~E. Cohen, R.~C. Farias, J.~Chanussot, and P.~Comon,
  ``Nonnegative tensor {CP} decomposition of hyperspectral data,'' {\em IEEE
  Trans. Geosci. Remote Sens.}, vol.~54, no.~5, pp.~2577--2588, 2015.

\bibitem{LRMR}
H.~Zhang, W.~He, L.~Zhang, H.~Shen, and Q.~Yuan, ``Hyperspectral image
  restoration using low-rank matrix recovery,'' {\em IEEE Trans. Geosci. Remote
  Sens.}, vol.~52, no.~8, pp.~4729--4743, 2014.

\bibitem{LRTDTV}
Y.~{Wang}, J.~{Peng}, Q.~{Zhao}, Y.~{Leung}, X.~{Zhao}, and D.~{Meng},
  ``Hyperspectral image restoration via total variation regularized low-rank
  tensor decomposition,'' {\em IEEE J. Sel. Top. Appl. Earth Observ. Remote
  Sens.}, vol.~11, no.~4, pp.~1227--1243, 2018.

\bibitem{LRTF_HR}
F.~Xiong, J.~Zhou, and Y.~Qian, ``Hyperspectral restoration via $\ell _0$
  gradient regularized low-rank tensor factorization,'' {\em IEEE Trans.
  Geosci. Remote Sens.}, vol.~57, no.~12, pp.~10410--10425, 2019.

\bibitem{9372832}
J.~L. {Wang}, T.~Z. {Huang}, X.~L. {Zhao}, T.~X. {Jiang}, and M.~K. {Ng},
  ``Multi-dimensional visual data completion via low-rank tensor representation
  under coupled transform,'' {\em IEEE Trans. Image Process.}, vol.~30,
  pp.~3581--3596, 2021.

\bibitem{9314228}
Y.~Y. {Liu}, X.~L. {Zhao}, Y.~B. {Zheng}, T.~H. {Ma}, and H.~{Zhang},
  ``Hyperspectral image restoration by tensor fibered rank constrained
  optimization and plug-and-play regularization,'' {\em IEEE Trans. Geosci.
  Remote Sens.}, pp.~1--17, 2021.

\bibitem{Wycoff2013HSR}
E.~{Wycoff}, T.-H. {Chan}, K.~{Jia}, W.-K. {Ma}, and Y.~{Ma}, ``A non-negative
  sparse promoting algorithm for high resolution hyperspectral imaging,'' in
  {\em Proc. IEEE Int. Conf. Acoust. Speech Signal Process.}, pp.~1409--1413,
  2013.

\bibitem{HySime}
J.~M. {Bioucas-Dias} and J.~M.~P. {Nascimento}, ``Hyperspectral subspace
  identification,'' {\em IEEE Trans. Geosci. Remote Sens.}, vol.~46, no.~8,
  pp.~2435--2445, 2008.

\bibitem{Yokoya2012HSR}
N.~{Yokoya}, T.~{Yairi}, and A.~{Iwasaki}, ``Coupled nonnegative matrix
  factorization unmixing for hyperspectral and multispectral data fusion,''
  {\em IEEE Trans. Geosci. Remote Sens.}, vol.~50, no.~2, pp.~528--537, 2012.

\bibitem{Aggarwal2016HU}
H.~K. {Aggarwal} and A.~{Majumdar}, ``Hyperspectral unmixing in the presence of
  mixed noise using joint-sparsity and total variation,'' {\em IEEE J. Sel.
  Topics Appl. Earth Observ. Remote Sens.}, vol.~9, no.~9, pp.~4257--4266,
  2016.

\bibitem{Qian2017Unmixing}
Y.~{Qian}, F.~{Xiong}, S.~{Zeng}, J.~{Zhou}, and Y.~Y. {Tang}, ``Matrix-vector
  nonnegative tensor factorization for blind unmixing of hyperspectral
  imagery,'' {\em IEEE Trans. Geosci. Remote Sens.}, vol.~55, no.~3,
  pp.~1776--1792, 2017.

\bibitem{Xiong2019HUBTD}
F.~{Xiong}, Y.~{Qian}, J.~{Zhou}, and Y.-Y. {Tang}, ``Hyperspectral unmixing
  via total variation regularized nonnegative tensor factorization,'' {\em IEEE
  Trans. Geosci. Remote Sens.}, vol.~57, no.~4, pp.~2341--2357, 2019.

\bibitem{7410766}
C.~{Lanaras}, E.~{Baltsavias}, and K.~{Schindler}, ``Hyperspectral
  super-resolution by coupled spectral unmixing,'' in {\em Proc. IEEE Int.
  Conf. Comput. Vis.}, pp.~3586--3594, 2015.

\bibitem{8075419}
L.~{Loncan}, J.~{Chanussot}, S.~{Fabre}, and X.~{Briottet}, ``Hyperspectral
  pansharpening based on unmixing techniques,'' in {\em Workshop Hyperspectral
  Image Signal Proces.: Evol. Remote Sens.}, pp.~1--4, 2015.

\bibitem{7494936}
A.~{Karami}, R.~{Heylen}, and P.~{Scheunders}, ``Hyperspectral image
  compression optimized for spectral unmixing,'' {\em IEEE Trans. Geosci.
  Remote Sens.}, vol.~54, no.~10, pp.~5884--5894, 2016.

\bibitem{8077483}
Y.~{Zhao}, J.~{Yang}, C.~{Yi}, and Y.~{Liu}, ``Joint denoising and unmixing for
  hyperspectral image,'' in {\em Workshop Hyperspectral Image Signal Proces.:
  Evol. Remote Sens.}, pp.~1--4, 2014.

\bibitem{unet}
O.~Ronneberger, P.~Fischer, and T.~Brox, ``U-{N}et: Convolutional networks for
  biomedical image segmentation,'' in {\em Medical Image Computing and
  Computer-Assisted Intervention}, pp.~234--241, Springer International
  Publishing, 2015.

\bibitem{Adam}
D.~P. Kingma and J.~Ba, ``Adam: A method for stochastic optimization,'' in {\em
  Proc. Int. Conf. Learn. Representations}, 2015.

\bibitem{BP}
R.~Rojas, {\em The Backpropagation Algorithm}, pp.~149--182.
\newblock Berlin, Heidelberg: Springer Berlin Heidelberg, 1996.

\bibitem{soft}
D.~L. {Donoho}, ``De-noising by soft-thresholding,'' {\em IEEE Trans. Inf.
  Theory}, vol.~41, no.~3, pp.~613--627, 1995.

\bibitem{Meisam}
Q.~{Shi}, H.~{Sun}, S.~{Lu}, M.~{Hong}, and M.~{Razaviyayn}, ``Inexact block
  coordinate descent methods for symmetric nonnegative matrix factorization,''
  {\em IEEE Trans. Signal Process.}, vol.~65, no.~22, pp.~5995--6008, 2017.

\bibitem{snr}
H.~{Othman} and {Shen-En Qian}, ``Noise reduction of hyperspectral imagery
  using hybrid spatial-spectral derivative-domain wavelet shrinkage,'' {\em
  IEEE Trans. Geosci. Remote Sens.}, vol.~44, no.~2, pp.~397--408, 2006.

\bibitem{7055246}
X.~Fu, W.-K. Ma, T.-H. Chan, and J.~M. Bioucas-Dias, ``Self-dictionary sparse
  regression for hyperspectral unmixing: Greedy pursuit and pure pixel search
  are related,'' {\em IEEE J. Sel. Topics Signal Process.}, vol.~9, no.~6,
  pp.~1128--1141, 2015.

\bibitem{compress}
S.~{Ge}, Z.~{Luo}, S.~{Zhao}, X.~{Jin}, and X.~{Zhang}, ``Compressing deep
  neural networks for efficient visual inference,'' in {\em Proc. IEEE Int.
  Conf. Multimedia Expo}, pp.~667--672, 2017.

\bibitem{8760540}
B.~Rasti, P.~Ghamisi, and J.~A. Benediktsson, ``Hyperspectral mixed gaussian
  and sparse noise reduction,'' {\em IEEE Geosci. Remote Sens. Lett.}, vol.~17,
  no.~3, pp.~474--478, 2020.

\bibitem{8413118}
S.~Li, Q.~Hao, G.~Gao, and X.~Kang, ``The effect of ground truth on performance
  evaluation of hyperspectral image classification,'' {\em IEEE Trans. Geosci.
  Remote Sens.}, vol.~56, no.~12, pp.~7195--7206, 2018.

\end{thebibliography}
	\bibliographystyle{ieeetr}
	
	\begin{IEEEbiography}[{\includegraphics[width=1in,height=1.25in,clip,keepaspectratio]{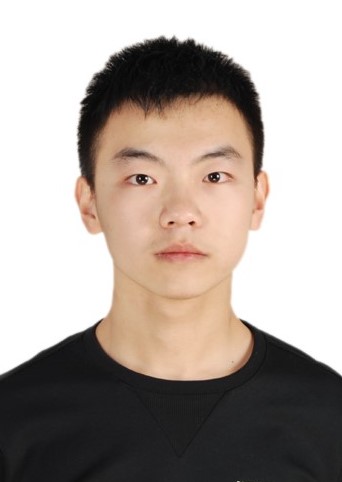}}]{Yu-Chun Miao} is currently a undergraduate majoring in Mathematics and Applied Mathematics from the University of Electronic Science and Technology of China (UESTC), Chengdu, China. His research interests include tensor learning and high-dimensional image processing.
	\end{IEEEbiography}
	
	\begin{IEEEbiography}[{\includegraphics[width=1in,height=1.25in,clip,keepaspectratio]{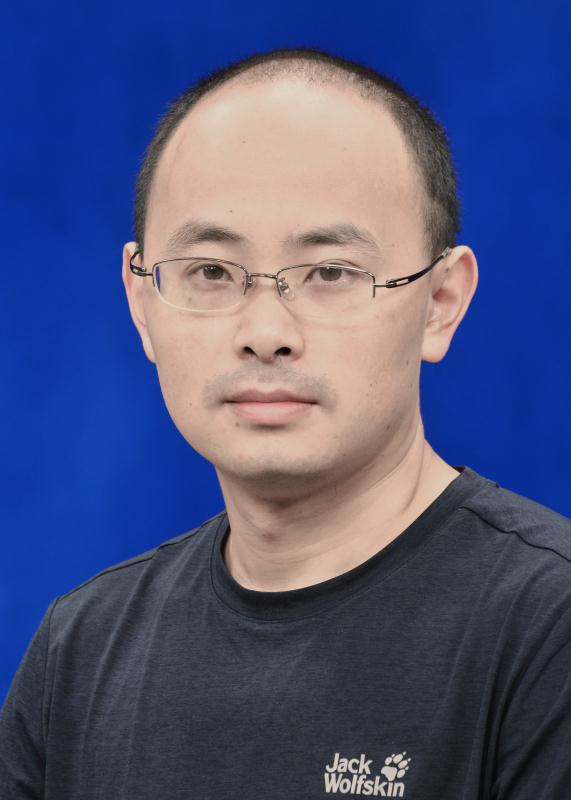}}]{Xi-Le Zhao} received the M.S. and Ph.D. degrees from the University of Electronic Science and Technology of China (UESTC), Chengdu, China, in 2009 and 2012. He is currently a Professor with the School of Mathematical Sciences, UESTC. His research interest mainly focuses on model-driven and data-driven methods for image processing problems. His homepage is \url{https://zhaoxile.github.io/}.
	\end{IEEEbiography}
	
	\begin{IEEEbiography}[{\includegraphics[width=1in,height=1.25in,clip,keepaspectratio]{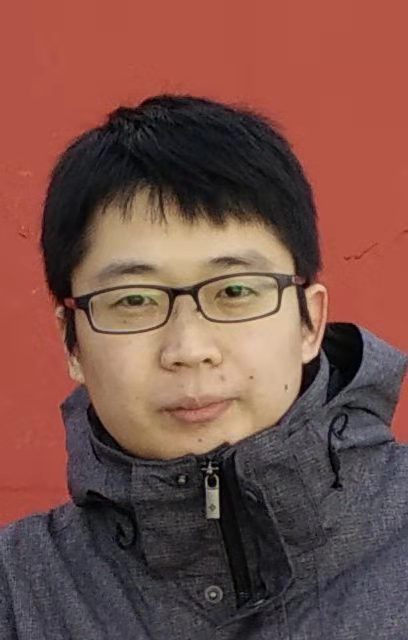}}]{Xiao Fu} (Senior Member, IEEE) received the B.Eng. and M.Sc. degrees from the University of Electronic Science and Technology of China (UESTC), Chengdu, China, in 2005 and 2010, respectively. He received the Ph.D. degree in Electronic Engineering from The Chinese University of Hong Kong (CUHK), Shatin, N.T., Hong Kong, in 2014. He was a Postdoctoral Associate with the Department of Electrical and Computer Engineering, University of Minnesota, Minneapolis, MN, USA, from 2014 to 2017. Since 2017, he has been an Assistant Professor with the School of Electrical Engineering and Computer Science, Oregon State University, Corvallis, OR, USA. His research interests include the broad area of signal processing and machine learning. 
		
		Dr. Fu received a Best Student Paper Award at ICASSP 2014, and was a recipient of the Outstanding Postdoctoral Scholar Award at University of Minnesota in 2016. His coauthored papers received Best Student Paper Awards from IEEE CAMSAP 2015 and IEEE MLSP 2019, respectively. He serves as a member of the Sensor Array and Multichannel Technical Committee  (SAM-TC) of the IEEE Signal Processing Society (SPS). He is also a member of the Signal Processing for Multisensor Systems Technical Area Committee (SPMuS-TAC) of EURASIP. He is the Treasurer of the IEEE SPS Oregon Chapter. He serves as an Editor of {\sc Signal Processing}. He was a tutorial speaker at ICASSP 2017 and SIAM Conference on Applied Linear Algebra 2021.
	\end{IEEEbiography}
	
	\begin{IEEEbiography}[{\includegraphics[width=1in,height=1.25in,clip,keepaspectratio]{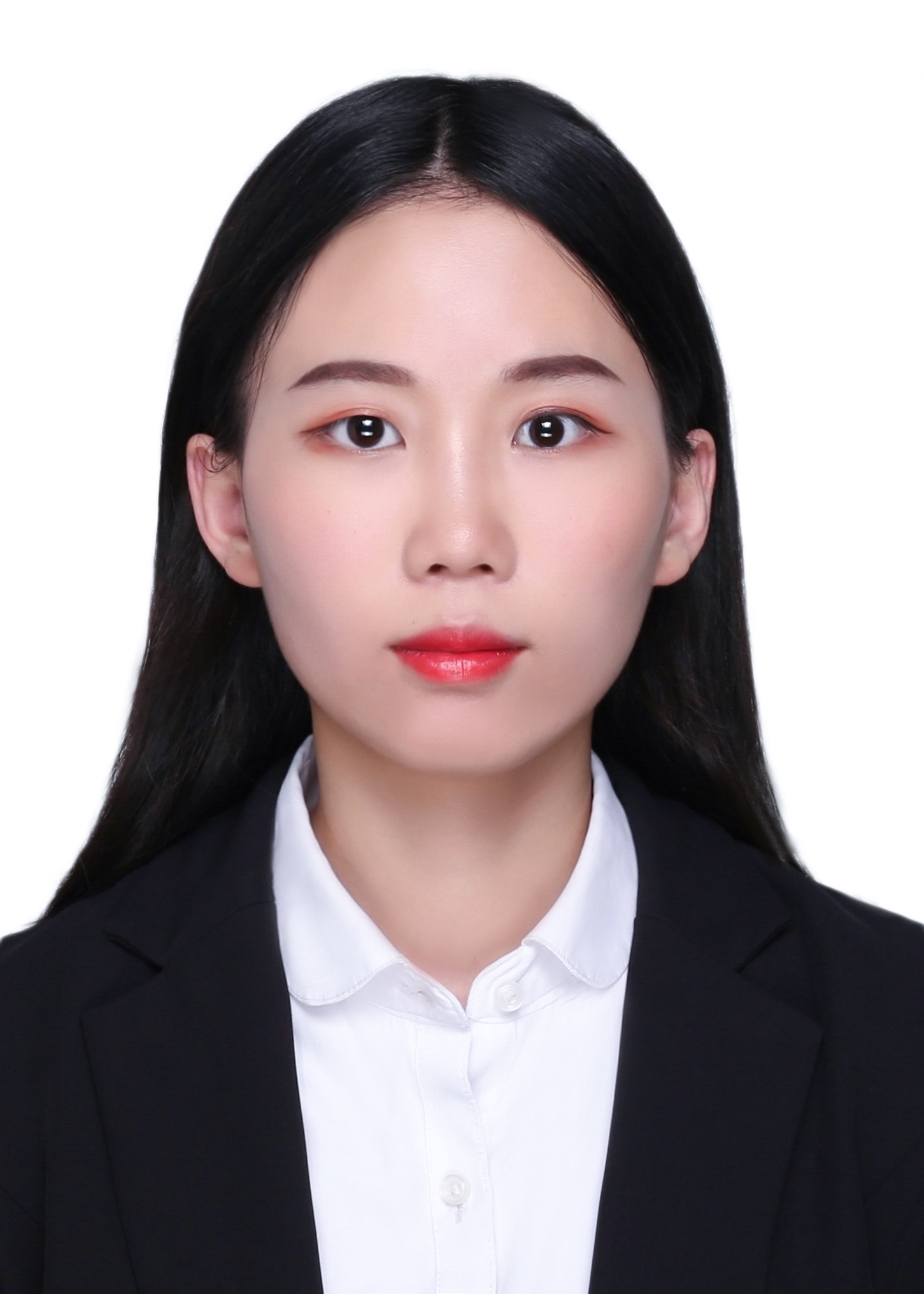}}]{Jian-Li Wang} received the B.S. degree in mathematics and applied mathematics from Neijiang Normal University, Neijiang, China, in 2017. She is currently working toward the Ph.D. degree in the School of Mathematical Sciences, University of Electronic Science and Technology of China, Chengdu, China. She is also a visiting student with the School of Electrical and Electronic Engineering (EEE), Nanyang Technological University (NTU), Singapore. Her current research interests include high-dimensional data processing, tensor modeling and computing, computer vision, and deep learning. More information can be found in her homepage \url{https://wangjianli123.github.io/homepage/}.
	\end{IEEEbiography}

	\begin{IEEEbiography}[{\includegraphics[width=1in,height=1.25in,clip,keepaspectratio]{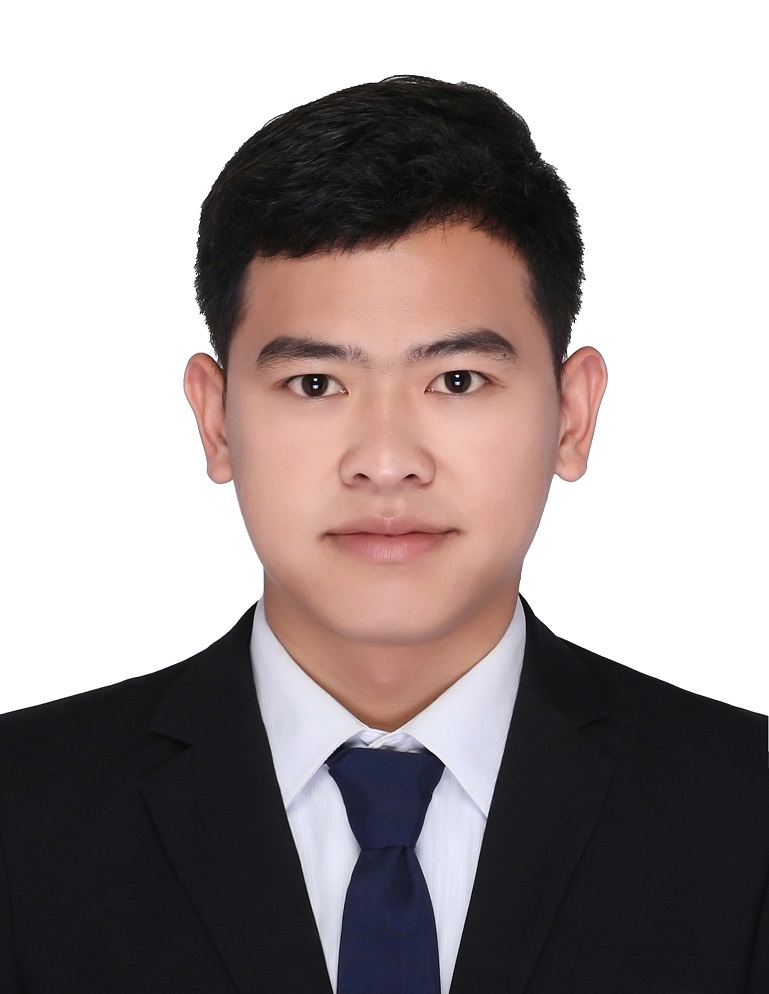}}]{Yu-Bang Zheng} received the B.S. degree in information and computing science from Anhui University of Finance and Economics, Bengbu, China, in 2017. He is currently working toward the Ph.D. degree in the School of Mathematical Sciences, University of Electronic Science and Technology of China, Chengdu, China. He is also a visiting student with the Tensor Learning Team, RIKEN Center for Advanced Intelligence Project, Tokyo, Japan. His current research interests include tensor modeling and computing, tensor learning, and high-dimensional data processing. More information can be found in his homepage \url{https://yubangzheng.github.io/}.
	\end{IEEEbiography}	
	
\end{document}